\pdfoutput=1
\documentclass[11pt]{article}
\usepackage[a4paper,margin=2.35cm]{geometry}
\usepackage[T1]{fontenc}
\DeclareUnicodeCharacter{03B8}{\ensuremath{\theta}}
\usepackage{newtxtext,newtxmath}
\usepackage{microtype}
\usepackage{graphicx}
\usepackage{booktabs,array}
\usepackage{amsmath}
\usepackage{siunitx}
\usepackage[super,sort&compress]{natbib}
\usepackage{xcolor}
\usepackage[hidelinks]{hyperref}
\usepackage[font=small,labelfont=bf]{caption}
\usepackage{authblk}
\usepackage{titlesec}
\usepackage{enumitem}
\usepackage{placeins}

\graphicspath{{figures/}}
\titleformat{\section}{\large\bfseries}{\thesection.}{0.5em}{}
\titleformat{\subsection}{\normalsize\bfseries}{\thesubsection.}{0.5em}{}
\renewcommand{\arraystretch}{1.18}
\title{A general sharp-interface simulation framework for microstructure evolution}
\author[1,*]{Albert Linda}
\author[1,*]{Dierk Raabe}
\affil[1]{Max Planck Institute for Sustainable Materials GmbH, D\"usseldorf, Germany}
\affil[*]{Correspondence: a.linda@mpi-susmat.de; d.raabe@mpi-susmat.de}
\date{}

\begin{document}
\maketitle

\begin{abstract}
Simulating microstructure evolution requires tracking many moving grain and phase boundaries coupled to thermodynamic and mechanical fields. We present Gaussian-projected kinetics (GPK), a sharp-interface framework in which every cell stores an integer grain or phase label. Near a boundary, a temporary Gaussian projection of the label map provides the interface normal and curvature and is then discarded. Capillarity, chemistry, stored energy, transport and elasticity move the boundaries through one topology-preserving update. Because the Gaussian measurement length is fixed in physical units, mesh refinement keeps a consistent capillary scale without a diffuse interface. GPK models grain growth, recrystallisation, multiphase evolution and elastic-coupled solidification in two and three dimensions. It is verified on analytical benchmarks and compared grain by grain with phase-field simulations, running 8.5 to 96 times faster at matched accuracy in our tests. An Al--Cu application with CALPHAD thermodynamic data connects it to measured recrystallisation and precipitation data.
\end{abstract}

\section{Introduction}
Microstructure evolution determines many structure--property relationships in metals and alloys. Important examples are isotropic and anisotropic grain growth, recrystallisation and multiphase boundary migration \citep{HumphreysHatherly2004,Khachaturyan1983}. All of them combine capillarity with chemical, stored-energy or mechanical driving forces. Computational methods differ mainly in how they represent the interfaces and how they enforce the local force balance at them \citep{Raabe1998,ProvatasElder2010}.

Diffuse-interface phase-field models are among the most general approaches for considering such microstructure scenarios with multiple interacting forces at mesoscopic scale \citep{Chen2002,ChenZhao2022,Tourret2022}. In a diffuse-interface model, the boundary between two grains or phases is not a sharp surface. Instead, a continuous order parameter changes smoothly from its value in one grain or phase to its value in the other. This change takes place over a finite width, the interface thickness. The interfacial energy and the boundary motion are computed from spatial derivatives of this smooth profile. These derivatives are accurate only when the computational grid places several points across the interface thickness. The interface thickness therefore often sets the grid spacing, and hence the computational cost. The size of the microstructural features of interest is often much larger. Phase-field models encode interfacial thermodynamics in continuous order parameters. They couple naturally to conserved and mechanical fields \citep{CahnHilliard1958,AllenCahn1979,Chen2002,Steinbach2009,Moelans2008}. Relevant applications include multiphase formulations \citep{SteinbachPezzolla1996} and grain growth in two and three dimensions \citep{FanChen1997,KrillChen2002}. They also include recrystallisation and chemically or elastically coupled phase evolution \citep{ChenKhachaturyan1991,WangJinKhachaturyan2002}. Quantitative calculations must resolve a finite interface thickness in phase-field model frameworks. In addition, polycrystalline models have to evolve by tracking multiple (grain-specific) order parameters. This renders three-dimensional and/or many-grain calculations computationally very costly and conceptually demanding.

Grid-based sharp-interface methods offer a different balance. They divide the specimen into computational cells. Each cell receives a discrete label, such as ``grain 17'' or the ``aluminium-rich phase''. Simple local transition rules can then move a boundary by changing a cell's label. This representation is computationally economical. It can also produce complex collective microstructure evolution. The rules must, however, be designed carefully. The reason is that a pixelated boundary does not directly provide a curvature or a unique normal direction. Cellular automata are a widely used example in recrystallisation modelling \citep{HesselbarthGobel1991,RollettRaabe2001,Raabe2002}. Their neighborhood rules are conceptually simple. They can, however, make curvature, anisotropy and thermodynamic calibration difficult to control, and they can introduce grid dependence. This renders proper time-step scaling challenging. It also makes it difficult to consider configurational and scalar driving forces at the same time. Level-set methods instead represent a boundary as the zero contour of a continuous, distance-like field. They provide accurate sharp-interface motion and kinetic evolution for grain growth and recrystallisation \citep{Sethian1999,Bernacki2011}. Threshold dynamics alternates convolution and relabelling to reproduce mean-curvature flow \citep{Merriman1992,EsedogluOtto2015}. These methods avoid resolving a physical diffuse layer. Additional machinery is, however, needed when several phases, thermodynamic driving forces and cell-wise mechanical fields interact.

\begin{figure}[!htbp]
  \centering
  \includegraphics[width=0.98\textwidth]{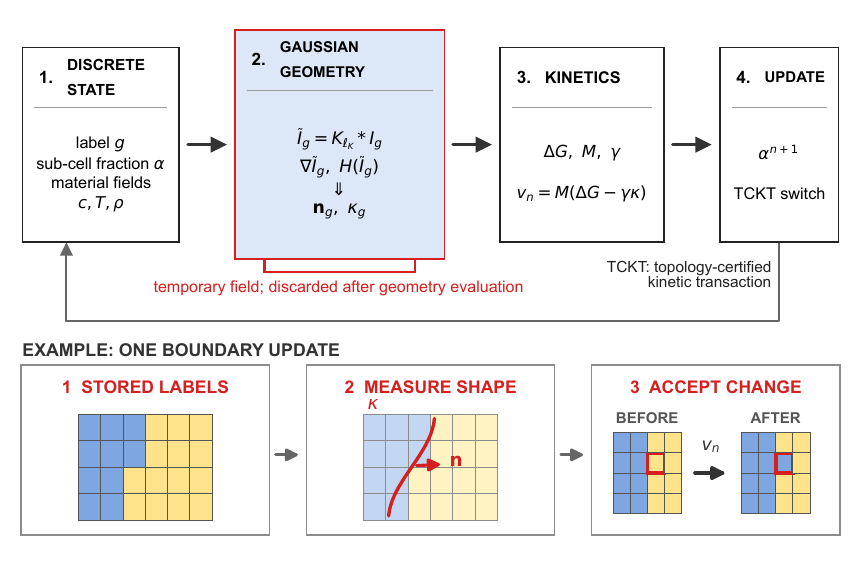}
  \caption{The GPK modelling cycle and one grid-level boundary update. The upper row shows the four computational stages. The lower row follows a boundary from stored grain labels, through a temporary Gaussian measurement of its shape, to an accepted cell-label change. Symbols: $g$ denotes the integer grain or phase label stored in each cell. $\alpha$ denotes the sub-cell fraction that locates the boundary within an interfacial cell. $c$ denotes the composition. $T$ denotes the temperature. $\rho$ denotes the dislocation density. $I_g$ denotes the binary indicator of label $g$ (one in cells carrying $g$ and zero elsewhere). $K_{\ell_K}$ denotes the Gaussian smoothing kernel with physical width (measurement length) $\ell_K$. $*$ denotes convolution. $\widetilde I_g=K_{\ell_K}*I_g$ denotes the temporary smooth field. $\nabla\widetilde I_g$ denotes its gradient (vector of first spatial derivatives). $H(\widetilde I_g)$ denotes its Hessian (matrix of second spatial derivatives). $\mathbf n_g$ (or $\mathbf n$) denotes the outward unit normal of the boundary of $g$. $\kappa_g$ (or $\kappa$) denotes the local boundary curvature, that is, the sum of the principal curvatures. $\Delta G$ denotes the free-energy difference that drives the boundary. $M$ denotes the boundary mobility. $\gamma$ denotes the interfacial energy. $v_n$ denotes the velocity of the boundary along its normal. $\alpha^{n+1}$ denotes the updated sub-cell fraction after timestep $n$. TCKT denotes the topology-certified kinetic transaction that checks a proposed label change (``switch'') before it is accepted.}
  \label{fig:overview}
\end{figure}

We develop Gaussian-projected kinetics (GPK) to combine the strengths of both approaches. GPK keeps the compact state of a grid-based label method. It adds the smooth geometry and the explicit material laws that are familiar from continuum models. The central idea is to separate what the microstructure \emph{is} from how its boundary geometry is \emph{measured}. Integer labels preserve grain or phase identity from one timestep to the next. A temporary Gaussian projection, computed separately for each label, converts the stepped boundary into a smooth measurement field. A sub-cell fraction records where the sharp boundary lies inside an interfacial cell. This fraction does not depend on the Gaussian width. Interchangeable material modules then provide the mobility, the free-energy difference and the interfacial energy. The projection therefore tackles the numerical capillarity problem. It supplies a stable local interface normal and curvature. It does not replace the constitutive model, and it does not prescribe how far the boundary moves. This is an important difference from Gaussian threshold dynamics. There, convolution and thresholding together define the curvature-flow timestep. In GPK, convolution only measures geometry, and a separate physical kinetic law determines the displacement. Figure~\ref{fig:overview} summarises this distinction and exemplifies a complete modelling cycle.

The study addresses three questions. First, can a common data structure hold grain and phase identity, optional material fields and temporary interface geometry, with one strictly defined role for each? In particular, the identity labels and material fields should be the only information carried from one timestep to the next. The temporary geometry field should only measure the boundary shape. It should never change a label or a material field by itself. Second, can one kinetic law and one Gaussian geometry measurement be stated with unambiguous signs, units and resolution limits? Third, can different material processes use the same geometry and update engine in both two and three dimensions? The evaluation therefore tests the complete modelling framework. It does not test the Gaussian function or the curvature formula in isolation.

The Results part begins with a plain description of two kinds of information: the information stored by the model and the information it reconstructs temporarily. We then derive the boundary-velocity law and the Gaussian geometry measurement. Next, we verify each numerical component on problems with a known outcome. Finally, we apply the combined framework to complete microstructure simulations.

We use four terms with specific meanings. \emph{Verification} compares an implemented calculation with an analytical identity or a controlled numerical reference. \emph{Validation} compares a model output with an observation that had not been used to calibrate or otherwise determine that output. \emph{Calibration} estimates an otherwise unknown parameter from a stated datum. Hence, agreement with that same datum is not validation. \emph{Conditioning} fixes part of a trajectory or population from observed data. Other quantities are left free for comparison. A \emph{qualitative demonstration} shows that a specific model variant produces the expected type of microstructure. Each model variant explicitly defines which variables, data fields and dependencies enter its driving forces. Examples are the stored energy, the composition or the elastic strain. The expected microstructure is, for example, a network of equiaxed grains in isotropic grain growth. In recrystallisation, it is a set of new grains that grow until they impinge. Such a demonstration does not carry a numerical accuracy claim. These categories keep three kinds of evidence apart: a code check, a fitted endpoint and an independent material comparison. The Supporting Information gives the complete test inventory, numerical tolerances, parameter sources and claim boundaries.

Figure~\ref{fig:overview} represents a closed update cycle, not a sequence of independent models. The persistent state is the only information carried from one timestep to the next. Geometry is reconstructed from that state. The constitutive module then evaluates the driving force between the resident and candidate regions. Finally, the common update returns the accepted sub-cell motion and label changes to the state. The return arrow highlights one point. Grain growth, recrystallisation, multiphase growth and elastic coupling differ only in the type of fields that enter the constitutive box. They do not need separate curvature estimators or topology rules. The diagram thus shows which part of GPK is reusable and where material-specific assumptions enter.

\section{Results}
\subsection{A general Gaussian-projected kinetics architecture for microstructure evolution}
GPK represents a microstructure model as four connected layers (Figure~\ref{fig:overview} and Table~\ref{tab:model-contract}). A reader can follow the layers as a repeated question-and-answer sequence. First, \emph{what material occupies each cell?} The persistent state answers this with grain or phase labels and any required material variables. Second, \emph{what is the local boundary shape?} The temporary geometry layer reconstructs a normal and a curvature. It does this only near cells whose neighbours have different labels. Third, \emph{how fast should that boundary move?} The constitutive layer combines the local material state with the resident and candidate labels. From these, it calculates the mobility $M$, the free-energy difference $\Delta G$ and the interfacial energy $\gamma$. Together with the measured curvature $\kappa$, these quantities give the final normal velocity $v_n$ of the interface segment. This is done through the kinetic law derived below (Equation~\ref{eq:master}). Fourth, \emph{which cell-level changes represent that motion without creating numerical holes or splits?} The update layer advances the sub-cell position and resolves competing candidates. It also checks every proposed label change. This separation allows the same geometry calculation to serve several material processes. At the same time, the thermodynamic assumptions of each process remain explicit and replaceable.

\begin{table}[!htbp]
\centering
\caption{The four layers of the GPK model and the rules that connect them. An additional process changes the state variables and constitutive functions that it defines, while retaining the geometry and update layers. Symbols are given in brackets after each quantity.}
\label{tab:model-contract}
\small
\renewcommand{\arraystretch}{1.18}
\begin{tabular}{@{}>{\raggedright\arraybackslash}p{0.16\textwidth}>{\raggedright\arraybackslash}p{0.25\textwidth}>{\raggedright\arraybackslash}p{0.31\textwidth}>{\raggedright\arraybackslash}p{0.17\textwidth}@{}}
\toprule
Layer & Principal quantities & Responsibility & Process dependence \\
\midrule
\textbf{Persistent state} & Grain or phase label ($g$), sub-cell fraction ($\alpha$), optional composition ($c$), temperature ($T$), dislocation density ($\rho$) and orientation & Region identity, conserved or internal variables, sub-cell interface position & Choice of material fields \\
\textbf{Temporary geometry} & Binary indicator ($I_g$), Gaussian-smoothed indicator ($K_{\ell_K}*I_g$), outward normal ($\mathbf n_g$), curvature ($\kappa_g$) & Differentiable geometry from discrete topology & Resolution and kernel scale only \\
\textbf{Constitutive physics} & Mobility ($M_{pq}$), free-energy difference ($\Delta G_{q-p}$), interfacial energy ($\gamma_{pq}(\mathbf n)$), transport coefficients such as diffusivity ($D$) & Thermodynamic driving force, capillarity, mobility and bulk coupling & Process-specific module \\
\textbf{Kinetics and topology} & Normal velocity ($v_n$), trial sub-cell fraction ($\alpha^*$), competitor selection and label switch & Interface displacement and consistent topology change & Shared algorithm \\
\bottomrule
\end{tabular}
\end{table}

This separation does not imply that all boundary-driven processes have identical physics. It means that the same numerical measurement of boundary shape can be combined with different material laws. In isotropic grain growth, the bulk free-energy difference is zero ($\Delta G=0$), and the interfacial energy $\gamma$ is constant. Anisotropic growth allows the mobility $M$ or the interfacial energy $\gamma$ to depend on the orientation. This can be the grain-to-grain misorientation or the boundary orientation relative to the crystal axes. Recrystallisation adds the stored deformation energy associated with dislocations. Multiphase growth compares the Gibbs energies of competing phases. Elastic coupling adds the change in mechanical energy. These inputs change why and how fast a boundary moves. They do not change how GPK detects the boundary or measures its curvature.

Stored and temporary quantities have different roles. The label $g$, the sub-cell coordinate $\alpha$ and the material fields $\mathbf u$ are carried from one timestep to the next. Other quantities exist only while the current state $S^n$ is being evaluated. These are the indicator $I_g$, the smooth projection $\widetilde I_g$, the normal $\mathbf n_g$, the curvature $\kappa_g$, the candidate label and the pair velocity. The superscript $n$ identifies the timestep. The outputs $M_{pq}$, $\Delta G_{q-p}$ and $\gamma_{pq}$ apply to one boundary between two regions. The \emph{resident} region $p$ is the grain or phase that currently occupies the cell. The \emph{candidate} region $q$ is a neighbouring grain or phase that could replace $p$ in this cell if the boundary moves across it. It is called a candidate because it is only a possible new owner of the cell. It becomes the owner only when the update step accepts the change. The three outputs are calculated by the selected material model. They are not used as hidden global fitting factors. Changing the material model therefore cannot silently change how an interface is detected or updated.

\paragraph{Sign conventions.}
Many results below depend on the signs of normals, curvatures and velocities. We therefore fix these signs once here (Table~\ref{tab:signs}). Let $p$ be the resident region that currently occupies a cell, and let $q$ be the neighbouring candidate region. The normal $\mathbf n_p$ always points outward from $p$ into $q$. The curvature $\kappa_p$ is positive when $p$ is locally convex. An example is a circular grain $p$ enclosed by $q$. A positive normal velocity $v_n$ means that $p$ grows. A negative value means that $p$ shrinks. Finally, $\Delta G_{q-p}$ is the free-energy density of the candidate minus that of the resident. A lower-energy candidate therefore gives $\Delta G_{q-p}<0$ and makes the resident recede. The same conventions are used in every derivation, in the implementation and in every process module.

\begin{table}[!htbp]
\centering
\caption{Sign and unit conventions for normals, curvatures and velocities. They are used in all derivations, in the implementation and in every process module.}
\label{tab:signs}
\small
\begin{tabular}{@{}lll@{}}
\toprule
Symbol & Definition & Consequence \\
\midrule
$\mathbf n_p$ & Outward unit normal of resident $p$ & Points from $p$ to $q$ \\
$\kappa_p=\nabla\!\cdot\!\mathbf n_p$ & Curvature (sum of principal curvatures) & $+1/R$ for a 2-D convex $p$ of radius $R$ \\
$v_n$ & Normal velocity along $\mathbf n_p$ & $v_n>0$ expands $p$ \\
$\Delta G_{q-p}=\psi_q-\psi_p$ & Candidate minus resident free-energy density & Lower-energy $q$ gives $\Delta G<0$ \\
$M$ & Sharp-interface boundary mobility & Unit $[M]=\mathrm{m^4\,J^{-1}\,s^{-1}}$ \\
\bottomrule
\end{tabular}
\end{table}

\paragraph{Where the quantities come from.}
GPK does not determine material parameters from the Gaussian projection. The inputs and computed quantities fall into four distinct classes. The first class contains persistent state variables supplied by the initialization or an auxiliary field solver. The second contains geometric quantities calculated from the current labels. The third contains constitutive quantities obtained from a material relation, a thermodynamic database, a literature value or an explicitly identified calibration. The fourth contains numerical controls selected by resolution or stability tests. Table~\ref{tab:parameter-provenance} gives this provenance at the point where the symbols first enter the algorithm.

Some letters need care. The Gaussian convolution kernel is written $K_{\ell_K}$. It has no thermodynamic meaning. The letter $G$ is reserved for free energies: $G_p(c,T)$ denotes the Gibbs energy of phase $p$ per unit volume, as a function of composition $c$ and temperature $T$. The symbol $\psi_p$ denotes the free-energy density of the resident $p$, including all contributions used by a given process. The label of a grain is always $g$, and its curvature is $\kappa$. The widths $\sigma_K$, $\sigma_r$ and $\sigma_v$ of the Gaussian kernels are numbers of cells. The interfacial energy is always denoted by $\gamma$. The difference between candidate and resident is $\Delta G_{q-p}=\psi_q-\psi_p$, measured in \si{J.m^{-3}}. Thus $\Delta G$ is normally computed from the current material state. It is not adjusted independently at each interface.

\begin{table}[!htbp]
\centering
\caption{Origin, units and role of the principal GPK quantities. Each quantity is named in words and followed by its symbol in brackets. ``Computed'' means recalculated from the current state at every geometry or constitutive evaluation. It does not denote a fitted parameter.}
\label{tab:parameter-provenance}
\small
\begin{tabular}{@{}>{\raggedright\arraybackslash}p{0.20\textwidth}>{\raggedright\arraybackslash}p{0.17\textwidth}>{\raggedright\arraybackslash}p{0.51\textwidth}@{}}
\toprule
Quantity (symbol) & Status and units & How it is obtained \\
\midrule
Label ($g_i$), sub-cell fraction ($\alpha_i$) and material fields ($\mathbf u_i$) of cell $i$ & Stored state & Labels and initial sub-cell fractions come from the initial microstructure. The material-field list $\mathbf u_i$ contains only enabled fields, such as composition ($c$), temperature ($T$), dislocation density ($\rho$), orientation or elastic variables. Each field is initialized or advanced by its own field equation. \\
Gaussian kernel ($K_{\ell_K}$), kernel width in cells ($\sigma_K$), physical measurement length ($\ell_K$) & Numerical geometry; $\ell_K=\sigma_K\Delta x$ [m] & The kernel weights are fixed by Equation~\ref{eq:discrete-gaussian}. The width is selected from the verified range $\Delta x\leq\ell_K\ll R_{\min}$, where $\Delta x$ is the grid spacing and $R_{\min}$ the smallest radius of interest. It is not obtained from material calibration. Refinement changes $\sigma_K$ in cells when necessary to keep $\ell_K$ fixed. \\
Outward normal ($\mathbf n_g$), curvature ($\kappa_g$) & Computed geometry; $\kappa$ [m$^{-1}$] & Evaluated from the gradient and Hessian (matrix of second derivatives) of the smoothed indicator $K_{\ell_K}*I_g$ using Equations~\ref{eq:geometry} and \ref{eq:curvature-hessian}. They are outputs of the geometry module, not user-supplied constants. \\
Free-energy density ($\psi_p$), free-energy difference ($\Delta G_{q-p}$) & Constitutive output; [J\,m$^{-3}$] & $\Delta G_{q-p}=\psi_q-\psi_p$. It is zero between equal-energy grains, follows Equation~\ref{eq:stored-energy} for recrystallisation, Equation~\ref{eq:precipitation-gibbs} or a CALPHAD table for multiphase chemistry, and receives the mechanical contribution in Equation~\ref{eq:elastic-scaling}. \\
Boundary mobility ($M_{pq}$) & Material kinetic input; [m$^4$\,J$^{-1}$\,s$^{-1}$] & Read from a stated material parameter set, evaluated from a temperature/misorientation law such as Equation~\ref{eq:read-shockley-mobility}, or estimated from an explicitly stated independent kinetic datum. Any calibration is identified as such. $M$ is not inferred from $K_{\ell_K}$ or $\kappa$. \\
Interfacial energy ($\gamma_{pq}$, or $\gamma(\mathbf n)$ when orientation dependent) & Interfacial thermodynamic input; [J\,m$^{-2}$] & Taken from a stated boundary/phase-pair value or an orientation-dependent energy relation such as Equations~\ref{eq:cubic-anisotropy} and \ref{eq:fourfold-anisotropy}. Strong precipitate anisotropy uses the variational Cahn--Hoffman form rather than $\gamma\kappa$. \\
Diffusivity ($D$), elastic stiffness tensor ($\mathbf C$), eigenstrain ($\boldsymbol\epsilon^*$) & Material inputs & Diffusivity ($D$), elastic stiffness tensor ($\mathbf C$) and eigenstrain ($\boldsymbol\epsilon^*$, the stress-free transformation strain) are supplied by the selected material module from literature or assessed data. The transport and mechanical solvers convert them and the current fields into fluxes or free-energy contributions. \\
Grid spacing ($\Delta x$), timestep ($\Delta t$), interface and diffusion safety factors ($C_I$, $C_D$) & Numerical controls & $\Delta x$ is chosen from the smallest required physical feature and checked by refinement. $\Delta t$ is computed from Equation~\ref{eq:interface-cfl}. $C_I$ and $C_D$ are stated safety factors verified by stability sweeps, not material parameters. \\
Diffuse-interface thickness ($\xi$) & Derivation only; [m] & The diffuse thickness appears only in the sharp-interface derivation. It cancels from Equation~\ref{eq:master} and is not an input to a GPK simulation. \\
\bottomrule
\end{tabular}
\end{table}

Consequently, the process modules differ only in how they supply the three inputs of the common kinetic law. Isotropic grain growth reads given constant values of $M$ and $\gamma$ and sets $\Delta G=0$. Anisotropic growth evaluates $M$ and/or $\gamma$ from the boundary orientation or the misorientation. Recrystallisation computes $\Delta G$ from stored-energy fields. Multiphase growth computes it from competing Gibbs energies. Elastic coupling adds the change in mechanical energy. The application sections state the origin of each numerical value. A value can be a literature transcription, a database result, a controlled benchmark input, a calibration or a conditioned experimental datum. The complete value-by-value provenance is collected in the Supporting Information and in the associated input files.

When a grain network is initialized synthetically, its topology is made consistent before the first kinetic step. Each seed occupies a unique cell. Each cell then belongs to its nearest seed (a Voronoi construction), measured with periodic distances on the same periodic domain used later for the evolution. Ties in integer distances can leave a diagonal cell that carries the same identity as a grain but is not connected to it. The initializer therefore keeps the largest face-connected piece of each seed identity. Every smaller piece is attached, layer by layer, to an already connected face neighbour. This cleaning step acts only on the initial raster. It transfers the neighbour orientation together with the reassigned cell, and it is complete before the state at $t=0$ is written. Evolved interfaces are never subjected to this operation. Their topology is governed by the kinetic transaction described below.

\FloatBarrier
\subsection{A complete timestep from stored state to committed topology}
\label{sec:timestep}
We now describe one complete timestep. Its input is the persistent state at time $t^n$, which collects everything the model stores:
\begin{equation}
S^n=\left(g_i^n,\ \alpha_i^n,\ \mathbf u_i^n\right)\ \text{for every cell } i=1,\ldots,N,
\qquad
\mathbf u_i^n=(c_i^n,T_i^n,\rho_i^n,\text{orientation}_i,\ldots).
\label{eq:state-vector}
\end{equation}
Here $S^n$ denotes the complete stored state at timestep $n$, and $N$ is the number of grid cells. The subscript $i$ identifies one cell. In that cell, $g_i$ is the resident grain or phase label. The sub-cell fraction $\alpha_i$ is the fraction of the cell occupied by that resident label, measured along the boundary normal. The vector $\mathbf u_i$ lists the material fields required by the selected process, for example the composition $c_i$, the temperature $T_i$ and the dislocation density $\rho_i$. Each cell therefore stores only a label, a fraction and a short list of material values. The next state, $S^{n+1}$, is produced through the following eight operations (Figure~\ref{fig:timestep}):

\begin{figure}[!htbp]
  \centering
  \includegraphics[width=0.90\textwidth]{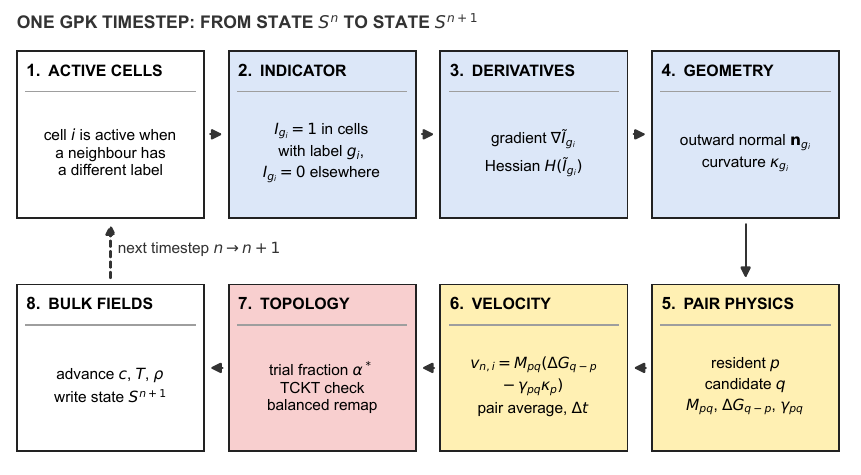}
  \caption{Schematic of the eight operations of one GPK timestep, from the stored state $S^n$ to the next state $S^{n+1}$. Blue boxes measure geometry, yellow boxes evaluate the material physics and the red box changes labels. White boxes hold the stored state. Symbols: $i$ denotes the index of an active (interfacial) cell. $g_i$ denotes its resident label. $I_{g_i}$ denotes the binary indicator of $g_i$. $\widetilde I_{g_i}$ denotes the Gaussian-smoothed indicator. $\nabla$ denotes the gradient. $H$ denotes the Hessian (matrix of second derivatives). $\mathbf n_{g_i}$ denotes the outward unit normal. $\kappa_{g_i}$ and $\kappa_p$ denote the curvature of the resident label. $p$ and $q$ denote the resident and candidate labels. $M_{pq}$ denotes the boundary mobility. $\Delta G_{q-p}$ denotes the free-energy difference. $\gamma_{pq}$ denotes the interfacial energy. $v_{n,i}$ denotes the normal velocity at cell $i$. $\Delta t$ denotes the timestep. $\alpha^*$ denotes the trial sub-cell fraction. TCKT denotes the topology-certified kinetic transaction. $c$, $T$ and $\rho$ denote the composition, temperature and dislocation density.}
  \label{fig:timestep}
\end{figure}

\begin{enumerate}[leftmargin=2.1em,itemsep=0.25em,topsep=0.35em]
\item \textbf{Detect and compact active cells.} Mark cell $i$ as interfacial (active) when at least one neighbour that shares a face or an edge has a different label. Then collect all active cells, in increasing index order, into a compact list. Bulk cells bypass geometry and interface kinetics.
\item \textbf{Condition the local indicator.} At each active cell $i$, take its resident label $g_i$ and evaluate the binary indicator $I_{g_i}$. This indicator equals one in cells that carry label $g_i$ and zero in all other cells. It is evaluated only over the compact Gaussian support, that is, the finite neighbourhood of cell $i$ in which the Gaussian weights are non-zero. A junction of three or more grains is therefore never treated as one combined boundary of all labels. Each resident label sees only its own boundary.
\item \textbf{Apply fused Gaussian derivatives.} Combine (convolve) the one-dimensional Gaussian weights once with the fourth-order finite-difference stencils for the first and second derivatives. Products of the resulting one-dimensional kernels give the gradient $\nabla\widetilde I_{g_i}$ of the smoothed indicator $\widetilde I_{g_i}$ directly at the active cell. They also give its Hessian $H(\widetilde I_{g_i})$, the matrix of second derivatives. A full smoothed field is never stored.
\item \textbf{Reconstruct geometry.} Evaluate the outward unit normal $\mathbf n_{g_i}$ and the curvature $\kappa_{g_i}$, the sum of the principal curvatures, from this gradient and Hessian. The sign convention of Table~\ref{tab:signs} is fixed once for all process modules.
\item \textbf{Evaluate pair-wise physics.} For each active cell with resident label $p$, list the different labels $q$ of its neighbours. These are the candidates. The selected constitutive module returns the mobility $M_{pq}$, the free-energy difference $\Delta G_{q-p}$ and the interfacial energy $\gamma_{pq}$ for each pair. An optional phase-selection step chooses the thermodynamically admissible candidate.
\item \textbf{Compute and project motion.} Compute the normal velocity of the interface at cell $i$, $v_{n,i}=M_{pq}(\Delta G_{q-p}-\gamma_{pq}\kappa_p)$. Here $\kappa_p$ is the curvature measured for the resident label $p$. Then average (project) that velocity separately on each active label pair with a Gaussian of width $\sigma_v$. This width is 0.85 cell in two dimensions and up to 3 cells in three dimensions. The reasons for these values are given below. Finally, choose the timestep $\Delta t$ from the interface-crossing bound and, when enabled, the bulk-transport bound.
\item \textbf{Certify, commit and anti-alias.} First compute every trial fraction $\alpha^*$ from the unchanged state $S^n$. A trial fraction may cross zero for a persistent region identity, such as an individual grain. Such a change must then pass the connectivity rules of the topology-certified kinetic transaction (TCKT) described below. A phase-class label may change its number of separate pieces through resolved coalescence or pinch-off. After the accepted crossings, a Gaussian exchange removes unresolved one-cell protrusion/indentation pairs (anti-aliasing). This exchange preserves the number of cells of every label exactly. It is also topology certified.
\item \textbf{Advance bulk fields.} Apply the selected diffusion, recovery or other bulk-field modules, write $S^{n+1}$ and repeat until the requested physical time is reached.
\end{enumerate}

\begin{figure}[!htbp]
  \centering
  \includegraphics[width=0.98\textwidth]{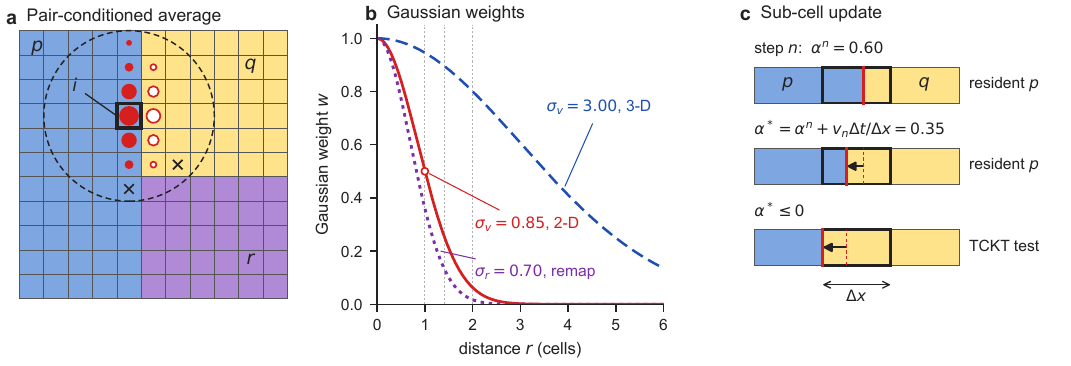}
  \caption{Velocity averaging and sub-cell update in one timestep. (a) Cells that enter the average for cell $i$ (thick outline) at a junction of the grains $p$ (blue), $q$ (yellow) and $r$ (purple). Only cells on the $p$--$q$ boundary inside the averaging radius (dashed circle) are used. Filled circles mark $s_j=+1$ (resident $p$) and open circles $s_j=-1$ (resident $q$), assuming that $p$ has the smaller label number. The circle size indicates the Gaussian weight. Crosses mark boundary cells of the other two label pairs, which are excluded. (b) Gaussian weight against distance. The curves show the two-dimensional velocity width $\sigma_v=0.85$ cell (solid red) and the remap width $\sigma_r=0.70$ cell (dotted purple). The dashed blue curve is the three-dimensional velocity width $\sigma_v=3.00$ cells for the production kernel $\sigma_K=8$ cells. The open circle marks the weight of one half at one cell. Vertical lines mark a face neighbour, a diagonal neighbour and a cell two away. (c) Sub-cell fraction $\alpha$ of the outlined cell while the resident $p$ recedes. The boundary (red bar) moves by $v_n\Delta t$ in each step, from its previous position (dashed line). When the trial fraction $\alpha^*$ reaches zero, the TCKT decides whether the cell is relabelled to $q$.}
  \label{fig:velocity-subcell}
\end{figure}

Step 6 averages the velocity of each interface cell. It replaces the raw velocity of each interface cell by a weighted average. The average is taken over nearby cells that lie on the same boundary between labels $p$ and $q$. This removes cell-to-cell noise from the velocity. Two complications must be handled. First, the same physical boundary is viewed from opposite resident sides. The raw velocities on the two sides therefore have opposite signs. To combine them without cancellation, GPK gives all cells on the $p$--$q$ boundary one common sign convention. We define the sign $s_i=+1$ when the label $g_i$ of cell $i$ is the smaller of the two label numbers $p$ and $q$. Otherwise, $s_i=-1$. Second, at a junction, where three or more grains meet, cells on other boundaries must not enter the average. Only cells $j$ near cell $i$ that also lie on the $p$--$q$ boundary are therefore included. The average is a Gaussian-weighted mean over these cells, with a width $\sigma_v$ of 0.85 cell in two dimensions and up to 3 cells in three dimensions. The averaged velocity is
\begin{equation}
\widehat v_i=s_i\,\frac{\sum_j w_{ij}\,s_j\,v_j}{\sum_j w_{ij}},
\qquad
w_{ij}=\exp\!\left(-\frac{r_{ij}^2}{2\sigma_v^2}\right).
\label{eq:pair-velocity}
\end{equation}
Both sums run over the cells $j$ within the averaging radius $r_v$ around cell $i$ that lie on the $p$--$q$ boundary. The radius is $r_v=3\sigma_v$, rounded up to a whole number of cells. $\widehat v_i$ is the pair-averaged normal velocity at cell $i$, which enters the sub-cell update and the timestep limit. $v_j$ is the raw normal velocity at cell $j$, computed from Equation~\ref{eq:master}. The weight $w_{ij}$ is a Gaussian of the distance $r_{ij}$ between the two cells, measured in cells. Its width is $\sigma_v=0.85$ cell in two dimensions (in three dimensions, $0.375\sigma_K$, limited to 0.85--3.00 cells). Nearby cells therefore count more than distant ones. Equation~\ref{eq:pair-velocity} is an ordinary weighted mean: the numerator sums the weighted, signed velocities and the denominator sums the weights. The final factor $s_i$ converts the result back to the sign convention of cell $i$. The average is evaluated only on the $N_\Gamma$ interfacial cells. Its cost is therefore proportional to $N_\Gamma r_v^d$, where $d$ is the spatial dimension. Figure~\ref{fig:velocity-subcell}(a) shows which cells enter the average at a junction, and Figure~\ref{fig:velocity-subcell}(b) shows the weights.

\paragraph{Choice of the averaging widths.}
The two-dimensional width $\sigma_v=0.85$ cell is not a fitted parameter. It is the width at which a face neighbour, one cell away, receives exactly half the weight of the centre cell. Setting $\exp[-1/(2\sigma_v^2)]=1/2$ gives $\sigma_v=1/\sqrt{2\ln2}\approx0.85$ cell. A diagonal neighbour then receives one quarter of the centre weight, and a cell two away only 6\%. Beyond the radius $r_v=3\sigma_v$, the weights are below 1.2\% and are omitted. The average therefore removes the scatter between neighbouring cells, but it hardly changes a velocity that varies smoothly along the boundary. A velocity pattern with a wavelength of ten cells, for example, keeps 87\% of its amplitude. A wider average would blur real variations of the curvature, and a narrower one would leave the cell-to-cell scatter. In three dimensions, a boundary is a surface with two tangential directions. A wider average is then needed so that neighbouring cells on one face receive consistent crossing times. The value $\sigma_v=3.00$ cells was selected for the production kernel $\sigma_K=8$ cells on a separate calibration ensemble of 80 grains. Widths of 2, 3 and 4 cells were tested with one or two remap sweeps. Only three cells with two sweeps lowered the boundary-face count in both the isotropic and the anisotropic run. It also lowered the number of poorly supported boundary cells by at least 11\% in both runs (Supplementary Table~S1). These are cells with at most two face neighbours of the same label. The factor $0.375=3/8$ reproduces this value for $\sigma_K=8$ cells and scales the width with the geometry kernel at other resolutions. The lower limit of 0.85 cell keeps the three-dimensional average at least as wide as the two-dimensional one. The upper limit of 3.00 cells keeps it below half the width of the production geometry kernel. The velocity is therefore never averaged over a wider region than the one used to measure the curvature.

The sub-cell update in step 7 follows directly from the interface displacement. Consider a locally planar interface. Let $\ell_p$ be the length of the cell occupied by the resident $p$, measured along the normal. The sub-cell fraction is then $\alpha=\ell_p/\Delta x$, where $\Delta x$ is the cell size. Here $v_n$ is the pair-averaged normal velocity of step 6. When the interface moves with this velocity, the length $\ell_p$ changes at the rate $v_n$. Positive $v_n$ expands $p$, so the fraction changes at the rate
\begin{equation}
\dot\alpha=\frac{v_n}{\Delta x}.
\label{eq:alpha-ode}
\end{equation}
Here the dot denotes the time derivative. During one timestep $\Delta t$, we take $v_n$ as constant (the forward Euler method). The new fraction is then
\begin{equation}
\alpha^*=\alpha^n+\frac{v_n\Delta t}{\Delta x},
\qquad
\alpha^{n+1}=\min(\alpha^*,1).
\label{eq:alpha}
\end{equation}
For example, if $v_n\Delta t$ equals one fifth of $\Delta x$, the fraction increases by 0.2. Figure~\ref{fig:velocity-subcell}(c) follows this update for a receding resident. The fraction is capped at one because a cell cannot be more than fully occupied. If the trial fraction $\alpha^*$ reaches zero or becomes negative, the resident label would leave the cell.

\paragraph{Why the topology test (TCKT) is needed.}
A label change is an all-or-nothing event: one cell switches from grain $p$ to grain $q$. The kinetic law decides only when the boundary has crossed the cell. It has no information about how the cells of a grain are connected. On a grid, however, one cell can be the only link between two parts of a grain, or the only gap between two separate regions. Switching such a cell would split a grain into two pieces that carry the same label, or merge two separate regions into one. These events are artefacts of the pixelated boundary. A smooth boundary that moves by curvature does not split a grain within one step. Spurious splits and mergers would corrupt the quantities that GPK reports: grain counts, grain-size distributions, neighbour counts and the grain-by-grain identity used in the comparison with phase field. GPK therefore treats each crossing as a transaction. It is committed only when it keeps the connectivity of both grains intact, and it is otherwise deferred to the next step. This is the meaning of the name: the topology of each kinetically requested change is certified before it is committed. Physical topology changes, such as the disappearance of a grain or the switching of neighbours at a junction, remain possible.

\begin{figure}[!htbp]
  \centering
  \includegraphics[width=0.98\textwidth]{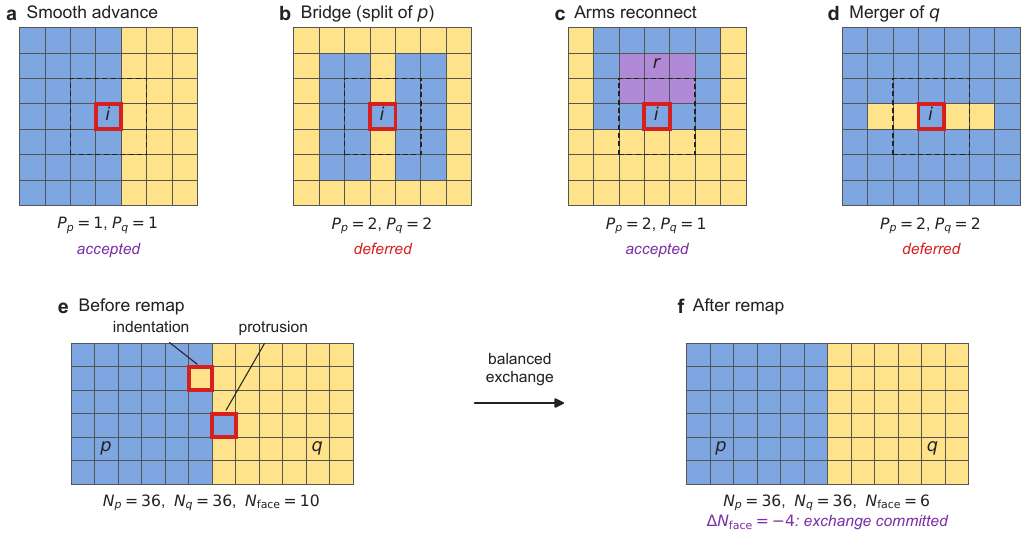}
  \caption{Topology-certified kinetic transaction (TCKT) and balanced Gaussian remap (BGR). (a)--(d) A proposed change of the centre cell $i$ (red outline) from the resident $p$ (blue) to the candidate $q$ (yellow). The dashed square is the $3\times3$ block. $P_p$ and $P_q$ are the numbers of separate face-connected pieces of $p$ and $q$ in this block without the centre cell. (a) A smooth advance leaves one piece of each label and is accepted. (b) Cell $i$ is the only link between two arms of $p$, which would split. The change is deferred. The two pieces of $q$ connect outside the block, so only the resident rule fails. (c) The two local arms of $p$ reconnect around a ring that encloses a third grain $r$ (purple). The change opens the ring and is accepted. (d) Adding cell $i$ would join two separate pieces of $q$, so the change is deferred. The two pieces of $p$ connect around them, so only the candidate rule fails. (e) A one-cell protrusion of $p$ and a one-cell indentation of $q$. (f) After the balanced exchange, the cell counts $N_p$ and $N_q$ are unchanged, and the number of boundary faces $N_{\rm face}$ falls from 10 to 6. All counts in the figure are computed from the drawn label arrays.}
  \label{fig:tckt-bgr}
\end{figure}

GPK applies this test whenever the label denotes a persistent region identity, such as an individual grain. The test asks whether changing this one cell would create an artificial split or merger. It looks at the small block of cells centred on cell $i$. This block contains $3\times3$ cells in two dimensions and $3\times3\times3$ cells in three. Cells are considered connected when they share a face. Each cell thus has four possible neighbours in two dimensions and six in three. We now imagine that the centre cell has been removed from the resident grain $p$. Let $P_p$ be the number of separate pieces of $p$ that remain in the block. Similarly, let $P_q$ be the number of separate pieces of the candidate grain $q$ in the block, again without the centre cell. The change of cell $i$ from $p$ to $q$ is accepted only if both of the following rules hold:
\begin{enumerate}[leftmargin=2.1em,itemsep=0.15em,topsep=0.3em,label=(\roman*)]
\item \emph{Resident rule.} $P_p=1$. Alternatively, $P_p>1$ and all pieces of $p$ reconnect outside the block, or $P_p=0$ and cell $i$ is the last cell of $p$.
\item \emph{Candidate rule.} $P_q=1$. Alternatively, $P_q>1$ and all pieces of $q$ are already connected outside the block, so that adding cell $i$ does not merge separate pieces of $q$.
\end{enumerate}
The first rule prevents an artificial split of $p$, and the second prevents an artificial merger of $q$. A count larger than one means that the local block contains several apparent arms of the same grain. GPK then searches a wider region, still without cell $i$, to determine whether those arms reconnect elsewhere. Arms that reconnect belong to one grain. The change may then represent a resolved cavity, tunnel or neighbour-switching event, and it is accepted. Arms that do not reconnect would produce an artificial split or merger, so the change is postponed. A phase-class label is treated differently, because several disconnected particles may legitimately share one phase label. Physical coalescence or pinch-off may therefore change the number of pieces of that phase. This distinction is set by the meaning assigned to a label in the input, not by the material name. The construction follows concepts of digital topology \citep{KongRosenfeld1989,Bertrand1994}. Unlike a strict local rule, however, it permits resolved physical contact and pinch-off events, which such a rule would freeze. Figure~\ref{fig:tckt-bgr}(a)--(d) shows four typical cases. A smooth advance is accepted, and a change that would split $p$ or merge separate pieces of $q$ is deferred. A ring of $p$ around a third grain may, however, be opened, because its two local arms reconnect around the ring.

Transactions are evaluated in a fixed order. Each proposal sees the state left by the proposals accepted before it. A postponed event keeps the small positive fraction $\alpha=10^{-6}$. This value is also the threshold below which a fraction counts as zero. It lies well above the round-off error of the single-precision arithmetic used for $\alpha$, about $10^{-7}$ for fractions of order one. It is, however, far smaller than any displacement that matters physically. The cell therefore stays at the boundary, and the crossing is attempted again in the next step. Any overshoot beyond the cell is discarded, but it is at most $C_I=1/8$ of a cell per step (Methods).

\paragraph{Why a balanced remap (BGR) is needed.}
Cells switch one at a time, each at its own crossing time, and the TCKT defers some crossings. After the kinetic step, the label field can therefore contain isolated one-cell protrusions, that is, single cells of $p$ that stick into $q$, and matching one-cell indentations. Such defects are smaller than the Gaussian measurement scale. They add cell-scale noise to the measured normal and curvature, increase the pixelated boundary area and can stall an otherwise smooth front. A simple smoothing or majority filter would remove them, but it would also change the volume of each grain and act as an extra, non-physical curvature flow. The balanced Gaussian remap (BGR) removes the defects without either side effect. ``Balanced'' means that protrusions and indentations are exchanged in equal numbers for each label pair, so every label keeps its number of cells. ``Gaussian'' means that the candidates are ranked with a narrow Gaussian score of the local neighbourhood. ``Remap'' means that cells are relabelled, not moved by a velocity. An exchange is accepted only when it shortens the pixelated boundary and passes the topology test.

BGR runs after the kinetic crossings and corrects pixel-scale geometry only. It is not part of the physical kinetics. It uses a narrow Gaussian kernel $K_r$ with a width of $\sigma_r=0.70$ cell in two dimensions. In three dimensions, the width is $\sigma_r=0.125\sigma_K$, but never less than 0.70 cell and never more than 1.00 cell. These values restrict the remap to the immediate neighbourhood of a cell. For $\sigma_r=0.70$ cell, a face neighbour receives 36\% and a diagonal neighbour 13\% of the centre weight, and a cell two away less than 2\% (Figure~\ref{fig:velocity-subcell}(b)). The remap therefore detects one-cell protrusions and indentations, but not the curvature of a resolved boundary. For the same reason, $\sigma_r$ is narrower than $\sigma_v$. The factor $0.125=1/8$ gives $\sigma_r=1$ cell for the production kernel $\sigma_K=8$ cells. The upper limit of 1.00 cell keeps the weight of a cell two away below 14\%. The convolution $K_r*I_q$ is the weighted fraction of neighbouring cells that carry label $q$. Consider an interface cell $i$ that currently carries label $p$. It proposes the label $q$ only when the weighted fraction for $q$ exceeds the fraction $K_r*I_p$ for $p$. Consider a pair of labels $(p,q)$ within the same chemical phase class. For this pair, GPK accepts the same number of the most favourable proposals in the $p\rightarrow q$ and $q\rightarrow p$ directions. Boundaries between different phases are excluded, because their motion must remain controlled by the thermodynamic kinetic law. Consequently, the number of cells $N_p$ of every label $p$ is unchanged:
\begin{equation}
N_p^{\rm after}=N_p^{\rm before}\quad\text{for every label }p.
\label{eq:balanced-remap}
\end{equation}
In effect, a one-cell protrusion is exchanged with a one-cell indentation (Figure~\ref{fig:tckt-bgr}(e) and (f)). To decide whether an exchange improves the boundary, we count the cell faces whose two adjacent cells carry different labels. This count, $N_{\rm face}$, is a simple measure of the pixelated interface area in the periodic domain. An exchange is committed only when it reduces this count ($\Delta N_{\rm face}<0$). Both moves must also pass the local topology test. If the $3^d$ test finds several branches, BGR requires these branches to reconnect inside a slightly larger block of $5^d$ cells. This block is just large enough to contain an oblique boundary that crosses the $3^d$ block. It is too small, however, to permit a distant topology change, which remains the task of the kinetic update. Otherwise, neither move is committed. Three-dimensional steps use a second sweep. The reason is that a surface has two tangential directions, and the first sweep can leave a one-cell ledge in the other direction. Each sweep keeps the number of cells of every label unchanged and strictly decreases $N_{\rm face}$. The remap is therefore isotropic and does not change any volume. It acts only after a kinetic crossing. Its physical width tends to zero as the grid spacing $\Delta x$ tends to zero. It is therefore not an additional growth law. A final resolution check allows very small regions to disappear into their resolved face neighbours. These are regions smaller than a $d$-dimensional ball with a radius of $\sigma_K$ cells. Without this check, the connectivity test would freeze grains that are too small for the Gaussian geometry measurement to represent. As this criterion depends only on geometric resolution, it is independent of material, phase class and crystallographic direction.

Five invariant rules make the algorithm reusable. First, labels and material fields are persistent, whereas Gaussian projections are temporary. Second, constitutive physics enters only through pair-wise functions that each process module defines explicitly. Third, all trial motions are computed from one unchanged state at the start of the step. Fourth, all label changes are checked. Crossings of a region identity use one connectivity test. Crossings of a phase-class label follow the same kinetic law but may change the number of separate particles. Every anti-alias exchange is locally topology certified and exactly balanced in label counts. Fifth, the geometry and interface-update path is identical in two and three dimensions. Only the dimensional stencils and the summed principal curvature change.

\paragraph{What changes between two and three dimensions.}
The kinetic law, the sub-cell update, the TCKT rules and the label-count balance are the same in both cases. Four numerical settings differ. First, the Gaussian kernel is chosen by the same rule, Equation~\ref{eq:kernel-window}, but in three dimensions the two dependent widths scale with it. The velocity average uses $\sigma_v=0.85$ cell in two dimensions and $\sigma_v=0.375\sigma_K$, limited to 0.85--3.00 cells, in three. A surface has two tangential directions, and neighbouring cells along both must receive consistent crossing times. Second, the remap width grows from 0.70 cell to $0.125\sigma_K$, but never more than 1.00 cell. Third, the remap runs one sweep in two dimensions and two sweeps in three. A one-cell ledge on a surface can run along either tangential direction, and the second sweep removes the ledge that the first one leaves. Fourth, the topology test uses a $3\times3$ block, in which each cell has four face neighbours, in two dimensions. In three dimensions it uses a $3\times3\times3$ block with six face neighbours per cell, and the remap reconnection block grows from $5\times5$ to $5\times5\times5$ cells. Finally, the curvature is the sum of two principal curvatures in three dimensions. A sphere of radius $R$ therefore has $\kappa=2/R$, whereas a circle of the same radius has $\kappa=1/R$.

Each accepted step produces two outputs. The first is the updated persistent state $S^{n+1}$. The second is an audit record that counts the accepted, deferred and balanced-exchange events. During initialization, a label can change only through the connectivity cleaning described above. During evolution, it can change only after a sub-cell crossing or through the balanced exchange that follows such a crossing. Gaussian projection alone never relabels a cell. BGR exchanges equal numbers of cells, so it cannot change the volume of any label. This distinction is important when interpreting rendered contours. A smooth line or surface shows the temporary geometry used to measure a categorical interface. In contrast, all topology, phase-fraction and grain statistics are computed from the persistent state and its sub-cell fractions.

\subsection{Thermodynamic foundation of the shared sharp-interface law}
The reduction of a diffuse-interface model to the sharp-interface law used here was developed in several steps. These steps differ in their mathematical rigour. Some are formal approximations, whereas others are rigorous proofs of convergence. Allen and Cahn connected nonconserved order-parameter kinetics to curvature-driven antiphase-boundary motion in 1979 \citep{AllenCahn1979}. Caginalp recovered the Gibbs--Thomson condition for a finite-thickness solidification model in 1986 \citep{Caginalp1986}. Rubinstein, Sternberg and Keller then constructed formal inner expansions leading to curve shortening in 1989 \citep{RubinsteinSternbergKeller1989}.

Rigorous convergence results followed. In 1991, Bronsard and Kohn proved that thin Allen--Cahn layers move by mean curvature when the interface is smooth \citep{BronsardKohn1991}. In 1992, Evans, Soner and Souganidis extended this result to a generalized form of mean-curvature motion \citep{EvansSonerSouganidis1992}. In 1993, Ilmanen proved convergence to a weak form of curvature flow, the Brakke flow \citep{Ilmanen1993}. In 1995, de Mottoni and Schatzman supplied higher-order expansions and error estimates for smooth interfaces \citep{deMottoniSchatzman1995}. Table~\ref{tab:sharp-history} summarizes these results and how they entered materials modelling.

\begin{table}[!t]
\centering
\caption{Milestones and uses of the sharp-interface limit relevant to the present model. ``Formal'' denotes an approximate construction by asymptotic expansion. The later entries are rigorous proofs of different kinds of convergence.}
\label{tab:sharp-history}
\small
\renewcommand{\arraystretch}{1.18}
\begin{tabular}{@{}>{\raggedright\arraybackslash}p{0.08\textwidth}>{\raggedright\arraybackslash}p{0.39\textwidth}>{\raggedright\arraybackslash}p{0.43\textwidth}@{}}
\toprule
Year & Mathematical or physical result & Subsequent role in microstructure modelling \\
\midrule
1979--89 & Curvature-controlled antiphase-boundary kinetics and formal diffuse-layer reduction to curve shortening \citep{AllenCahn1979,RubinsteinSternbergKeller1989}. & Basis for local curvature-driven interface laws used in grain growth and nonconserved phase transformations. \\
1986 & Phase-field/free-boundary analysis recovers the Gibbs--Thomson condition for solidification \citep{Caginalp1986}. & Connects diffuse solidification models to Stefan-type sharp-interface thermodynamics. \\
1991--92 & Rigorous convergence to smooth and then generalized mean-curvature motion \citep{BronsardKohn1991,EvansSonerSouganidis1992}. & Justifies Allen--Cahn regularization of moving boundaries, including continuation beyond classical singularities. \\
1993--95 & Convergence to Brakke flow and high-order smooth-interface expansions with error control \citep{Ilmanen1993,deMottoniSchatzman1995}. & Supplies weak-flow and quantitative asymptotic foundations for curvature-driven numerical models. \\
\bottomrule
\end{tabular}
\end{table}

In materials simulation, these results underpin polycrystalline phase-field models \citep{FanChen1997,KrillChen2002} and multiphase formulations \citep{SteinbachPezzolla1996}. They also underpin quantitative solidification models \citep{WarrenBoettinger1995,KarmaRappel1998} and sharp-interface solvers for grain growth and recrystallisation \citep{Bernacki2011}. No new convergence theorem is claimed here. Instead, we use these results for three practical purposes. They fix the normalization of the interfacial energy, the units of the mobility and the sign convention needed by GPK.

The scope of the resulting local law is important. For conserved composition (Cahn--Hilliard) dynamics, the sharp-interface limit is not local. It is a Mullins--Sekerka or Hele--Shaw problem, in which diffusion in the bulk controls the interface motion \citep{AlikakosBatesChen1994}. Composition-controlled multiphase growth therefore needs a transport model in addition to Equation~\ref{eq:master}.

The boundary-velocity law used by GPK follows from a simple energy balance. Consider a sharp boundary $\Gamma$ between a resident region $p$ and a candidate region $q$. The total free energy is the interface energy plus the bulk energy \citep{Modica1987}:
\begin{equation}
F_\Gamma=\int_\Gamma\gamma\,\mathrm dA+\int_\Omega \psi\,\mathrm dV.
\label{eq:sharp-energy}
\end{equation}
Here $\mathrm dA$ is an element of interface area, $\mathrm dV$ an element of volume and $\psi$ the bulk free-energy density. Now let the boundary move with the local normal velocity $v_n$. For a constant interfacial energy $\gamma$, the total energy then changes at the rate
\begin{equation}
\frac{\mathrm dF_\Gamma}{\mathrm dt}
=\int_\Gamma(\gamma\kappa_p-\Delta G_{q-p})\,v_n\,\mathrm dA.
\label{eq:first-variation}
\end{equation}
The first term in the bracket reflects the change in interface area. Shifting a convex boundary outward increases its area at a rate proportional to its curvature $\kappa_p$. The second term reflects the replacement of candidate material by resident material. The bracket, taken with a minus sign, is therefore the net driving force per unit area, $\Delta G_{q-p}-\gamma\kappa_p$. A moving boundary also dissipates energy, for example through atomic jumps across it. As for viscous friction, the dissipation is proportional to the square of the velocity, $v_n^2/M$ per unit area. A large mobility $M$ therefore means little friction. The boundary moves at the speed at which the released energy, driving force times velocity, is exactly consumed by this friction. This gives the central evolution law of GPK:
\begin{equation}
v_n=M\left(\Delta G_{q-p}-\gamma\kappa_p\right).
\label{eq:master}
\end{equation}
In words, the speed equals the boundary mobility $M$ times the net driving force. Both terms of the driving force are energies per unit volume, in \si{J.m^{-3}}. A positive $v_n$ means that the resident $p$ grows. A candidate with higher energy ($\Delta G_{q-p}>0$) therefore favours growth of the resident, and a candidate with lower energy makes it recede. Positive resident curvature always opposes resident expansion, because a convex grain lowers its boundary area by shrinking. Inserting Equation~\ref{eq:master} into Equation~\ref{eq:first-variation} gives
\begin{equation}
\frac{\mathrm dF_\Gamma}{\mathrm dt}
=-\int_\Gamma\frac{v_n^2}{M}\,\mathrm dA\leq0.
\label{eq:energy-dissipation}
\end{equation}
This shows that the total free energy can never increase under this motion. For an orientation-dependent $\gamma(\mathbf n)$, the exact energy change contains the interface stiffness instead of $\gamma$ (Section~\ref{sec:constitutive}). A boundary at rest ($v_n=0$) has $\Delta G_{q-p}=\gamma\kappa_p$ according to Equation~\ref{eq:master}. This is the Gibbs--Thomson equilibrium condition, in which the bulk and capillary driving forces balance. The same law, with exactly this meaning of $\gamma$ and $M$, also follows from a diffuse-interface model in the limit of a thin interface. This derivation is given below. Supplementary Note~1 supplies a symbolic check of every identity used in it.

\subsubsection*{Derivation from a diffuse-interface model}
\label{sec:thin-limit}
We now show that Equation~\ref{eq:master}, including the meaning of $\gamma$, also follows from a diffuse-interface model when its interface becomes thin. The sign conventions of Table~\ref{tab:signs} apply. Let $p$ be the resident region, represented by $\phi=1$, and let $q$ be the candidate region, represented locally by $\phi=0$. The signed distance $s$ increases from $p$ into $q$, along the outward normal of $p$.

The strategy is as follows. We start from a diffuse-interface model whose interface has a small thickness $\xi$. Step~1 determines the shape of the thin transition layer between the two regions. Step~2 lets this layer move and bend while it keeps its shape. A simple energy balance across the layer then gives the interface velocity. In the end, $\xi$ cancels out, which is why GPK never needs it.

We begin from the total free energy of the diffuse model,
\begin{equation}
F_\xi=\int_\Omega\left[\frac{6\gamma}{\xi}
\left(\frac{\xi^2}{2}|\nabla\phi|^2+W(\phi)\right)+\psi_{\mathrm{bulk}}(\phi)\right]\,\mathrm dV,
\qquad W(\phi)=\frac{1}{2}\phi^2(1-\phi)^2 .
\label{eq:free-energy}
\end{equation}
Here $\Omega$ is the simulated volume, and $F_\xi$ is its total free energy. The order parameter $\phi$ changes smoothly from one in the resident region $p$ to zero in the candidate region $q$. The small length $\xi$ is the diffuse thickness, which is used only in this derivation. The integrand contains three energy contributions. The gradient term $\tfrac{\xi^2}{2}|\nabla\phi|^2$ penalizes rapid spatial changes of $\phi$. The double-well function $W(\phi)$ penalizes intermediate values of $\phi$, because it is zero at the two bulk states $\phi=0$ and $\phi=1$ and positive in between. Together, these two terms produce the interfacial energy. The third term, $\psi_{\mathrm{bulk}}(\phi)$, is the bulk free-energy density. It equals $\psi_p$ in the resident and $\psi_q$ in the candidate region. The prefactor $6\gamma/\xi$ is chosen such that the interface energy per unit area equals $\gamma$ (this is checked in Equation~\ref{eq:surface-normalization}). The value six follows from the shape of the double well $W$.

The order parameter evolves such that the free energy decreases as fast as possible. With the kinetic coefficient $M/(6\xi)$, this gives the Allen--Cahn equation
\begin{equation}
\frac{\partial\phi}{\partial t}
=M\gamma\left(\nabla^2\phi-\frac{W_\phi(\phi)}{\xi^2}\right)
-\frac{M}{6\xi}\psi_\phi(\phi).
\label{eq:allen-cahn-normalized}
\end{equation}
Here $\partial\phi/\partial t$ is the rate of change of $\phi$ at a fixed point. The Laplacian $\nabla^2\phi$ is the sum of the second spatial derivatives of $\phi$. $W_\phi=\mathrm dW/\mathrm d\phi$ is the derivative of the double well, and $\psi_\phi=\partial \psi_{\mathrm{bulk}}/\partial\phi$. The right-hand side is the local change of $F_\xi$ caused by a small change of $\phi$, taken with a minus sign. The order parameter therefore always changes in the direction that lowers the energy. Both $\psi_\phi$ and the capillary term $\gamma\kappa$ that will appear below are energy densities, in $\mathrm{J\,m^{-3}}$. The mobility $M$ converts either of them into a velocity.

\subsubsection*{Step 1: shape of the transition layer}
Inside the thin layer, $\phi$ changes rapidly across the interface but only slowly along it. It is therefore convenient to measure distance across the layer in units of its thickness. We use the stretched coordinate $z=s/\xi$ (not the Cartesian coordinate $z$), where $s$ is the signed distance from the interface. A prime denotes the derivative with respect to $z$. Consider first a flat interface at rest without a bulk driving force. Then $\phi$ depends only on $z$, and we call this profile $\phi_0(z)$. Equation~\ref{eq:allen-cahn-normalized} reduces to
\begin{equation}
\phi_0''-W_\phi(\phi_0)=0,
\qquad
\phi_0(-\infty)=1,\quad\phi_0(+\infty)=0.
\label{eq:profile-ode}
\end{equation}
This equation states that the gradient term and the double-well term are in balance across the layer. The boundary conditions state that $\phi_0$ approaches the bulk values far from the interface. Multiplying by $\phi_0'$ and integrating once gives $\tfrac12(\phi_0')^2=W(\phi_0)$. For the selected double well, the solution is
\begin{equation}
\phi_0(z)=\frac12\left[1-\tanh\left(\frac z2\right)\right],
\qquad \phi_0'=-\phi_0(1-\phi_0).
\label{eq:profile}
\end{equation}
This is the familiar hyperbolic-tangent profile. It falls smoothly from one to zero over a few interface thicknesses. We will need one number that characterizes this profile:
\begin{equation}
I_1=\int_{-\infty}^{\infty}(\phi_0')^2\,\mathrm dz=\frac16.
\label{eq:profile-integral}
\end{equation}
The value $1/6$ follows directly from Equation~\ref{eq:profile} by using $\phi_0$ itself as the integration variable. It is exactly the factor that the prefactor six in Equation~\ref{eq:free-energy} was designed to cancel.

\subsubsection*{Step 2: interface velocity}
Now let the layer move with the normal velocity $v_n$ and let it be gently curved, with curvature $\kappa$. To leading order, it keeps the shape $\phi_0$. Two terms of Equation~\ref{eq:allen-cahn-normalized} then change. First, the moving profile passes a fixed point in space, so that $\partial\phi/\partial t=-v_n\phi_0'/\xi$ there. Second, the Laplacian of a curved layer contains an extra term:
\begin{equation}
\nabla^2\phi=\frac{1}{\xi^2}\phi_0''+\frac{\kappa}{\xi}\phi_0'.
\label{eq:local-laplacian}
\end{equation}
The extra term appears because the area of a curved interface changes when it is shifted along its normal. Inserting both expressions into Equation~\ref{eq:allen-cahn-normalized} gives
\begin{equation}
-\frac{v_n}{\xi}\phi_0'
=\frac{M\gamma}{\xi^2}\left(\phi_0''-W_\phi(\phi_0)\right)
+\frac{M\gamma\kappa}{\xi}\phi_0'
-\frac{M}{6\xi}\psi_\phi(\phi_0).
\label{eq:motion-order}
\end{equation}
The bracket is zero because of Equation~\ref{eq:profile-ode}. The remaining terms do not balance at every point inside the layer. The layer adjusts its shape slightly to make up for this. Such a small shape adjustment, however, does not change the overall energy balance across the layer \citep{RubinsteinSternbergKeller1989}. We therefore multiply Equation~\ref{eq:motion-order} by $\xi\phi_0'$ and integrate across the layer. This gives
\begin{equation}
-v_nI_1=M\gamma\kappa I_1-\frac{M}{6}\int_{-\infty}^{\infty}\psi_\phi(\phi_0)\,\phi_0'\,\mathrm dz.
\label{eq:fredholm}
\end{equation}
The remaining integral can be evaluated without knowing the details of the bulk energy. Using $\phi$ as the integration variable gives
\begin{equation}
\int_{-\infty}^{\infty}\psi_\phi(\phi_0)\,\phi_0'\,\mathrm dz
=\int_1^0 \psi_\phi(\phi)\,\mathrm d\phi
=\psi_q-\psi_p=\Delta G_{q-p}.
\label{eq:bulk-integral}
\end{equation}
In words, this integral is simply the difference between the bulk free-energy densities of the candidate and the resident. We now insert $I_1=1/6$ into Equation~\ref{eq:fredholm}. The factor $1/6$ then cancels against the factor six in the energy. The result is exactly the boundary-velocity law of Equation~\ref{eq:master}. The diffuse thickness $\xi$ no longer appears. Finally, we check that $\gamma$ really is the energy per unit interface area. We integrate the interfacial part of Equation~\ref{eq:free-energy} across the layer. Across the layer, $\xi^2|\nabla\phi|^2=(\phi_0')^2$ and $\mathrm ds=\xi\,\mathrm dz$. Using $\tfrac12(\phi_0')^2=W(\phi_0)$ from Step~1, the interface energy $E_\Gamma$ per unit area $A$ is
\begin{equation}
\frac{E_\Gamma}{A}=6\gamma\int_{-\infty}^{\infty}
\left[\tfrac12(\phi_0')^2+W(\phi_0)\right]\mathrm dz
=6\gamma\int_{-\infty}^{\infty}(\phi_0')^2\,\mathrm dz
=6\gamma I_1=\gamma.
\label{eq:surface-normalization}
\end{equation}
Thus neither an empirical prefactor nor a leftover factor of six remains. This normalization allows every constitutive module to supply the same quantities $M$, $\Delta G$ and $\gamma$ without process-specific rescaling.

Equation~\ref{eq:master} is the constitutive core of GPK. Different processes change only the three inputs $M$, $\Delta G_{q-p}$ and $\gamma$. For curvature-driven grain growth, $\Delta G_{q-p}=0$. This agrees with classical theories of curvature flow and grain growth \citep{Mullins1956,Hillert1965,MacPhersonSrolovitz2007}. Recrystallisation uses stored deformation energy, for instance in the form of a dislocation-density metric. Multiphase growth uses differences in the Gibbs energies of the phases. Elastic coupling adds a linear elastic free-energy contribution. A misorientation-dependent mobility or energy may be supplied through Read--Shockley-type relations \citep{ReadShockley1950}. Crystallographic anisotropy may be supplied through an orientation-dependent interfacial energy $\gamma(\mathbf n)$.

\subsection{Transient Gaussian geometry from discrete topology}
\label{sec:geometry}
The geometry module answers one local question. What are the normal and the curvature of a boundary that is represented only by discrete labels? The module does not move the interface. Each cell stores a label, a sub-cell fraction $\alpha$ between zero and one, thermodynamic fields and optional orientation variables. The fraction $\alpha$ is measured along the boundary normal. For a boundary aligned with the grid, it equals the volume fraction of the cell occupied by the resident. For an oblique boundary, it is used as an approximation of this volume fraction. Bulk cells have $\alpha=1$ for their current label. Interface cells are those with at least one neighbour of a different label. Phase volumes used for diagnostics are obtained by summing these fractions. The fraction $\alpha$ is not used to reconstruct a detailed interface shape inside the cell, as a volume-of-fluid method would do.

\paragraph{Reason for using a Gaussian projection.}
The Gaussian is a temporary measuring tool, not a physical interface profile. Differentiating a stepped label field directly gives grid-scale noise in the normal and in the curvature. Gaussian smoothing removes that noise over one controlled length scale. Because the kernel is positive and normalized, the smoothed indicator remains between zero and one. Because it is symmetric and isotropic, it does not favour any direction. It is also separable, that is, a three-dimensional Gaussian is the product of three one-dimensional Gaussians. This keeps the calculation inexpensive. The projection is rebuilt in each step and then discarded. Its width therefore controls only the scale of the geometry measurement. It does not create a diffuse interface, select a phase or set the boundary speed. Other smooth kernels could be used. The Gaussian, however, provides all of these properties with one well-defined scale.

\begin{figure}[!htbp]
  \centering
  \includegraphics[width=0.98\textwidth]{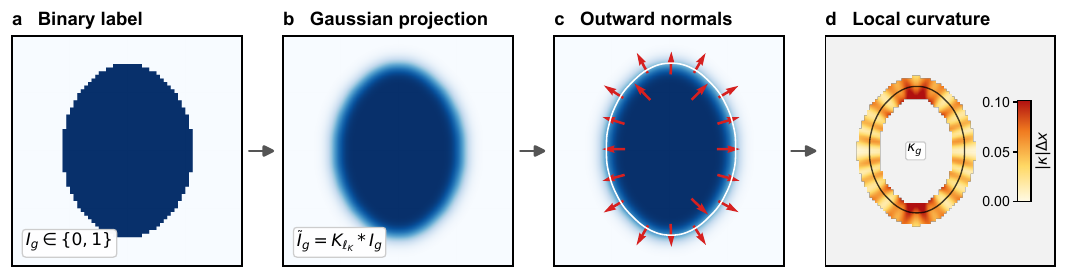}
  \caption{The transient geometry layer of GPK. (a) A binary resident-label indicator $I_g$ (one inside label $g$, zero outside). (b) It is projected into a temporary smooth field $\widetilde I_g=K_{\ell_K}*I_g$, where $K_{\ell_K}$ is the Gaussian kernel of width $\ell_K$ and $*$ denotes convolution. (c) Differentiation gives the outward normals. (d) The local curvature $\kappa_g$ is computed from the first and second derivatives of $\widetilde I_g$, and $|\kappa|\Delta x$ is its magnitude in units of the inverse grid spacing. Normals and curvature are passed to the shared kinetic law. The projection is discarded after evaluation and never becomes a persistent order parameter or a process model by itself.}
  \label{fig:geometry-pipeline}
\end{figure}

For every resident label $g$ on the interface, we form the binary indicator $I_g$. It equals one in cells with label $g$ and zero elsewhere. We smooth it by convolution with a Gaussian of width $\sigma_K$ cells, which gives $\widetilde I_g=K_{\ell_K}*I_g$. Convolution means that each cell receives a weighted average of its neighbours' values, with Gaussian weights. We need the first and second derivatives of $\widetilde I_g$. Smoothing and differentiation can be applied in either order:
\begin{equation}
\frac{\partial\widetilde I_g}{\partial x}=\frac{\partial K_{\ell_K}}{\partial x}*I_g,
\qquad
\frac{\partial^2\widetilde I_g}{\partial x\,\partial y}=\frac{\partial^2 K_{\ell_K}}{\partial x\,\partial y}*I_g,
\label{eq:gaussian-derivative}
\end{equation}
and likewise for all other first and second derivatives. Each derivative of $\widetilde I_g$ is therefore computed by applying Gaussian weights combined with fourth-order finite-difference formulas to $I_g$ (Section~\ref{sec:geometry-methods}), instead of differentiating the stepped label field. Together, the first derivatives form the gradient $\nabla\widetilde I_g$, and the second derivatives form the Hessian $H(\widetilde I_g)$. The derivatives are therefore well defined, even though $I_g$ itself jumps from one to zero. The outward unit normal is
\begin{equation}
\mathbf n_g=-\frac{\nabla\widetilde I_g}{|\nabla\widetilde I_g|}.
\label{eq:geometry}
\end{equation}
The gradient of $\widetilde I_g$ points towards increasing $\widetilde I_g$, that is, into the region $g$. The minus sign therefore makes the normal point outward from $g$. Dividing by the length $|\nabla\widetilde I_g|$ makes it a unit vector. Far from any boundary, $\widetilde I_g$ is nearly constant and its gradient is almost zero. There, the direction of the gradient is meaningless and is dominated by round-off. Cells with $|\nabla\widetilde I_g|\,\Delta x\leq10^{-4}$ are therefore excluded, and their normal is set to zero. Because this threshold is applied to the dimensionless product $|\nabla\widetilde I_g|\Delta x$, it does not change when $\Delta x$ changes. At a resolved boundary, this product is close to $1/(\sqrt{2\pi}\,\sigma_K)$ (Equation~\ref{eq:erf-profile}). It is 0.2 for $\sigma_K=2$ cells and still 0.04 for the widest kernel used here, $\sigma_K=10.667$ cells. The threshold therefore lies more than two orders of magnitude below any real boundary signal. The curvature is evaluated directly from the gradient and the Hessian through Equation~\ref{eq:curvature-hessian}. It is not obtained by differentiating the normal field a second time, which would add noise. In three dimensions, $\kappa_g$ is the sum of the two principal curvatures, so a sphere of radius $R$ has $\kappa_g=2/R$. Features with radii of curvature smaller than $\ell_K=\sigma_K\Delta x$, the characteristic width of the Gaussian smoothing kernel, cannot be resolved accurately by the kernel. The code therefore caps the measured curvature at $|\kappa_g|=1/\ell_K$. In three dimensions, this cap corresponds to a sphere of radius $2\ell_K$. Figure~\ref{fig:geometry-pipeline} shows the complete reconstruction.

The four panels in Figure~\ref{fig:geometry-pipeline} separate the stored state from the measurement. Panel (a) contains the stored labels and therefore the exact region identity. Panel (b) shows that the Gaussian projection spreads only the local measurement support. It does not move the boundary. The arrows in panel (c) point from the resident label toward its neighbour, as required by the sign in Equation~\ref{eq:geometry}. Panel (d) plots $\kappa$ only inside the accepted interface band. The bulk is therefore intentionally blank. Colour along the ring shows the local magnitude $|\kappa|\Delta x$. Its small variation with direction is caused by the grid. The panels show that a discontinuous label field can provide smooth geometry without being replaced by a diffuse order parameter.

Two properties must be checked before this geometry can be used. The first is the sign of the normal, and the second is the bias caused by smoothing. To check the sign, consider a locally flat boundary. Let $\zeta$ be the distance measured inward from the boundary, into region $g$. The indicator is then a step function, $I_g=1$ for $\zeta>0$ and $I_g=0$ for $\zeta<0$. Smoothing this step with a Gaussian of width $\ell_K=\sigma_K\Delta x$ gives
\begin{equation}
\widetilde I_g(\zeta)=\frac12\left[1+\operatorname{erf}\left(\frac{\zeta}{\sqrt2\,\ell_K}\right)\right],
\qquad
\frac{\mathrm d\widetilde I_g}{\mathrm d\zeta}
=\frac{e^{-\zeta^2/(2\ell_K^2)}}{\sqrt{2\pi}\,\ell_K}>0.
\label{eq:erf-profile}
\end{equation}
Here $\operatorname{erf}$ is the error function, familiar from diffusion problems. The result is a smooth S-shaped curve. The smoothed step therefore rises gradually from zero to one over a distance of a few $\ell_K$. Its slope is itself a Gaussian bell curve and is therefore positive everywhere. The gradient therefore points inward, and its negative is the outward normal, as used in Equation~\ref{eq:geometry}. Next, we need the curvature. The curvature measures how quickly the normal turns along the boundary, $\kappa=\nabla\cdot\mathbf n$. It is computed from the first and second derivatives of $\widetilde I_g$ with the standard curvature formula of level-set methods (Equation~\ref{eq:curvature-hessian}). In three dimensions, it gives the sum of the two principal curvatures. As a check, consider a circular region of radius $R$. Its outward normal points radially outward, and $\kappa=1/R$ in two dimensions. For a sphere, $\kappa=2/R$.

The smoothing bias can be understood from two complementary views of the Gaussian. The first view considers a wavy pattern of wavelength $\lambda$. Gaussian smoothing does not change the shape of such a wave. It only multiplies its amplitude by the damping factor
\begin{equation}
\text{damping factor}=\exp\!\left(-\frac{\ell_K^2k^2}{2}\right),\qquad k=\frac{2\pi}{\lambda}.
\label{eq:gaussian-spectrum}
\end{equation}
Here $k$ is the wavenumber. Short waves have a large $k$ and are damped strongly. As an example, take the default width $\ell_K=2\Delta x$. The shortest wave on the grid, with $\lambda=2\Delta x$, is then multiplied by about $3\times10^{-9}$, that is, it is removed completely. A wave with $\lambda=30\Delta x$, comparable to the size of a resolved grain, keeps about 92\% of its amplitude. Smoothing therefore removes the grid-scale staircase before differentiation, while the real boundary shape passes almost unchanged. The second view applies to smooth, well-resolved geometry. Here, the smoothed field equals the original field plus small corrections:
\begin{equation}
K_{\ell_K}*u=u+\frac{\ell_K^2}{2}\nabla^2u+\frac{\ell_K^4}{8}\nabla^4u+\ldots
\label{eq:gaussian-moment}
\end{equation}
Here $u$ is any smooth field, $\nabla^2u$ is its Laplacian and $\nabla^4u$ is the Laplacian applied twice. It has the same form as a short diffusion step. Smoothing with a Gaussian is indeed equivalent to letting the field diffuse for a short, fictitious time. No physical diffusion or heat is introduced in GPK. The series shows two useful facts. First, the leading correction contains only even derivatives, so smoothing introduces no directional bias. Second, the first error term is proportional to $\ell_K^2$. Relative to a feature of radius $R$, the curvature error is therefore of order $(\ell_K/R)^2$. Equivalently, the absolute curvature error is of order $\ell_K^2/R^3$. These estimates hold away from junctions and topology changes.

Together, these two views define a practical trade-off. A wider kernel suppresses differentiation noise. However, it also averages the curvature over a larger region. The useful range of kernel widths is therefore
\begin{equation}
\Delta x\leq\ell_K\ll R_{\min},
\label{eq:kernel-window}
\end{equation}
Here $R_{\min}$ is the smallest radius of curvature that must be computed accurately, and $\Delta x$ is the grid spacing. The left inequality requires the Gaussian to span enough grid cells to suppress cell-scale noise. The right inequality requires it to remain much narrower than the smallest feature whose curvature is to be trusted. The default width $\sigma_K=2$ cells lies in this window for the resolved tests used here. This window also states the main geometric limit of the method. Features comparable to or smaller than the Gaussian support cannot have quantitatively reliable curvature. In addition, nearby boundaries or junctions may influence one another through the kernel. Increasing $\ell_K$ reduces cell-scale noise but increases smoothing bias. Decreasing it has the opposite effect. The projection is reconstructed from the labels in every step. It is therefore neither stored nor advanced in time, and it does not need to resolve an equilibrium diffuse profile.

\begin{figure}[!htbp]
  \centering
  \includegraphics[width=0.98\textwidth]{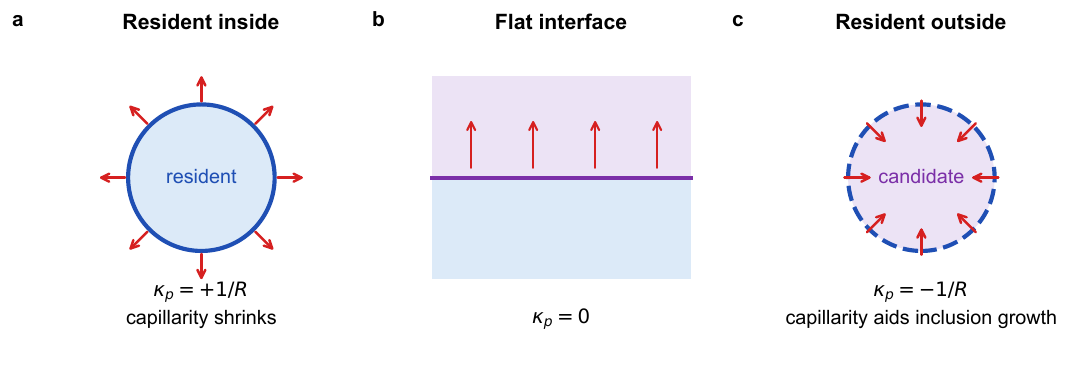}
  \caption{Curvature and velocity signs for the resident label. The same circular interface has positive curvature when the resident occupies the interior and negative curvature when it surrounds an inclusion. $\kappa_p$ is the curvature of the resident $p$, and $R$ is the radius of the circle.}
  \label{fig:curvature-signs}
\end{figure}

Figure~\ref{fig:curvature-signs} illustrates the resident-label sign convention for three geometries: a convex grain, a flat boundary and a matrix around an inclusion. This prevents a common error. That error is to insert the positive curvature of a small embryo into a law written for the surrounding matrix.

Figure~\ref{fig:curvature-signs} also fixes the physical interpretation of the signs used by every constitutive module. In panel (a), the resident is the enclosed grain. Thus $\kappa=+1/R$, and the capillary term $-M\gamma\kappa$ makes the grain shrink. The flat interface in panel (b) has $\kappa=0$. It therefore isolates any imposed bulk driving force. In panel (c), the resident is the matrix that surrounds the inclusion. Its outward normal points away from the matrix and therefore into the inclusion. Hence $\kappa=-1/R$. Different grains or phases may thus define ``outward'' in opposite directions. A physical law may prescribe a velocity in the inclusion's own outward direction. Gibbs--Thomson and stored-energy-driven growth laws are usually written this way. Such a law must first be re-expressed in the sign convention of the resident, that is, of the cell that currently owns the boundary. Only then can it be inserted into the shared velocity update. Showing all three cases explicitly prevents a swap of resident and candidate labels from silently reversing the sign of a physically meaningful velocity.

\begin{figure}[!htbp]
  \centering
  \includegraphics[width=0.98\textwidth]{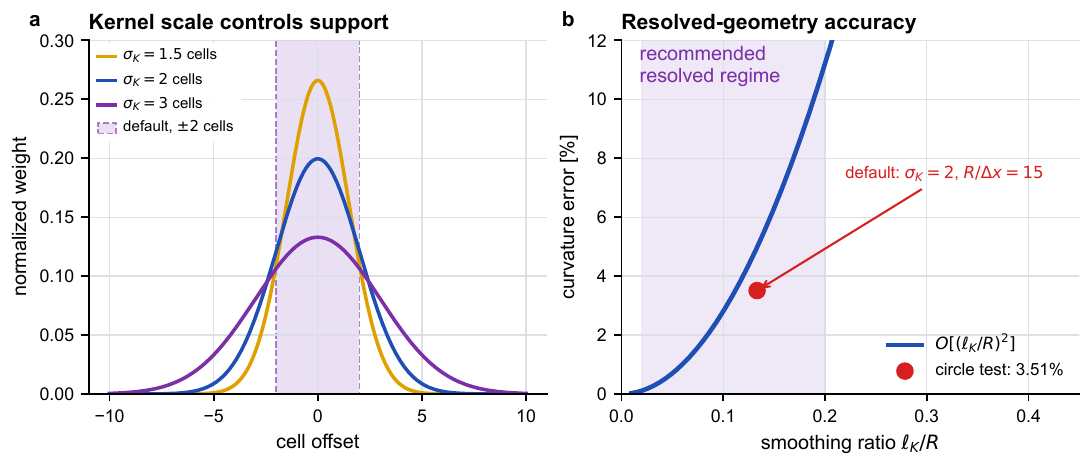}
  \caption{Gaussian-kernel scale selection and curvature accuracy. The kernel regularizes derivatives of the discrete indicator, while the circle benchmark quantifies the increasing smoothing bias as the kernel support approaches the geometric radius. The useful operating window lies between pixel-scale noise and feature-scale averaging. (a) Gaussian weights for kernel widths $\sigma_K$ of 1.5, 2 and 3 cells; the shaded band marks $\pm2$ cells around the centre. (b) Curvature error against the smoothing ratio $\ell_K/R$, where $\ell_K=\sigma_K\Delta x$ is the kernel length, $R$ the circle radius and $\Delta x$ the grid spacing. The curve shows the expected error of order $(\ell_K/R)^2$.}
  \label{fig:gaussian-accuracy}
\end{figure}

Figure~\ref{fig:gaussian-accuracy} then turns the scale inequality in Equation~\ref{eq:kernel-window} into a practical kernel-selection plot and reports the circle benchmark used below.

The two panels of Figure~\ref{fig:gaussian-accuracy} explain why the Gaussian is a controlled geometric scale rather than a cosmetic filter. Panel (a) shows that increasing $\sigma_K$ suppresses cell-scale oscillations in the derivatives of the binary indicator. However, it also broadens the region over which a boundary is sampled. Panel (b) measures the resulting trade-off on circles. The error is small when $\ell_K/R$ lies in the shaded operating interval. It rises once the kernel averages over an appreciable fraction of the radius. The marked default point gives the reported 3.51\% error at $R/\Delta x=15$. Thus $\sigma_K$ is selected from the smallest radius that must remain quantitative, using Equation~\ref{eq:kernel-window}. It is not tuned separately to obtain a favourable morphology.

\subsection{Verification on basic geometries and elementary kinetics}
\begin{figure}[!htbp]
  \centering
  \includegraphics[width=0.98\textwidth]{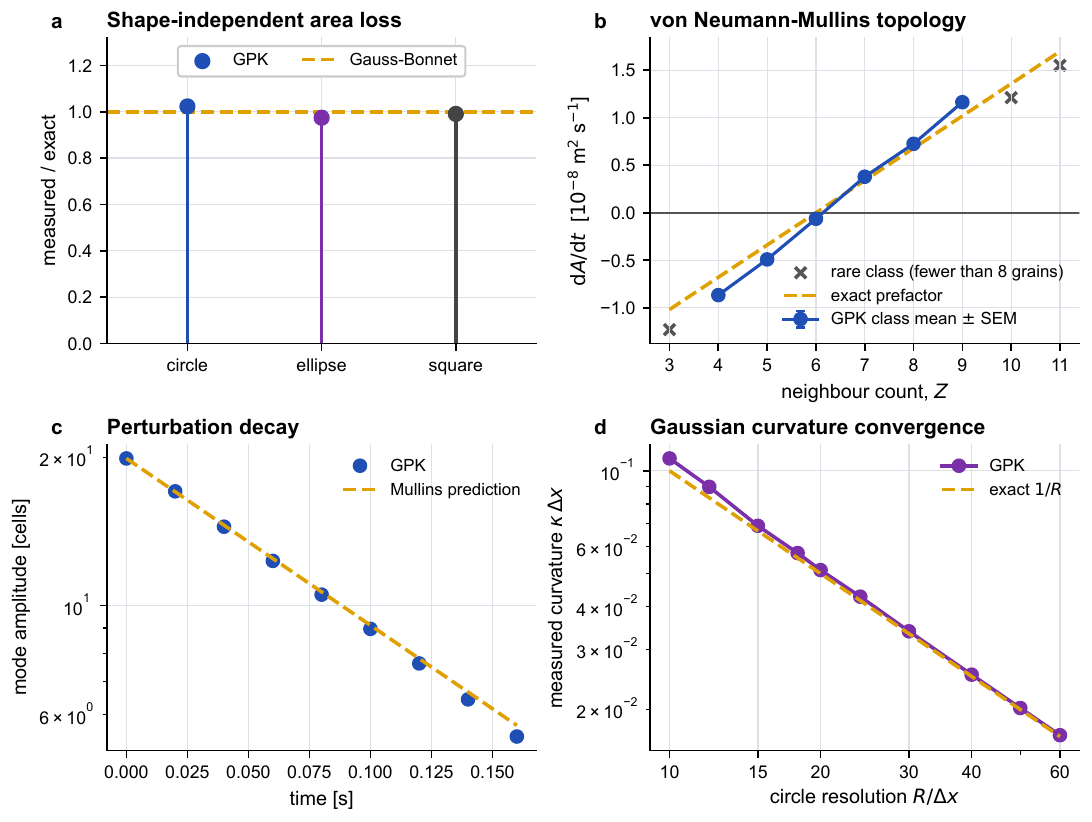}
  \caption{Foundational verification of the shared GPK layers. (a) Shape-independent area loss, measured relative to the exact rate $2\pi M\gamma$. (b) Von Neumann--Mullins topology trend from 1165 topology-stable grains across five independent networks. $\mathrm dA/\mathrm dt$ is the rate of change of the grain area and $Z$ the number of neighbours. Class means with at least eight grains show standard errors. Rare classes with fewer grains are shown as unconnected crosses. (c) Nine-point sinusoidal perturbation decay. (d) Measured Gaussian--Hessian curvature $\kappa\Delta x$ against the exact $1/R$ value at ten circle resolutions $R/\Delta x$, where $R$ is the circle radius and $\Delta x$ the grid spacing. Exact values and limitations are reported in the text and Supplementary Tables~S2 and S3.}
  \label{fig:foundations}
\end{figure}

Verification proceeds from the simplest shared operator to the complete update (Figure~\ref{fig:foundations}). The first test is a circle, whose curvature magnitude and sign are known at every point. We use $R=\SI{15}{\micro\metre}$, $\Delta x=\SI{1}{\micro\metre}$ and $\sigma_K=2$ cells. GPK then gives $\kappa=6.9005\times10^4\ \mathrm{m^{-1}}$, compared with the exact value $1/R=6.6667\times10^4\ \mathrm{m^{-1}}$. This is an error of 3.51\%. We then increase $R/\Delta x$ from 15 to 60 while holding $\ell_K/R=2/15$ fixed. Holding this ratio fixed keeps the smoothing bias constant. The error nevertheless decreases to 0.89\%. Most of the error at $R/\Delta x=15$ is therefore caused by sampling the boundary on the grid, and this part decreases as the grid is refined. Two simpler alternatives were also tested: counting labels in a local $3\times3$ block and using integer distances. Both give an incorrect sign or a much larger error in the same test. The remaining tests use curvature inside integral and kinetic laws. They therefore ask whether the local geometry remains consistent at the scale of whole grains and grain networks.

Each benchmark has a defined scope. The circle isolates the local curvature magnitude and sign. Planar bicrystals isolate the sign of the bulk driving force and the mobility scale. Curved bicrystals test integrated capillary motion and visible shape relaxation. The von Neumann--Mullins network checks curvature in a many-grain topology. Sinusoidal decay tests the orientation-independent linear response. Refinement separates grid error from the bias caused by the finite kernel width. Passing one row is never used as evidence for another. Numerical acceptance criteria, diagnostic-only quantities and excluded under-resolved cases are listed separately in the Supporting Information.

Planar bicrystals provide the complementary test without curvature. The suite contains eight flat interfaces. In three of them, $\Delta G=0$, and the interfacial energy is isotropic, dependent on the phase pair or dependent on orientation. The other five are driven by stored energy, undercooling, reduction free energy, supersaturation and a multiphase Gibbs-energy difference, respectively. A flat interface without driving force must not move. The three stationary cases indeed change exactly 0 of 6,400, 0 of 7,680 and 0 of 7,680 cells. For the five driven cases, we compare the measured velocity with the exact value $v=M\Delta G$. The relative deviations $v_{\rm GPK}/v_{\rm exact}-1$ range from $-1.10\times10^{-7}$ to $+3.49\times10^{-8}$. Supplementary Figure~S6 shows the initial and final microstructures for all eight cases. The accompanying table records the constitutive term and the expected response. These bicrystals test the sign of the bulk force and the mobility scaling. Curvature, junctions and topology changes do not interfere.

Curved bicrystals then place one grain inside a second grain. The inner grain is a circle, a $1.83{:}1$ ellipse or a square. All three are evolved on $360^2$ grids with $\Delta x=\SI{0.1}{\micro\metre}$. The runs continue until only 35.7--40.8\% of the initial inclusion area remains. For the circle, the slope of $R^2(t)$ differs from the exact capillary law by 2.30\%. For the ellipse and the square, the area-loss rates differ from the shape-independent value derived below by 2.58\% and 0.93\%. The ellipse aspect ratio relaxes from 1.84 to 1.37. The square rounds to a nearly circular final boundary. Supplementary Figure~S7 shows both the regressions and the true initial and final label fields. Together with the planar suite, these tests cover zero curvature, smoothly varying curvature and initially sharp corners.

The exact reference for these curved tests follows from a classical geometric result. Consider a closed two-dimensional grain boundary under pure capillarity. Equation~\ref{eq:master} then gives $v_n=-M\gamma\kappa$. The area $A$ of the grain changes at a rate equal to the normal velocity integrated around the boundary, $\dot A=\oint_\Gamma v_n\,\mathrm ds$. Here $\Gamma$ is the closed boundary, $\oint_\Gamma$ denotes an integral once around it and $\mathrm ds$ is an element of arc length. The curvature integrated around any smooth closed curve equals $2\pi$, the total turning angle of its tangent. Therefore,
\begin{equation}
\frac{\mathrm dA}{\mathrm dt}
=-M\gamma\oint_\Gamma\kappa\,\mathrm ds
=-2\pi M\gamma.
\label{eq:gauss-bonnet}
\end{equation}
This is a consequence of the Gauss--Bonnet theorem. In words, every smooth inclusion loses area at the same constant rate $2\pi M\gamma$, whatever its shape. For a circle with $A=\pi R^2$, this gives the equivalent law
\begin{equation}
R^2(t)=R_0^2-2M\gamma t.
\label{eq:circle-shrinkage}
\end{equation}
Here $R_0$ is the initial radius. The square of the radius therefore decreases linearly in time. The refined circular, elliptical and square bicrystals reproduce the corresponding rate with errors of 2.30\%, 2.58\% and 0.93\%, respectively. These errors are not universal constants. They refer to the refined $360^2$ grids with $\Delta x=\SI{0.1}{\micro\metre}$. They depend on the ratio of the inclusion size to the grid spacing and to the Gaussian measurement length $\ell_K$. The refinement tests below quantify this dependence.
Figure~\ref{fig:foundations}(a) compares all three shapes with the Gauss--Bonnet prediction.

The same idea extends to a grain inside a network. Consider a grain $i$ with $Z_i$ neighbours, and hence $Z_i$ triple junctions. At equilibrium, the three boundaries at each junction meet at $120^\circ$. The boundary of the grain therefore turns by $60^\circ$, or $\pi/3$, at each junction. The total turning angle around the grain is still $2\pi$. It is now shared between the curved boundary segments and the $Z_i$ corners. The corners take $Z_i\pi/3$, so the curved segments turn by only
\begin{equation}
\oint_{\Gamma_i}\kappa\,\mathrm ds=2\pi-Z_i\frac{\pi}{3}=\frac{\pi}{3}(6-Z_i),
\end{equation}
where $\Gamma_i$ is the boundary of grain $i$. Using this instead of $2\pi$ in Equation~\ref{eq:gauss-bonnet} gives the von Neumann--Mullins law \citep{Mullins1956} for the area $A_i$ of grain $i$:
\begin{equation}
\frac{\mathrm dA_i}{\mathrm dt}=\frac{\pi}{3}M\gamma(Z_i-6).
\label{eq:von-neumann-mullins}
\end{equation}
In words, grains with more than six neighbours grow, grains with fewer than six shrink, and six-sided grains keep their area. The rate depends only on the number of neighbours, not on the grain size or shape. We test this law on all grains of large networks. Five independent periodic $640^2$ networks are initialized with 300 grains each. They use $\Delta x=\SI{1}{\micro m}$, $M=1.0\times10^{-8}\,\mathrm{m^4\,J^{-1}\,s^{-1}}$ and $\gamma=\SI{0.324}{J.m^{-2}}$. The initial characteristic radius is 20.85 cells, or 5.21 Gaussian widths. A grain is included only when its set of neighbours does not change during the 32-step measurement window. This gives 1165 grains with stable topology. A linear fit of their area rates against $Z-6$ gives a coefficient of $4.070\times10^{-9}\,\mathrm{m^2\,s^{-1}}$. This is 19.95\% above the ideal value $\pi M\gamma/3=3.393\times10^{-9}\,\mathrm{m^2\,s^{-1}}$. The correlation between topology and area rate is $r=0.914$. The $Z=9$ class contains 29 grains. Its mean rate is $1.163\times10^{-8}\,\mathrm{m^2\,s^{-1}}$, compared with $1.018\times10^{-8}\,\mathrm{m^2\,s^{-1}}$ from Equation~\ref{eq:von-neumann-mullins}. The $Z=10$ class contains only five grains. It is therefore shown only as an unconnected diagnostic point. Its mean is $1.212\times10^{-8}\,\mathrm{m^2\,s^{-1}}$, compared with the theoretical $1.357\times10^{-8}\,\mathrm{m^2\,s^{-1}}$. Class means with at least eight grains are plotted with standard errors. The all-grain coefficient remains the primary verification metric.
The sign change at six neighbors and the fitted trend are shown in Figure~\ref{fig:foundations}(b).

The last kinetic test is the decay of a small wave on a flat boundary. Let the boundary be described by its height $y=h(x,t)$ above a flat reference line, with the resident $p$ below the boundary. Its outward normal then points upward. For small slopes, the curvature of $p$ is $\kappa\approx-h_{xx}$, where the subscript $xx$ denotes the second derivative with respect to $x$. The minus sign means that a crest of the boundary, where $h_{xx}<0$, is convex for $p$. The normal velocity is approximately $h_t$, the time derivative of the height. Pure curvature flow therefore gives the diffusion-like equation $h_t=M\gamma h_{xx}$. We insert a cosine wave $h=h_1(t)\cos(kx)$ with amplitude $h_1(t)$ and wavenumber $k$. This gives
\begin{equation}
\frac{\mathrm dh_1}{\mathrm dt}=-M\gamma k^2h_1,
\qquad
h_1(t)=h_1(0)\exp(-M\gamma k^2t).
\label{eq:mullins-decay}
\end{equation} In words, the wave amplitude decays exponentially. Short waves (large $k$) decay much faster than long ones. The refined benchmark with nine measured times gives a 3.78\% error in the decay rate. The fit of the logarithmic amplitude has $1-R^2=2.0\times10^{-4}$. Here and below, $R^2$ of a fit is its coefficient of determination, not a squared radius (Figure~\ref{fig:foundations}(c)). Equation~\ref{eq:energy-dissipation} states that the continuum energy can only decrease. To check this in the simulations, we estimate the total boundary length with the Crofton method. This method counts how often the boundaries cross grid lines in several directions. The estimate shows small jumps when junctions rearrange or the topology changes. Over the tested relaxed-network window, the net drift was $-4.65\%$. This result satisfies our predefined acceptance criterion. That criterion requires only that the measured boundary length does not increase systematically over the window. It does not establish a strict decrease at every step for this grid-based length estimate.

Spatial refinement is performed at a fixed physical Gaussian measurement length $\ell_K=\sigma_K\Delta x$ (Table~\ref{tab:parameter-provenance} and Equation~\ref{eq:kernel-window}). When the grid spacing $\Delta x$ becomes smaller, the width $\sigma_K$ in cells must therefore increase. In this way, the same physical neighbourhood is sampled. Figure~\ref{fig:foundations}(d) shows that the local curvature error decreases from 3.51\% at $R/\Delta x=15$ to 0.89\% at $R/\Delta x=60$. A separate pore-closure test keeps an error of 14.17--14.30\% in $\mathrm dR^2/\mathrm dt$ on three grids. In this test, $\ell_K$ is fixed in physical units while the pore shrinks. The ratio $\ell_K/R$ therefore grows during each run, and the smoothing bias, which does not depend on the grid, dominates the error. These two tests separate two effects. The first is the convergence of the local curvature estimator. The second is the bias of the complete relabelling dynamics caused by the finite kernel width.

Accuracy deteriorates when the physical radius approaches the Gaussian support, as anticipated by Equation~\ref{eq:kernel-window}. For the same reason, the recrystallisation analysis excludes nuclei whose initial radius is smaller than two cells (Methods). The JMAK runs use the default kernel, $\sigma_K=2$ cells or $\ell_K=2\Delta x$, and nuclei with an initial radius of three cells. Nuclei smaller than two cells would lie below the resolution limit of this kernel. This analysis rule is separate from the removal of unresolved regions during the evolution, described in Section~\ref{sec:timestep}. Figure~\ref{fig:gaussian-accuracy} reports the geometric-resolution dependence and makes the kernel-width trade-off explicit.

Figure~\ref{fig:foundations} links the local geometry to increasingly global consequences. Panel (a) shows that the integrated area-loss law is recovered for three shapes, despite their different initial curvature distributions. Panel (b) tests the same operator inside a many-grain network. The mean area rate changes sign near six sides, as required by von Neumann--Mullins theory. The isolated crosses mark side classes with too few grains. They are not used to extend the fitted trend. Panel (c) demonstrates that a small boundary perturbation decays at the predicted rate at all nine sampled times. Panel (d) returns to the local estimator. It shows a systematic approach to the exact circle curvature as $R/\Delta x$ increases. Agreement across these panels is stronger evidence than any single circle test. It remains, however, a verification of the shared capillary and topology layers. It is not a validation of a material-specific mobility.

\subsection{Constitutive coupling of grain-boundary motion to material driving forces}
\label{sec:constitutive}
The tests in this section ask whether five processes can share the state, geometry and update layers. These are isotropic and anisotropic grain growth, recrystallisation, multiphase boundary growth and elastic coupling. Only the constitutive fields defined by each process may change. The code provides one common interface between a resident and a candidate region. Through it, each process supplies $M$, $\Delta G$ and $\gamma$ without replacing the numerical engine.

\subsubsection*{Constitutive reductions used by the models}
Every bulk driving force is expressed in the same way, as the difference in free energy per unit volume between candidate and resident. Superscripts such as $\Delta G^{\rho}$ or $\Delta G^{\rm chem}$ only name the physical origin of a contribution. In general,
\begin{equation}
\Delta G_{q-p}=\psi_q-\psi_p.
\label{eq:delta-g-definition}
\end{equation}
Here $p$ is the resident label and $q$ the candidate label. For a candidate with lower energy, $\Delta G_{q-p}<0$. According to Equation~\ref{eq:master}, the resident then recedes. The following modules differ only in how they compute $\psi_p$ and $\psi_q$.

\subsubsection*{Stored-energy, multiphase and elastic modules}
\emph{Stored energy.} A deformed grain stores energy in its dislocations. Each dislocation carries an elastic line energy of roughly $\mu b^2/2$ per unit length. Multiplying by the dislocation density $\rho$, the dislocation length per unit volume, gives the stored energy per unit volume. This estimate provides the bulk driving force used by recrystallisation models based on stored energy \citep{HesselbarthGobel1991,Bernacki2011}:
\begin{equation}
E_\rho=\frac12\beta_d\mu b^2\rho,
\qquad
\Delta G^{\rho}=E_\rho(q)-E_\rho(p),
\qquad
\dot\rho=-k_{\mathrm{rec}}\rho.
\label{eq:stored-energy}
\end{equation}
Here $E_\rho$ is the stored energy per unit volume, and $\beta_d$ is a dimensionless constant of order one. It corrects the simple line-energy estimate for the character and arrangement of the dislocations. The symbol $\mu$ is the shear modulus, and $b$ is the magnitude of the Burgers vector. The middle relation states that the driving force is the difference in stored energy between candidate and resident. A dislocation-free recrystallised grain ($q$) growing into a deformed grain ($p$) therefore has $\Delta G^{\rho}<0$, and the deformed resident recedes. The last relation is an optional law for static recovery. It states that the dislocation density decays exponentially in time, with rate constant $k_{\mathrm{rec}}$.

\emph{Multiphase chemistry.} The multiphase module uses simple polynomial Gibbs energies. They serve as a compact, replaceable substitute for tabulated or CALPHAD free-energy landscapes \citep{ChenKhachaturyan1991}:
\begin{equation}
G_q(c,T)=a_q+b_qc+e_qc^2+d_q(T-T_{\mathrm{ref}}),
\qquad
\Delta G_{q-p}=G_q-G_p.
\label{eq:precipitation-gibbs}
\end{equation}
Here $c$ is the composition, $T$ the temperature and $T_{\mathrm{ref}}$ a reference temperature. The numbers $a_q$, $b_q$, $e_q$ and $d_q$ are the polynomial coefficients of phase $q$. All energies are per unit volume, in \si{J.m^{-3}}, so that $G_q$ plays the role of $\psi_q$. In words, each phase has a free energy that depends quadratically on composition and linearly on temperature. Among the neighbouring candidates, the module selects the admissible one with the lowest $G_q$. The polynomial is used here only to test the coupling and the phase competition. A calibrated application can replace it with a CALPHAD query. The geometry and topology update do not change.

\emph{Elasticity.} A transformed region often has a slightly different natural shape than the surrounding matrix. This stress-free shape change is described by the eigenstrain $\epsilon^*$. Its size is set by the lattice misfit $\delta$. If the material cannot adopt this shape freely, elastic energy is stored:
\begin{equation}
E_{\mathrm{el}}=\tfrac12\,C\,\left(\epsilon-\epsilon^*\right)^2,
\qquad
\Delta G^{\mathrm{eff}}_{q-p}=\Delta G^{\mathrm{chem}}_{q-p}+E_{\mathrm{el},q}-E_{\mathrm{el},p}.
\label{eq:elastic-scaling}
\end{equation}
Here $\epsilon$ is the total strain and $C$ the elastic stiffness. The first relation is the familiar elastic energy density, ``one half times stiffness times elastic strain squared''. The elastic strain is the total strain minus the eigenstrain. For an anisotropic crystal, $C$ and the strains have several components, and the product is summed over all of them. In tensor form, these are the stiffness tensor $\mathbf C$ and the eigenstrain $\boldsymbol\epsilon^*$ listed in Table~\ref{tab:parameter-provenance}. The second relation adds the difference in elastic energy between candidate and resident to the chemical driving force $\Delta G^{\mathrm{chem}}_{q-p}$. For a fixed particle shape and fixed elastic constants, the elastic strain is proportional to the misfit $\delta$. The energy is then proportional to $\delta^2$. This quadratic scaling is used as the regression check for the elastic module. It is not presented as a full-field Eshelby solution.

\subsubsection*{Boundary mobility and anisotropy modules}
\emph{Misorientation-dependent mobility.} Low-angle grain boundaries usually move more slowly than high-angle boundaries. The misorientation dependence of the boundary mobility follows the Read--Shockley form, which describes a boundary as an array of dislocations \citep{ReadShockley1950}. Let $\Delta\theta$ be the misorientation angle between the two grains, and let $\theta_m$ be the high-angle threshold. With the normalized angle $\vartheta=\Delta\theta/\theta_m$, the mobility is
\begin{equation}
f_{\mathrm{RS}}(\vartheta)=\begin{cases}
\vartheta(1-\ln\vartheta),&0<\vartheta<1,\\
1,&\vartheta\geq1,\\
0,&\vartheta=0,
\end{cases}
\qquad
M(\Delta\theta,T)=M_0f_{\mathrm{RS}}(\vartheta)\exp\left(-\frac{Q}{R_gT}\right).
\label{eq:read-shockley-mobility}
\end{equation}
Here $M_0$ is the mobility prefactor, $Q$ the activation energy, $R_g$ the gas constant and $T$ the temperature. The function $f_{\mathrm{RS}}$ rises from zero at $\vartheta=0$ to one at $\vartheta=1$ and stays at one above. The exponential factor is the usual Arrhenius temperature dependence. The function $f_{\mathrm{RS}}$ has the shape of the Read--Shockley energy of a low-angle dislocation array. Here it is used to interpolate the mobility between low and high angles. It saturates at the high-angle threshold $\theta_m=15^\circ$.

\emph{Orientation-dependent interfacial energy.} The interfacial energy can also depend on how the boundary is oriented relative to the crystal axes of the grain. Let $\mathbf n_c$ be the boundary normal expressed in the crystal frame of the resident grain $p$. The code evaluates $\gamma$ in this frame. The two sides of the same boundary may therefore use slightly different values, and the pair-averaged velocity of Section~\ref{sec:geometry-methods} combines both sides. The code accepts a general orientation-dependent energy. All anisotropic simulations reported here use the following cubic form:
\begin{equation}
\gamma=\gamma_0\left[1+\chi_4\left(n_{c,x}^4+n_{c,y}^4+n_{c,z}^4-\frac35\right)\right],
\label{eq:cubic-anisotropy}
\end{equation}
where $\gamma_0$ is the mean interfacial energy, $\chi_4$ is the anisotropy strength, and $n_{c,x}$, $n_{c,y}$ and $n_{c,z}$ are the components of $\mathbf n_c$. The bracket is largest when the normal points along a cube axis and smallest when it points along a cube diagonal. The term $-3/5$ makes its average over all directions in three dimensions zero, so $\gamma_0$ remains the mean energy. In two dimensions, the code uses the four-fold form
\begin{equation}
\gamma=\gamma_0\left[1+\chi_4\cos4\theta_c\right],
\label{eq:fourfold-anisotropy}
\end{equation}
where $\theta_c$ is the angle of the boundary normal relative to the crystal axes. Its average over all directions is also $\gamma_0$, and $\gamma$ varies between $\gamma_0(1-\chi_4)$ and $\gamma_0(1+\chi_4)$. In three dimensions, the variation of Equation~\ref{eq:cubic-anisotropy} is smaller, from $\gamma_0(1-\tfrac{4}{15}\chi_4)$ to $\gamma_0(1+\tfrac{2}{5}\chi_4)$. The present implementation inserts this local $\gamma(\mathbf n)$ into the capillary term $\gamma\kappa$. For the four-fold form, the interface stiffness is $\gamma+\mathrm d^2\gamma/\mathrm d\theta_c^2=\gamma_0(1-15\chi_4\cos4\theta_c)$. It becomes negative for some orientations when $\chi_4>1/15$. The two-dimensional anisotropic runs use $\chi_4=0.10$, which exceeds this limit. Because the local $\gamma$ stays positive, the motion with $\gamma\kappa$ remains well posed. It is, however, not the exact energy-decreasing motion of the anisotropic energy, which would contain the stiffness instead of $\gamma$. The missing orientations of a strongly anisotropic equilibrium shape are, however, not reproduced. A fully variational formulation for strong anisotropy would instead need the surface stiffness or the Cahn--Hoffman term \citep{CahnHoffman1974}. It would also need a regularization at orientations that are missing from the equilibrium shape.

\subsubsection*{Focused process-level checks}
The real-material suite provides quantitative anchors. At a triple junction, the boundary tensions must balance: $\sum_{j=1}^3\gamma_j\mathbf t_j=0$. Here $\mathbf t_j$ is the unit tangent of boundary $j$, pointing away from the junction. For equal boundary energies, three equal vectors can sum to zero only when they are separated by $120^\circ$. This is the Herring condition, which we test with pure-Al parameters \citep{Herring1951}. The mean grain size follows the parabolic coarsening law of normal grain growth \citep{BurkeTurnbull1952,Hillert1965,Anderson1984,FanChen1997}. In this law, the square of a characteristic grain radius grows linearly in time with a rate constant $K_{\rm gg}$. In a grain network, the prefactor of this law depends on the dimension, the topology and the chosen size measure. The polycrystal test therefore checks only that the square of a characteristic radius grows linearly in time. This radius is computed from the mean grain area. The test reports the normalized rate constant $K_{\rm gg}/(M\gamma)$. It does not equate this coefficient with the exact result for an isolated circle (Equation~\ref{eq:circle-shrinkage}). In two dimensions, the von Neumann--Mullins relation (Equation~\ref{eq:von-neumann-mullins}) states that the rate of area change of a grain depends only on its number of sides. The three-dimensional topology results are interpreted against the MacPherson--Srolovitz generalization of this relation \citep{MacPhersonSrolovitz2007}. It links the rate of volume change of a grain to the integral of the mean curvature over its surface. It also links this rate to the topology of the network of faces and edges.

Recrystallisation is compared with the Johnson--Mehl--Avrami--Kolmogorov (JMAK) law \citep{JohnsonMehl1939,Avrami1939}. Taking the logarithm twice turns this law into a straight line:
\begin{equation}
X(t)=1-\exp(-K_{\rm J}t^m),
\qquad\text{so that}\qquad
\ln[-\ln(1-X)]=\ln K_{\rm J}+m\ln t.
\label{eq:jmak-linear}
\end{equation} Here $X$ is the recrystallised volume fraction, $K_{\rm J}$ a rate constant and $m$ the Avrami exponent. On a plot of $\ln[-\ln(1-X)]$ against $\ln t$, the data therefore fall on a straight line whose slope is $m$. For nuclei that are all present from the start (site saturation), the ideal exponent equals the dimension: 2 in two dimensions and 3 in three. The calculations contain 21 saved states in 2-D and 16 in 3-D. Of these, 19 and 14 states lie in the accepted range of transformed fractions. They give $m=1.528$ with a fit quality of $R^2=0.978$ in two dimensions and $m=1.975$ with $R^2=0.980$ in three dimensions. The deviation from the ideal values has several causes. These are the finite size of the nuclei, the grid-resolved fronts, impingement at junctions and the finite fitting window. It is not a fitted material constant (Supplementary Figure~S5).

In the multiphase regression test, the lower-energy candidate is selected while three competing labels coexist during the transient. The fraction of the favoured phase increases, as expected. The elastic regression test recovers the quadratic scaling of the misfit energy in Equation~\ref{eq:elastic-scaling}. These checks show that the constitutive inputs are routed correctly. They do not replace material-specific calibration of CALPHAD data, mobilities or elastic constants.

\subsection{Focused two- and three-dimensional microstructure tests}
\begin{figure}[!htbp]
  \centering
  \includegraphics[width=0.96\textwidth]{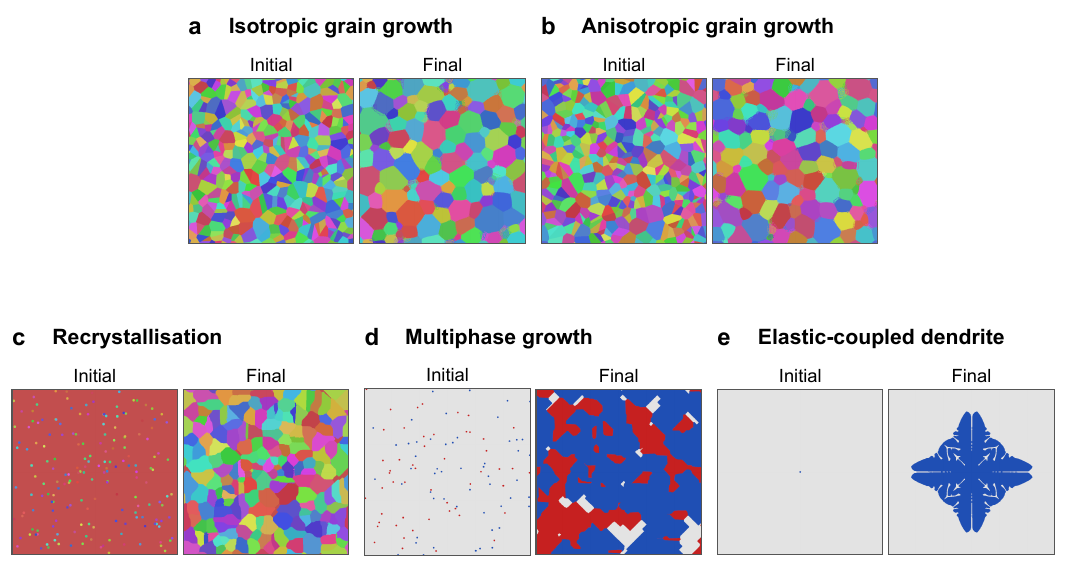}
  \caption{True initial and final states from focused two-dimensional GPK simulations: (a) isotropic grain growth, (b) anisotropic grain growth, (c) stored-energy-driven recrystallisation, (d) multiphase boundary growth and (e) elastic-coupled dendritic solidification. In (a), the boundary mobility $M$ and the interfacial energy $\gamma$ are both constant. In (b), $M$ has the same constant value. Only $\gamma$ depends on the boundary orientation relative to the crystal axes of each grain (four-fold anisotropy, Equation~\ref{eq:fourfold-anisotropy}, with anisotropy strength $\chi_4=0.10$). Panel (e) shows the central \SI{400}{\micro m} of a \SI{512}{\micro m} domain at $t=0$ and at the final time $t=\SI{80}{ms}$ (9603 timesteps). The initial seed has a radius of only \SI{2}{\micro m}. Persistent identities use categorical colours. Each panel is a PyVista rendering of the stored cell labels. No boundary line, smoothing, label regularisation or contour reconstruction is added. All measurements use the same stored labels.}
  \label{fig:qualitative-2d}
\end{figure}

The two-dimensional gallery in Figure~\ref{fig:qualitative-2d} shows the five capabilities retained in this study. These are isotropic grain growth, anisotropic grain growth, recrystallisation, multiphase boundary growth and elastic-coupled dendritic solidification. Each panel pairs the initial field with the end state of the same simulation. The first four cases use $768\times768$ cells. Grain growth and recrystallisation use $\Delta x=\SI{0.25}{\micro m}$ on a $192\times192~\si{\micro m^2}$ domain. The multiphase test uses $\Delta x=\SI{0.125}{\micro m}$ on a $96\times96~\si{\micro m^2}$ domain. The dendrite needs a much larger box. Its arms must grow far beyond the diffusion length without reaching the periodic edges. It therefore uses $1024\times1024$ cells with $\Delta x=\SI{0.5}{\micro m}$, a $512\times512~\si{\micro m^2}$ domain. The densities of grains and nuclei are kept fixed as the domain is enlarged. This gives 360 initial grains, 180 recrystallisation nuclei and 108 multiphase seeds. In all five cases, $\ell_K=\sigma_K\Delta x=\SI{1}{\micro m}$. This corresponds to $\sigma_K=4$ cells for grain growth and recrystallisation, $\sigma_K=8$ cells for the multiphase case and $\sigma_K=2$ cells for the dendrite. The two grain-growth cases use the same boundary mobility, $M=5.0\times10^{-11}\,\mathrm{m^4\,J^{-1}\,s^{-1}}$, and the same mean interfacial energy, $\gamma_0=\SI{0.324}{J.m^{-2}}$. In the isotropic case, both $M$ and $\gamma$ are constant. In the anisotropic case, the mobility remains constant. Only the interfacial energy depends on the orientation of the boundary normal relative to the crystal axes of each grain. This dependence follows the four-fold anisotropy of Equation~\ref{eq:fourfold-anisotropy} with $\chi_4=0.10$. This value exceeds $1/15$, the limit above which the interface stiffness of the four-fold form becomes negative for some orientations (Section~\ref{sec:constitutive}). The model inserts the local $\gamma(\mathbf n)$ into $\gamma\kappa$, which keeps the motion well posed. This run therefore shows the effect of an orientation-dependent energy, but it does not reproduce the missing orientations of a strongly anisotropic equilibrium shape.

After \SI{10}{s}, the isotropic and anisotropic cases retain 152 and 156 grains, respectively. In these runs, 61.8\% and 60.3\% of the cells have changed identity. In both end states, every surviving grain is one connected piece, and no piece consists of only one or two cells. The dense initial networks therefore become genuinely coarser networks. They do not simply undergo a uniform change of scale. Recrystallisation reaches $X_{\rm RX}=0.9915$ at \SI{20.02}{s}. The multiphase case changes 92.7\% of its labels by \SI{8}{ms}. It retains all three phases, with phase fractions $(f_0,f_1,f_2)=(0.0648,0.2897,0.6454)$. The dendrite grows from a circular seed of radius \SI{2}{\micro m} in a liquid with $c_0=0.02$ and an undercooling of \SI{10}{K}. It uses $M=2.0\times10^{-9}\,\mathrm{m^4\,J^{-1}\,s^{-1}}$, $D=3.0\times10^{-9}~\si{m^2.s^{-1}}$ and four-fold anisotropy with $\chi_4=0.04$. Solute transport follows the partitioning model of Equation~\ref{eq:partitioned-diffusion} with $k=0.3$, and elastic coupling is active. The run ends at $t=\SI{80}{ms}$ after 9603 timesteps, with a solid fraction of 0.145. The opposite tips are then \SI{290}{\micro m} apart. They remain at least \SI{110}{\micro m}, or 220 cells, from every domain edge. The displayed morphology has therefore not interacted with a periodic boundary. Panel (e) shows the central $400\times400~\si{\micro m^2}$ of the domain. PyVista \citep{SullivanKaszynski2019} displays the stored label cells directly. It adds no interface line and does not smooth the label map. Any visible staircase or isolated cell is therefore part of the stored state. The kinetics, in contrast, use the smooth Gaussian geometry of Section~\ref{sec:geometry}.

The visual changes in Figure~\ref{fig:qualitative-2d} correspond to different terms in the same velocity law. Panels (a) and (b) start from equally dense Voronoi networks. Isotropic capillarity produces comparatively equiaxed grains. In contrast, the orientation-dependent interfacial energy, acting with the same constant mobility, keeps a broader range of elongated boundary segments. The grain counts fall from 360 to 152 and 156, respectively. In panel (c), numerous small recrystallised nuclei grow into the deformed matrix with its stored energy. They grow until they impinge, at $X_{\rm RX}=0.9915$. Panel (d) shows competition between three phases rather than coarsening of a single phase. The lower-energy phases form connected domains, while all three phase fractions remain non-zero. Panel (e) shows a fully developed dendrite. Four primary arms grow along the preferred crystal directions, and secondary side branches form along their flanks. Rejected solute is trapped in thin liquid channels between the branches. This morphology requires both the interface law and solute partitioning. The rejected solute slows the flat parts of the front, while the tips reach fresh liquid and grow faster. The run is stopped long before the arms reach a periodic boundary. The panels display the stored label cells. These distinctions are therefore properties of the simulated states, not of a line overlay or plotting interpolation.
\FloatBarrier

\begin{figure}[!htbp]
  \centering
  \includegraphics[width=0.93\textwidth]{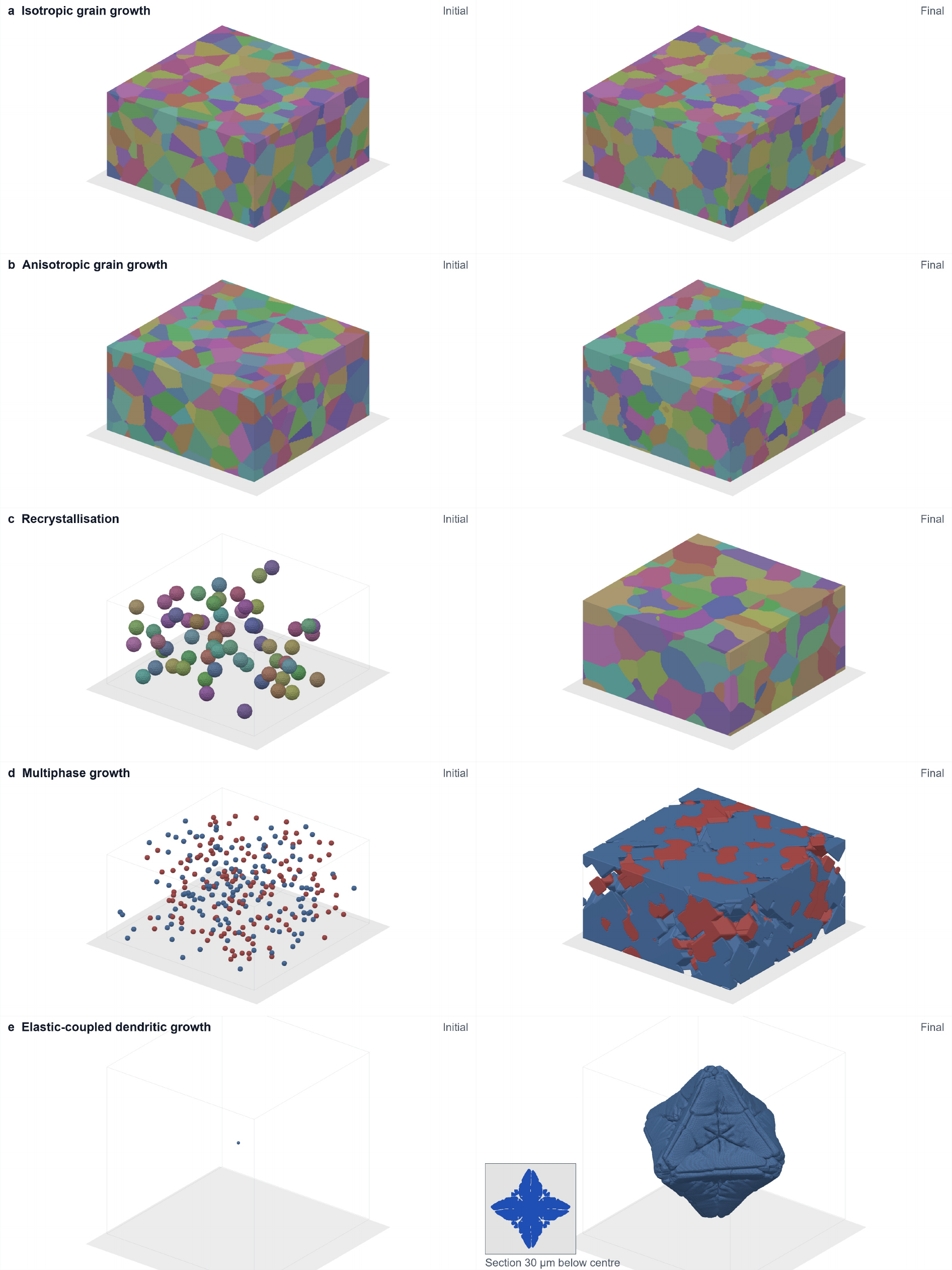}
  \caption{True initial and final states from three-dimensional GPK simulations: (a) isotropic grain growth, (b) anisotropic grain growth, (c) recrystallisation, (d) multiphase boundary growth and (e) elastic-coupled dendritic solidification. In (a), the boundary mobility $M$ and the interfacial energy $\gamma$ are both constant. In (b), $M$ has the same constant value. Only $\gamma$ depends on the boundary orientation relative to the crystal axes of each grain (cubic anisotropy, Equation~\ref{eq:cubic-anisotropy}, with anisotropy strength $\chi_4=0.06$). Rows (a)--(d) use $256\times256\times128$ cells ($64\times64\times32~\si{\micro m^3}$). Row (e) uses $512^3$ cells (a \SI{256}{\micro m} cube) and ends at $t=\SI{60}{ms}$ after 10805 timesteps. Panels (a) and (b) show the complete specimen exterior. Panel (c) shows all 64 initial nuclei and the complete final recrystallised exterior. Panels (d) and (e) show the complete transformed phases. The inset in (e) is the stored label section \SI{30}{\micro m} below the crystal centre. Row (e) adds ambient occlusion, which darkens the grooves between side branches. Surfaces come from raw VTK label cells without a boundary overlay, contour smoothing or interpolation.}
  \label{fig:qualitative-3d}
\end{figure}

The three-dimensional runs in Figure~\ref{fig:qualitative-3d} operate on full volumes. They are not extrusions of two-dimensional results. Isotropic and anisotropic grain growth start from periodic Voronoi fields with 240 grains on a fixed $64\times64\times32~\si{\micro m^3}$ domain. Both use $M=5.0\times10^{-11}\,\mathrm{m^4\,J^{-1}\,s^{-1}}$ and $\gamma_0=\SI{0.324}{J.m^{-2}}$. In the isotropic case, $M$ and $\gamma$ are constant. In the anisotropic case, $M$ remains constant. Only $\gamma$ depends on the boundary orientation relative to the crystal axes of each grain, through the cubic anisotropy of Equation~\ref{eq:cubic-anisotropy} with $\chi_4=0.06$. The fields first evolve on $192\times192\times96$ cells up to \SI{4.0}{s}. The labels and orientations are then copied, without filtering, onto a finer grid of $256\times256\times128$ cells ($\Delta x=\SI{0.250}{\micro m}$). Each fine cell takes the value of the nearest coarse cell. A GPK relaxation of \SI{0.10}{s} follows. The Gaussian width changes from 8 to 10.667 cells. The physical measurement length therefore stays fixed at \SI{2.667}{\micro m}. Three isotropic grains and one anisotropic grain disappear during the evolution, leaving 237 and 239 grains. The dense starting population makes simultaneous boundary migration visible without reducing the displayed specimen size.

Several diagnostics confirm the quality of the fine-grid end states. Relative to the initial fields, the number of cell faces between different grains decreases by 0.44\% and 1.54\%. Cells with at most two face neighbours of the same label are a sign of pixel noise. Their share falls from 1.97\% and 1.98\% initially to 0.69\% in both end states. Every grain is one connected piece, and no piece consists of only one or two voxels. No grain is completely enclosed by a single other grain. The smallest final grains contain 6526 and 7123 voxels. All initial, prolonged and final diagnostics are listed under ``Qualitative-gallery run definitions'' in the Supporting Information. Recrystallisation and multiphase growth run directly on the same $256\times256\times128$ grid with $\Delta x=\SI{0.25}{\micro m}$. Their domain is therefore the same $64\times64\times32~\si{\micro m^3}$ volume as that of the grain-growth cases. The recrystallisation case with 64 nuclei reaches $X_{\rm RX}=0.9949$ at the final time of \SI{20.02}{s}. Multiphase growth with 320 seeds reaches the phase fractions $(f_0,f_1,f_2)=(0.133,0.265,0.603)$ at the final time of \SI{6}{ms}. The three-dimensional dendrite uses the same material parameters, partitioning transport and elastic coupling as the two-dimensional dendrite. A three-dimensional crystal needs a larger volume before its arms and side branches separate. The run therefore uses $512^3$ cells with $\Delta x=\SI{0.5}{\micro m}$, a cube of edge \SI{256}{\micro m}, and $\sigma_K=2$ cells. It ends at the final time of \SI{60}{ms} after 10805 timesteps, with a solid fraction of 0.146. The opposite tips are then \SI{229}{\micro m} apart. They remain at least 26 cells, or \SI{13}{\micro m}, from every face. This is about nine diffusion lengths $D/V$ at the final tip speed.

Three-dimensional simulation adds information that a planar section cannot provide. It resolves both principal curvatures and the connectivity of grain faces, edges and vertices. It also captures topological events that a planar section can hide or misclassify. In addition, anisotropic and elastic driving forces can act through out-of-plane orientations and constraints. Panels (a) and (b) show the complete exterior of the full grain-growth volumes. All saved grains therefore contribute to the visible surface of the polycrystal. Panel (c) shows all 64 initial recrystallisation nuclei and the complete exterior after impingement, not a selected subset. Panels (d) and (e) show the complete transformed phases. No panel adds a boundary overlay, a smoothed display surface or geometric smoothing. Lighting from several directions only improves the visual impression of depth. All connectivity, volume and boundary-quality measurements use the stored three-dimensional labels.

These refined volumes are the main morphology results. Smaller grids are used only where they answer a controlled numerical question. The grain-by-grain comparison between GPK and phase field (PF) uses identical grids of $192^2$ cells and identical initial labels. The cost comparison, in contrast, first selects a grid for each method using accuracy criteria defined before the runs. The curvature and mesh-convergence studies use several resolutions. In every refinement comparison, the physical domain size and the Gaussian measurement length $\ell_K=\sigma_K\Delta x$ are held fixed. The kernel therefore samples the same physical neighbourhood, even though its width in cells changes with the grid spacing.

The sharp label representation allows this refinement without a prescribed number of grid points across a diffuse interface. In practice, the resolution is set by the required geometric scale. It is not set by an imposed interface thickness. The cost comparison in Section~\ref{sec:positioning} therefore compares the present implementations only after explicit accuracy criteria have been applied. These are morphology criteria for both methods and, for PF, an additional criterion on the energy of the diffuse profile. Both methods run on the same hardware. The result depends on these observables and tolerances. It is not a claim of lower cost than every phase-field formulation.

Figure~\ref{fig:qualitative-3d} should be read in terms of changes of surfaces and connectivity, not only of projected size. In panels (a) and (b), the three visible faces form one continuous polycrystalline specimen. They are not a collection of isolated particles. Colour changes indicate which grain occupies each cell, without superimposed boundary lines. A comparison of the initial and final states of the same specimen face shows that many boundary segments migrate simultaneously. Meanwhile, the full volumes lose three and one grains, respectively. The anisotropic case, in which only the interfacial energy depends on boundary orientation, retains more elongated surface sections. Panel (c) first shows all 64 nuclei inside the deformed volume. It then shows the continuous mosaic of recrystallised grains after three-dimensional growth and impingement. This change from nuclei to a grain network distinguishes distributed nucleation from the growth of isolated particles. Panel (d) shows how the three competing phases are connected through the depth of the specimen. A single two-dimensional section would classify some of these paths incorrectly. Panel (e) shows a three-dimensional dendrite. Six primary arms grow along the cube axes. Side branches grow from the arms and form the ribbed relief on the exterior. The inset is a horizontal section \SI{30}{\micro m} below the centre. It shows four primary arms, their side branches and the liquid channels between them. The run is stopped while a layer of liquid still separates every tip from the domain faces. The three-dimensional calculation therefore adds three kinds of information: both principal curvatures, the connectivity of faces, edges and vertices, and the out-of-plane constitutive response. None of these can be inferred reliably from the two-dimensional gallery.
\FloatBarrier

\subsection{Computational positioning of the general framework}
\label{sec:positioning}
We now compare the computational cost of GPK with that of phase field. Let $N$ be the number of cells. The stored data of GPK and the work for updating bulk fields are both proportional to $N$. The interface cells are first collected into a list of $N_\Gamma$ active cells. Gaussian smoothing and the evaluation of gradient and Hessian are then combined into one operation. The one-dimensional Gaussian is combined once with the fourth-order derivative stencils. The resulting products are evaluated only for the resident label at each active cell. Let $r\approx3.5\sigma_K$ be the radius of the truncated Gaussian, in cells. For a fixed dimension $d$, the work is proportional to $N_\Gamma(2r+5)^d$. The term $2r+5$ is the width of the combined smoothing and derivative stencil: the Gaussian spans $2r+1$ cells, and the derivative formula adds two cells on each side. The storage remains proportional to $N$. It does not depend on the number of OpenMP threads, and no full-domain buffer is needed for each grain. Wide three-dimensional kernels switch to a shared separable workspace of four arrays when this is cheaper. This alternative also needs storage proportional to $N$ and is independent of the thread count. Velocity averaging and the construction of topology proposals are likewise limited to the interface. Topology-changing commits are processed in a fixed order, because an accepted move changes the connectivity test of nearby proposals. In contrast, the tested multi-order-parameter phase-field (PF) implementation stores and evolves data proportional to $N_\eta N$ for $N_\eta$ order parameters. It must also resolve the diffuse interface. These structural differences do not by themselves prove a speed or memory advantage. We therefore time the two implementations directly.

Table~\ref{tab:method-positioning} summarizes the practical differences and their limits. The key distinction from phase-field methods is not that individual grid cells are more accurate. Rather, it lies in how the grid spacing is chosen. In phase-field models, the grid must be fine enough to resolve a prescribed diffuse-interface thickness. This is the artificial width over which the order parameter changes between phases. In GPK, the grid spacing is instead chosen to achieve the desired geometric resolution of the sharp interface. As a result, GPK can afford a finer mesh or a larger three-dimensional domain for a given memory budget. This holds whenever a sharp-interface description is appropriate. Relative to cellular automata, the key physical quantities retain their standard physical meanings and units. These are the boundary mobility $M$, the thermodynamic driving force $\Delta G$, the interfacial energy $\gamma$ and the curvature $\kappa$. They are not absorbed into dimensionless transition probabilities. Relative to level-set and front-tracking methods, the labels make grain identity and topology changes automatic. Relative to threshold dynamics, convolution only measures geometry. It neither sets the timestep nor replaces the constitutive velocity. These features are most useful for problems with many regions and with CALPHAD, stored-energy or elastic fields. Phase field remains preferable when the diffuse-interface structure or non-equilibrium order-parameter profiles are themselves the physical quantities of interest.

\begin{table}[!htbp]
\centering
\caption{Positioning of GPK relative to common microstructure-evolution techniques. Statements describe modelling structure, not universal performance rankings. $M$ is the boundary mobility, $\Delta G$ the thermodynamic driving force and $\gamma$ the interfacial energy.}
\label{tab:method-positioning}
\small
\begin{tabular}{@{}p{0.15\textwidth}p{0.19\textwidth}p{0.23\textwidth}p{0.34\textwidth}@{}}
\toprule
Method & Persistent interface state & Geometry/topology & Principal trade-off for the present applications \\
\midrule
GPK & One categorical label and a lineal sub-cell coordinate & Interface-local Gaussian geometry; certified automatic relabelling & Compact bulk state and explicit $M,\Delta G,\gamma$ coupling; requires a known sharp-interface law and a reported Gaussian scale \\
Multi-order-parameter phase field \citep{Chen2002,Moelans2008,Steinbach2009} & Several continuous fields & Diffuse interfaces; topology is implicit & Versatile free-energy coupling, but interface thickness and multiple fields must be resolved \\
Level set/front tracking \citep{Sethian1999,Bernacki2011} & Signed-distance or front functions & Accurate sharp transport; junction bookkeeping is additional & High geometric fidelity, with more machinery for many labelled regions and multiphase competition \\
Cellular automaton \citep{HesselbarthGobel1991} & Categorical cell state & Neighborhood-rule topology & Inexpensive and simple, but curvature and kinetic calibration can depend on the lattice rule \\
Threshold dynamics \citep{Merriman1992,EsedogluOtto2015} & Categorical regions & Convolution and thresholding constitute the motion step & Elegant mean-curvature flow; direct non-capillary constitutive velocities require extensions \\
\bottomrule
\end{tabular}
\end{table}

\begin{figure}[!htbp]
  \centering
  \includegraphics[width=0.98\textwidth]{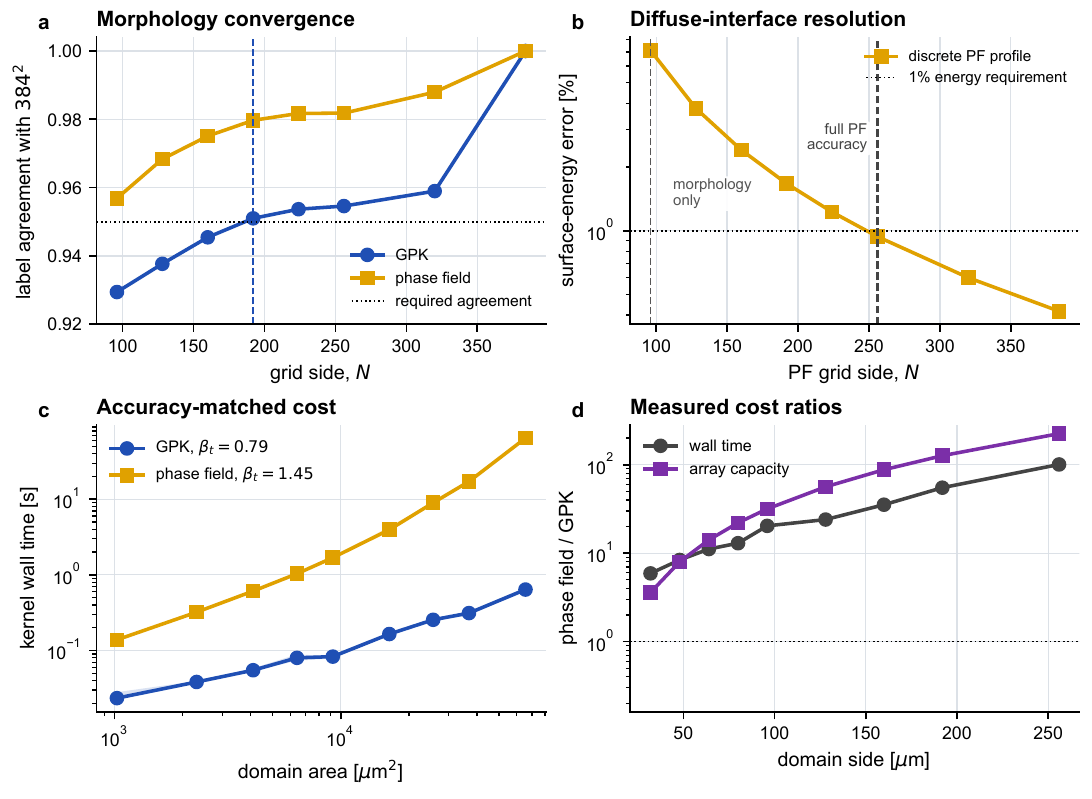}
  \caption{Accuracy-matched GPK--phase-field cost audit. (a) Morphology convergence against each method's $384^2$ result. $N$ is the number of cells along one side of the grid. The vertical line marks the first GPK grid that satisfies all four morphology requirements. (b) Relative error of the discrete PF planar-interface energy. PF passes the categorical morphology tests at $96^2$ but requires $256^2$ to satisfy the additional 1\% diffuse-profile requirement. (c) Median wall times of the evolution kernel for the selected $192^2/256^2$ base discretisations. These were transferred at fixed spacing and grain density to nine physical domain sizes from 32 to \SI{256}{\micro m}. All runs end at \SI{0.05}{s}. Shading gives the range of three runs, and $\beta_t$ is the exponent of a power-law fit against domain area. (d) Measured PF/GPK ratios of wall time and array capacity on a logarithmic ordinate. Wall time is the elapsed real (clock) time of a run, and array capacity is the memory allocated for the numerical arrays of each solver. Values above unity favour GPK. The largest run contains $768^2/1024^2$ cells and 1280 grains/order parameters. All runs use the same compiler, hardware and 24-thread configuration. The cost advantage is conditional on resolving the PF diffuse field, not a universal solver ranking.}
  \label{fig:benchmark}
\end{figure}

The main cost comparison is made at equal, predefined accuracy rather than at equal grid size. Both methods evolve the same periodic Voronoi microstructure with 80 grains on a \SI{64}{\micro m} square up to \SI{0.25}{s}. The resolution is varied from $96^2$ to $384^2$ cells. Meanwhile, the physical Gaussian measurement length and the PF diffuse-interface width are both held at \SI{0.667}{\micro m}. The PF time scale (``clock'') is recalibrated at each resolution with the shrinking-circle test. Each result is compared with the $384^2$ reference on the same lattice. Four requirements were defined before the runs. The mean absolute relative error (MAE) of the grain areas must be at most 0.030. The correlation of the grain areas must be at least 0.995. At least 95\% of the cells must carry the same grain label as in the reference. The boundary F1 score must be at least 0.900, with a matching distance of two reference cells. The F1 score is defined in Section~\ref{sec:gpk-pf}. GPK first passes all four criteria at $192^2$ cells ($\Delta x=\SI{0.333}{\micro m}$). Figure~\ref{fig:benchmark}(a) shows this convergence for both methods.

For a diffuse-interface calculation, accuracy cannot be judged from morphology alone. To compare morphologies, each PF cell is assigned to the grain whose order parameter $\eta_g$ has the largest value at that location. In phase-field models, these order parameters are continuous fields that distinguish the different grains or phases. In addition, a separate test checks the energy of a relaxed flat PF interface. The excess free energy per unit area of a flat interface scales with $\sqrt{\kappa_{\rm PF}}$. Dividing the discrete excess energy by this factor therefore gives a dimensionless value. For $\omega_{\rm PF}=1.5$, its exact continuum value is $\sqrt{2}/3$. We require this energy error to be below 1\%, a limit defined in advance. With this requirement, the coarsest acceptable PF grid moves from $96^2$ (morphology only) to $256^2$ ($\Delta x=\SI{0.250}{\micro m}$). Its measured error is 0.939\%. The full-accuracy comparison therefore uses 1.33 times more PF points per direction, or 1.78 times as many cells. This advantage applies only when the PF profile must also be resolved. If the grain morphology is the only observable, the $96^2$ PF result already passes the morphology tests. No coarse-mesh advantage can then be claimed.

The selected grid spacings, $192^2$ for GPK and $256^2$ for PF on the \SI{64}{\micro m} domain, are then kept fixed. With these spacings, we simulate nine domains with sides of 32, 48, 64, 80, 96, 128, 160, 192 and \SI{256}{\micro m}. The grain density remains $80/(\SI{64}{\micro m})^2$, which gives 20--1280 grains. For every size, we time three fresh runs of the evolution up to a common physical time of \SI{0.05}{s}. The order of the two solvers alternates between runs. All runs use the same Intel Core Ultra 9 285 processor with 24 OpenMP threads. The largest case contains $768^2$ GPK cells and $1024^2$ PF cells with 1280 PF order parameters. Here, wall time is the elapsed real (clock) time of the evolution kernel. Array capacity is the memory allocated for the numerical arrays of each solver. For the central \SI{64}{\micro m} case (Table~\ref{tab:gpk-pf-cost}), GPK needs 27 steps and 0.0693 s. PF needs 72 steps and 0.5888 s, a factor of 8.50 more. At \SI{256}{\micro m}, the median wall times are 0.758 and 72.677 s, a factor of 95.87. The array capacities increase from 2.85 MiB (GPK) and 40.25 MiB (PF) at \SI{64}{\micro m} to 45.56 MiB and 10.00 GiB at \SI{256}{\micro m}. In a separate storage test on $64^2$--$512^2$ grids, GPK needs 81 bytes per cell, independent of the number of OpenMP threads. The PF memory grows with both the number of cells and the number of order parameters.

We fit power laws, $t_{\rm wall}\propto A_{\rm dom}^{\beta_t}$, of the wall time $t_{\rm wall}$ against the physical domain area $A_{\rm dom}$ over the tested range. The exponents are $\beta_t=0.807$ for GPK (fit quality $R^2=0.985$) and $\beta_t=1.502$ for PF ($R^2=0.981$). The corresponding exponents for the array capacity are 1.00 and 1.995. In words, the GPK cost grows more slowly than the domain area, whereas the PF cost grows faster than the area. This is consistent with two facts. GPK works only near interfaces, whereas PF stores and updates $N_\eta$ dense order-parameter fields, and $N_\eta$ grows with the domain. These speed-up factors are empirical measurements obtained with the specific solver implementations, hardware and accuracy criteria used in this study. They should not be interpreted as fundamental theoretical limits on the relative computational cost of the two approaches. Different phase-field solvers may yield different comparisons. Examples are semi-implicit, implicit, spectral, GPU-accelerated or sparse active-set schemes. The comparison on identical $192^2$ grids is therefore given in the Supporting Information only as a diagnostic of the relative overhead of the current implementations. It is clearly separated from the main claim, which compares the two methods on grids chosen to achieve equivalent physical accuracy. The Supporting Information also contains the unprocessed results of the resolution study, all timing repeats, the memory counts and the independent analysis of kinetic accuracy.

\begin{table}[!htbp]
\centering
\caption{Measured computational cost for the central accuracy-matched \SI{0.05}{s}, \SI{64}{\micro m}, 80-grain comparison in the extended domain-size sweep. Wall time is the elapsed real (clock) time of the evolution kernel. Wall times are medians of three evolution-only runs, and parentheses give the observed range. Array capacity is the memory allocated for the numerical arrays owned by each solver. It excludes allocator metadata and runtime-library storage.}
\label{tab:gpk-pf-cost}
\small
\begin{tabular}{@{}lrrr@{}}
\toprule
Quantity & GPK & Phase field & PF/GPK \\
\midrule
Grid side & 192 & 256 & 1.33 \\
Accepted steps & 27 & 72 & 2.67 \\
Kernel wall time [s] & 0.0693 (0.0654--0.0693) & 0.5888 (0.5877--0.5949) & 8.50 \\
Wall time per step [ms] & 2.57 & 8.18 & 3.19 \\
Allocated array capacity [MiB] & 2.85 & 40.25 & 14.13 \\
\bottomrule
\end{tabular}
\end{table}

The order of the panels in Figure~\ref{fig:benchmark} is essential to the cost claim. Panel (a) first finds the coarsest grid that satisfies the four morphology requirements for each method. Panel (b) then adds the energy requirement for the diffuse profile, which applies only to PF. Only after these accuracy selections are fixed are the wall times compared in panel (c). The gap between the methods widens with domain size. This is consistent with GPK updating a compact label state near interfaces, while the tested PF solver evolves a growing number of dense order-parameter fields. Panel (d) shows that both the time ratio and the memory ratio exceed one and increase over the tested range. The figure therefore supports a comparison of these two implementations at equal accuracy. It does not establish a universal ranking over sparse, implicit, spectral or GPU phase-field solvers.
\FloatBarrier

\subsection{Grain-resolved comparison with phase field}
\label{sec:gpk-pf}
The comparison is resolved grain by grain. Agreement of the mean grain size alone could hide different growth or disappearance histories of individual grains. GPK is compared with the multi-order-parameter phase-field model of Fan and Chen \citep{FanChen1997,Chen2002}. In this model, each grain $g$ has its own order parameter $\eta_g$. It equals one inside grain $g$ and zero elsewhere. The dimensionless free energy and the kinetics of the model are
\begin{align}
F_{\rm PF}&=\int_\Omega\left[\sum_g\left(-\frac{\eta_g^2}{2}+\frac{\eta_g^4}{4}+\frac{\kappa_{\rm PF}}{2}|\nabla\eta_g|^2\right)
+\omega_{\rm PF}\sum_{g<g'}\eta_g^2\eta_{g'}^2\right]~\mathrm dV,\\
\frac{\partial\eta_g}{\partial\tau}&=L_{\rm PF}\left[\eta_g-\eta_g^3-2\omega_{\rm PF}\,\eta_g\sum_{g'\neq g}\eta_{g'}^2+\kappa_{\rm PF}\nabla^2\eta_g\right].
\label{eq:pf-comparison}
\end{align}
Here $\tau$ is dimensionless time, and $\kappa_{\rm PF}$, $\omega_{\rm PF}$ and $L_{\rm PF}$ are the dimensionless coefficients of the gradient energy, the grain interaction and the kinetics. In the energy, the first sum has its minima at $\eta_g=\pm1$ and penalizes gradients of each order parameter. The term with $\omega_{\rm PF}$ penalizes the overlap of two grains at the same point, and each pair of grains is counted once. It makes $\eta_g=0$ favourable wherever the order parameter of another grain is one, so that each point belongs to one grain. The second equation is the Allen--Cahn equation for each $\eta_g$, of the same form as Equation~\ref{eq:allen-cahn-normalized} for a single order parameter. Its sum runs over all other grains $g'$. It makes each $\eta_g$ evolve such that the energy decreases.
 Every physical grain receives its own order parameter. This is essential, because using one order parameter for two separate grains would merge their identities. Both solvers start from the same periodic Voronoi structure with 80 grains on $192^2$ cells and on the same \SI{64}{\micro\metre} square. The grid spacing is $\Delta x=64/192\,\si{\micro\metre}\approx\SI{0.333}{\micro\metre}$. Both calculations use $M=5.0\times10^{-11}\,\mathrm{m^4\,J^{-1}\,s^{-1}}$ and $\gamma=\SI{0.324}{\joule\per\square\metre}$. The characteristic initial grain radius is 12.1 cells. This is more than six times the Gaussian width of two cells. The PF model has three dimensionless coefficients (Equation~\ref{eq:pf-comparison}): the gradient-energy coefficient $\kappa_{\rm PF}=4$, the grain-interaction coefficient $\omega_{\rm PF}=1.5$ and the kinetic coefficient $L_{\rm PF}=1$. PF time is measured in the dimensionless unit $\tau$. The PF time scale is not fitted to the polycrystal. It is calibrated independently with a single shrinking circular grain. For this circle, the PF rate is $-\mathrm d(R/\Delta x)^2/\mathrm d\tau=7.9948$. Comparing it with the exact law $\mathrm dR^2/\mathrm dt=-2M\gamma$ maps one PF time unit to \SI{0.027417}{s}. With the PF step $\Delta\tau=0.045$ used in this comparison, one explicit PF step then corresponds to \SI{1.2338}{ms}.

\begin{figure}[!htbp]
  \centering
  \includegraphics[width=0.98\textwidth]{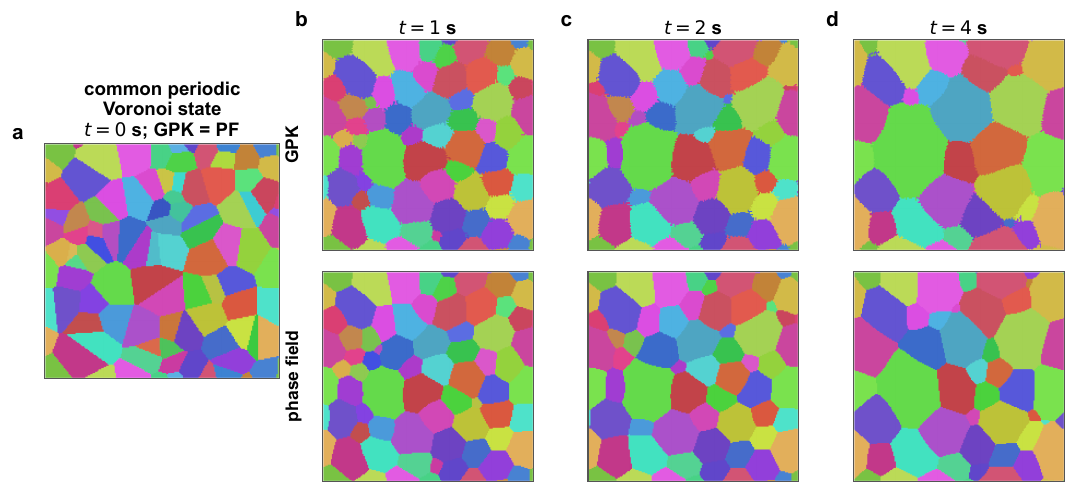}
  \caption{Grain-by-grain comparison of isotropic grain growth in GPK and phase field. Both start from the same periodic Voronoi structure with 80 grains on a $192^2$ grid. (a) Common initial microstructure used by both solvers. (b)--(d) GPK microstructure (upper row) and phase-field microstructure (lower row) at 1, 2 and \SI{4}{s}. In the phase-field rows, each cell takes the colour of the grain with the largest order parameter. These labels are elapsed physical times, not numerical integration-step sizes. A persistent grain identity uses the same colour in both methods. Both rows are PyVista renderings of the raw VTK cell images, with no line overlay or display smoothing. Quantitative overlap, boundary and topology metrics use those same categorical grids. The later separation reflects accumulated differences in grain-extinction and neighbour-switch timing.}
  \label{fig:gpk-pf}
\end{figure}

Figure~\ref{fig:gpk-pf} compares complete grain maps, not only boundary lines. Panel (a) is the periodic Voronoi structure passed unchanged to both solvers. Panels (b)--(d) show GPK above the PF result at 1, 2 and \SI{4}{s}. For PF, each cell is assigned to the grain whose order parameter $\eta_g$ has the largest value there. Each grain keeps the same colour in both rows. The quantitative comparison is evaluated at \SI{1}{s}, the time chosen for it before the runs. At this time, the areas of corresponding grains in the two methods have a correlation coefficient of $r=0.982$. Their mean absolute relative error (MAE) is 0.0979. Their mean intersection-over-union, the shared area divided by the combined area, is 0.853. In addition, 92.3\% of the cells carry the same grain identity in both methods. The boundary F1 score is 0.973 when boundaries within two pixels of each other are counted as matching. The F1 score combines two fractions. Precision is the fraction of GPK boundary pixels that match a PF boundary. Recall is the fraction of PF boundary pixels that are matched. It is their harmonic mean and ranges from zero for no agreement to one for perfect agreement. GPK and PF retain 60 and 62 grains. At \SI{2}{s}, the same five metrics are 0.963, 0.2126, 0.760, 86.2\% and 0.883, with 50 and 54 grains. At \SI{4}{s}, for which no acceptance criterion was set, they are 0.909, 0.4026, 0.724, 76.8\% and 0.655, with 29 and 40 grains. The longer-time difference shows that the two methods remove grains and switch neighbours in a different order. Complete metrics at all 17 saved times are tabulated in the Supporting Information.

Two constraints make this comparison fair. First, each of the 80 grains has one PF order parameter, and the PF time is fixed independently by the circular-shrinkage calibration. Second, GPK uses the same topology-certified update as in all other simulations, with no operation specific to this benchmark. The reported divergence therefore arises from the two numerical evolutions. It is not caused by merged labels or an uncalibrated time axis.

The labels 1, 2 and \SI{4}{s} in Figure~\ref{fig:gpk-pf} denote elapsed physical time. They are not the size of the numerical timestep. Both solvers use stable internal steps, and the PF time scale is fixed by the independent calibration. The growing difference instead follows from how the methods represent interfaces and topology changes. GPK moves a sharp boundary and accepts label changes through a topology-certified transaction. PF evolves order-parameter profiles of finite width. At a junction, several fields overlap and relax. Small differences in local boundary velocity, junction relaxation and the disappearance time of small grains therefore accumulate. Suppose one method removes a small grain or completes a neighbour switch first. The network of grain neighbours then changes. Later, the boundaries meet different neighbours and curvatures, even if the earlier microstructures agreed closely. Panel (d) therefore shows accumulated differences in the timing of topology changes at \SI{4}{s}. It is not evidence of a large-step instability, an initialization mismatch, a rendering artefact or recoloured labels. The quantitative equivalence claim is therefore restricted to the predefined comparison time of \SI{1}{s}. The states at 2 and \SI{4}{s} are reported as longer-time diagnostics.

Figure~\ref{fig:gpk-pf} allows the grain-by-grain comparison to be checked visually. Panel (a) removes the initial state as a source of disagreement, because the same Voronoi field is given to both solvers. In panel (b), at the predefined comparison time of \SI{1}{s}, most boundaries and surviving grains coincide. This agrees with the high label agreement and boundary F1 score. Panels (c) and (d) show where the agreement is lost later. Small grains disappear in a different order. Their neighbours then acquire different boundaries, and their histories diverge. The fixed colours allow these events to be followed grain by grain. The difference at the end is therefore reported as a diagnostic of the timing of topology changes. The quantitative claim is restricted to the early, predefined comparison time.

\subsection{Application to an Al--4 wt.\% Cu alloy: recrystallisation kinetics and $\theta'$ precipitate coarsening}
We selected Al--4 wt.\% Cu as a demanding but interpretable material case. It combines a network of $\alpha$-Al grains, recrystallisation driven by stored energy and the precipitation of stable $\theta$-Al$_2$Cu and metastable $\theta''$ and $\theta'$ phases. Coherency strains also contribute strongly to the plate shape of the precipitates. Experimental data on grain-boundary motion are available for Al--4 wt.\% Cu \citep{Rofman2009}. Recrystallisation kinetics are reported for Al--4.5 wt.\% Cu \citep{Matos2020AlCu}. Quantitative precipitate sizes and fractions are available, together with phase-field studies of the $\theta'$ shape \citep{RodriguezVeiga2018,Vaithyanathan2004,Ta2021}. The alloy therefore tests the common interface engine with real thermodynamic, kinetic and mechanical constraints.

The equilibrium calculation uses pycalphad \citep{OtisLiu2017} with the public, corrected COST-507 assessment of Al--Cu \citep{COST507}. At a Cu mole fraction of $c=0.017384$ (4 wt.\% Cu), the stable phase fractions are 0.9490 $\alpha$ and 0.0510 $\theta$ at \SI{473.15}{K}. At \SI{673.15}{K}, they are 0.9657 and 0.0343. At \SI{793.15}{K}, only $\alpha$ is stable. These equilibrium values fix the phase inventory. The metastable $\theta'$ evolution below instead uses the state Al--1.69 at.\% Cu at \SI{453}{K}. This state is supported by both experiments and phase-field studies \citep{Ta2021}.

For recrystallisation, the experiment on conventionally solidified material reports complete recrystallisation after five minutes at \SI{623.15}{K}. The material had been deformed to a 91\% area reduction \citep{Matos2020AlCu}. We use this single end point to calibrate an effective boundary mobility, $M=1.224\times10^{-13}\,\mathrm{m^4\,J^{-1}\,s^{-1}}$. The interfacial energy and the stored energy of the dislocations per unit volume (Section~\ref{sec:constitutive}) are fixed at $\gamma=\SI{0.30}{J.m^{-2}}$ and $E_\rho=\SI{1.0e6}{J.m^{-3}}$. The GPK simulation on $192^2$ cells reaches $X=0.9954$ at \SI{300.77}{s}. The S-shaped curve of $X(t)$ before this end point is a calibrated prediction. It is not an independent validation, because the paper provides micrographs rather than a measured $X(t)$ series. The unidirectionally solidified material in the same study recrystallises only partially. It is not represented by this run, which assumes nuclei present from the start and a uniform stored energy.

The grain-network comparison in Figure~\ref{fig:gpk-pf} uses the interfacial energy and mobility of Al in both solvers. It therefore tests the numerical interface evolution, not alloy-specific effects such as solute drag. The literature shows that individual grains in Al--4 wt.\% Cu can grow or shrink during hot deformation \citep{Rofman2009}. This is consistent with the grain-resolved histories. However, the experiment does not provide an initial condition cell by cell, so a quantitative overlay is not possible.

We also compare the largest usable timesteps. The PF test uses the explicit Fan--Chen update (Equation~\ref{eq:pf-comparison}). Here $\Delta\tau$ is the dimensionless PF timestep. The step-size sweep is compared with a run that uses a very small reference step, $\Delta\tau=0.005625$. The grain comparison and the timing runs use $\Delta\tau=0.045$, which lies inside the accurate range found below. Steps up to $\Delta\tau=0.055$ change the squared radius of a test circle by at most 0.061\%. A step of $\Delta\tau=0.0625$ produces an 11.3\% error and an order parameter of up to $\max|\eta|=1.20$, outside its physical range. Larger steps make the solution grow without bound. With the grid spacing of Figure~\ref{fig:gpk-pf} and the circle calibration, the last accurate and first inaccurate PF steps correspond to \SI{1.508e-3}{s} and \SI{1.714e-3}{s}, respectively. GPK accepts a stable timestep of \SI{3.008e-2}{s}. Larger requested steps are reduced through Equation~\ref{eq:interface-cfl}. The GPK step is therefore 17.6 times larger than the first inaccurate PF step and 19.9 times larger than the last accurate one. The complete sweep is tabulated in the Supporting Information. This comparison applies only to this explicit PF solver on the same grid and to this sharp-interface observable. Semi-implicit, fully implicit or spectral phase-field schemes can relax the step-size limit of the explicit method. We do not claim that they fail.

The precipitation module for this alloy adds the physics that the generic elastic regression test lacks. The total Cu concentration is advanced in time with conservative diffusion fluxes across the cell faces. ``Conservative'' means that the Cu leaving one cell enters its neighbour exactly. In COST-507, the relevant Al$_2$Cu compound has the Cu mole fraction $c_{\theta'}=1/3$. At \SI{453}{K}, the grand potentials of the FCC matrix and of Al$_2$Cu are equal at the matrix composition $c_m^{\rm eq}=3.719\times10^{-4}$. This is therefore the equilibrium matrix composition. Before the simulation, pycalphad tabulates the chemical driving force as a function of composition. GPK interpolates this table during the evolution. The $\theta'$ phase occurs in three orientation variants. Each variant has a tetragonal eigenstrain $(\epsilon_a,\epsilon_a,\epsilon_c)=(0.00746,0.00746,-0.051)$, with the axes permuted cyclically between variants. The resulting elastic fields are computed with a periodic Fourier solver for cubic Al ($C_{11}=107.07$, $C_{12}=63.08$, $C_{44}=28.52$ GPa). The interfacial energy $\gamma(\mathbf n)$ of the plates depends strongly on orientation. For such strongly anisotropic energies, the familiar capillary term $\gamma\kappa$ must be extended by a term that accounts for the change of $\gamma$ with orientation. We denote this extended capillary term by $P_{\rm cap}$. Like $\gamma\kappa$, it is an energy per unit volume. It is computed with the Cahn--Hoffman construction, which is the thermodynamically consistent form. In two dimensions, $P_{\rm cap}=(\gamma+\mathrm d^2\gamma/\mathrm d\theta_c^2)\,\kappa$, where $\theta_c$ is the angle of the boundary normal relative to the crystal axes and $\gamma+\mathrm d^2\gamma/\mathrm d\theta_c^2$ is the interface stiffness. The plate calculation here is three-dimensional. There, the same correction is computed from the Cahn--Hoffman vector, a vector built from $\gamma(\mathbf n)$ and its change with orientation, whose divergence along the interface replaces $\gamma\kappa$. For an isotropic energy, $P_{\rm cap}$ reduces to $\gamma\kappa$. The broad faces of the plates are observed to migrate much more slowly than their rims. This motivates an anisotropic mobility \citep{CahnHoffman1974,Ta2021}. Because the Gaussian measurement resolves orientations only approximately, this anisotropy is represented by a smooth effective mobility ratio of 51:1 between rim and face. This smooth form keeps slow broad faces and fast rims. It neither creates branches below the kernel scale nor prescribes a plate shape. Together, these terms give
\begin{equation}
v_n=M(\mathbf n)\left[\Delta G^{\rm chem}-\Delta E^{\rm el}-P_{\rm cap}\right],
\qquad \frac{\partial c_{\rm tot}}{\partial t}=\nabla\!\cdot\left[D(1-f_{\theta'})\nabla c_m\right].
\label{eq:alcu-kinetics}
\end{equation}
The first relation is Equation~\ref{eq:master} with three driving-force contributions. Here the plate ($\theta'$) is the resident, so a positive $v_n$ means plate growth. $\Delta G^{\rm chem}$ is the chemical driving force for this growth, that is, the free-energy density of the surrounding matrix minus that of $\theta'$. It corresponds to $\Delta G_{q-p}$ in Equation~\ref{eq:master}. $\Delta E^{\rm el}$ is the increase in elastic energy per unit volume when matrix is transformed into plate, computed from the elastic fields. It is the elastic energy of the resident minus that of the candidate, and it therefore enters with a minus sign. The term $-\Delta E^{\rm el}$ thus equals $E_{{\rm el},q}-E_{{\rm el},p}$ in Equation~\ref{eq:elastic-scaling}. The last term, $P_{\rm cap}$, is the anisotropic capillary term. The mobility $M(\mathbf n)$ depends on the orientation of the boundary. The second relation is the diffusion equation for Cu. Here $c_{\rm tot}$ and $c_m$ are the total and the matrix Cu concentrations, and $D$ is the Cu diffusivity. Diffusion takes place only in the matrix, which occupies the volume fraction $1-f_{\theta'}$, where $f_{\theta'}$ is the local $\theta'$ fraction. Each Cu flux between two neighbouring cells is computed once and shared by both. When a precipitate grows by $\Delta f_{\theta'}$, it needs the extra Cu amount $(c_{\theta'}-c_m)\Delta f_{\theta'}$. A local transfer supplies exactly this amount and removes it from the neighbouring matrix. This is the discrete form of the Stefan condition (Section~\ref{sec:transport}). The full formulation is given in Methods (Section~\ref{sec:alcu-module}), and the numerical checks in the Supporting Information.

The local voxel solver establishes the sharp-interface velocity, the selection of the plate habit, the conservative solute transfer and the topology properties. However, it cannot represent a statistically meaningful experimental volume with sub-nanometre cells. A \SI{1}{\micro m} cube with \SI{0.50}{nm} cells would contain $2000^3$ cells. We therefore add a population model for dilute particles. It uses the same persistent-particle representation and covers the experimentally studied interval from 120 to 168 h. Each particle carries a three-dimensional position, a crystallographic variant, a diameter $\ell_i$ and a thickness $H_i$ (the symbol $\ell_i$ is unrelated to the kernel length $\ell_K$). Its volume is that of a thin disc, $V_i=\pi \ell_i^2H_i/4$. The representative periodic volume is \SI{1}{\micro m^3}. This is large enough to contain the plates observed at 168 h. It also allows measured number densities to be converted directly into particle counts.

The 120 h state is initialized from the TEM statistics of Rodr\'iguez-Veiga et al. \citep{RodriguezVeiga2018}; all values are given as mean $\pm$ s.d. There are 131 $\theta'$ particles with a mean diameter of $142.5\pm25.1$ nm, a mean thickness of $4.9\pm0.6$ nm and $f_{\theta'}=0.8\pm0.1\%$. There are also 2883 $\theta''$ particles with a mean diameter of $45.1\pm2.1$ nm, a mean thickness of $2.1\pm0.2$ nm and $f_{\theta''}=1.0\pm0.2\%$. The correlation between diameter and thickness is estimated from the reported spread of the aspect ratio, using first-order error propagation. Particle sizes are then drawn from a two-dimensional normal distribution, keeping only positive values. The particles receive random periodic positions and equal numbers of each variant. All lengths are then multiplied by one common factor so that each phase has its measured 120 h volume fraction. This keeps the aspect ratios and the size ranking of the particles. The main realization contains 3014 precipitates. After this scaling, the mean $\theta'$ diameter is 127.67 nm, the mean thickness is 4.42 nm and the mean aspect ratio is 28.72.

The coarsening of $\theta'$ plates is controlled by the migration of ledges on their broad faces. Anisotropic mobility can substantially increase the aspect ratio of the plates \citep{MerleMerlin1981,Ta2021}. Capillarity removes the smallest particles first, in order of increasing volume. The measured number density at 168 h and the disappearance of $\theta''$ are used only to set the rate of particle loss. We introduce the normalized time $\tau_s=(t-120\,\mathrm{h})/48\,\mathrm{h}$. It runs from zero at 120 h to one at 168 h. When $\theta''$ (Al$_3$Cu) dissolves, its Cu is transferred to $\theta'$ (Al$_2$Cu). Per unit volume, $\theta''$ contains less Cu than $\theta'$, by the factor $(1/4)/(1/3)=0.75$. The particle number $N_{\theta'}$ and the volume fractions $f_{\theta''}$ and $f_{\theta'}$ then follow
\begin{equation}
N_{\theta'}=131\left(\frac{17}{131}\right)^{\tau_s},
\qquad
f_{\theta''}=0.010\,(1-\tau_s)^2,
\qquad
f_{\theta'}=0.008+0.75\,(0.010-f_{\theta''}).
\label{eq:alcu-population}
\end{equation}
In words, the number of $\theta'$ particles decreases exponentially from 131 at $\tau_s=0$ to the measured 17 at $\tau_s=1$. It is rounded to the nearest whole number. The $\theta''$ fraction falls smoothly from 1.0\% to zero. The $\theta'$ fraction starts at the measured 0.8\% and increases by the Cu released from $\theta''$. This approximation assumes equal atomic volumes and predicts $f_{\theta'}(168\,\mathrm{h})=1.55\%$. The measured value of $1.0\pm0.5\%$ is not used as an input. The shape change is taken from the verified local GPK calculation on $128^3$ cells. In that calculation, the aspect ratio of a plate increases from 8.0 to 10.417, by the factor $\Lambda_{\rm loc}=1.3021$. We transfer this dimensionless response to the population as $\Lambda=\Lambda_{\rm loc}^{\,\tau_s}$, so that $\Lambda$ grows from 1 at 120 h to $\Lambda_{\rm loc}$ at 168 h. Let $\ell_i^0$ and $H_i^0$ be the diameter and thickness of particle $i$ at 120 h. The sizes of the surviving particles are then
\begin{equation}
\ell_i=\Phi\,\Lambda^{1/3}\,\ell_i^0,
\qquad
H_i=\Phi\,\Lambda^{-2/3}\,H_i^0.
\label{eq:alcu-population-scale}
\end{equation} The factor $\Phi$ enlarges all surviving particles by the same amount. It is chosen at each time such that their total volume equals $f_{\theta'}L_{\rm box}^3$, where $L_{\rm box}=\SI{1}{\micro m}$ is the side of the volume. The factors $\Lambda^{1/3}$ and $\Lambda^{-2/3}$ change the aspect ratio $\ell_i/H_i$ by $\Lambda$. At the same time, they keep $\ell_i^2H_i$, and hence the particle volume, unchanged. The measured 168 h values of the $\theta'$ fraction, diameter, thickness and aspect ratio do not enter Equations~\ref{eq:alcu-population}--\ref{eq:alcu-population-scale}. They are nevertheless not fully independent tests. The measured 168 h particle count enters $N_{\theta'}$ and therefore controls the volume assigned to each survivor. They are therefore quantities conditioned on the measured data and used only as a check.

\begin{figure}[!t]
  \centering
  \includegraphics[width=0.98\textwidth]{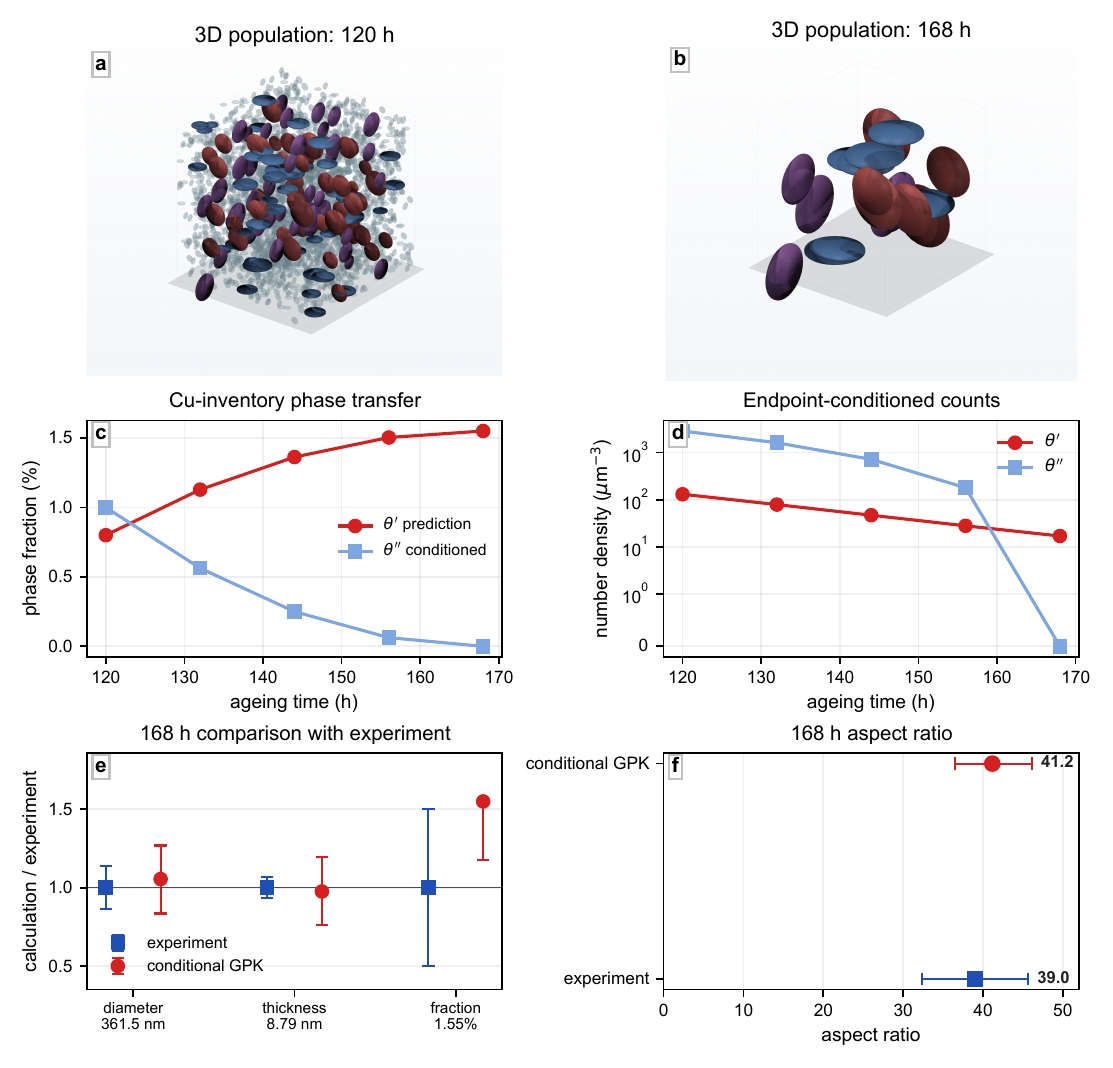}
  \caption{Quantitative three-dimensional Al--Cu population comparison at \SI{453}{K}. (a,b) High-resolution PyVista renderings of the \SI{1}{\micro m^3} VTK populations at 120 and 168 h. Colours identify the three $\theta'$ variants, and faint light-blue plates are $\theta''$. Soft lighting and floor shadows give an impression of depth. They do not change the particle geometry. (c) Stoichiometric Cu-inventory transfer computes the $\theta'$ volume fraction $f_{\theta'}$, while the disappearance of $\theta''$ is endpoint-conditioned. (d) Conditioned number-density trajectory. The 120 h volume contains 131 $\theta'$ and 2883 $\theta''$ particles. At 168 h, 17 $\theta'$ particles remain. (e,f) Conditional 168 h comparison of phase fraction, diameter, thickness and aspect ratio with TEM \citep{RodriguezVeiga2018}. Experimental error bars are the reported standard deviations, and GPK error bars span the structural sensitivity sweep. The phase-field statement is relative because Ta et al. did not tabulate an absolute aspect ratio \citep{Ta2021}.}
  \label{fig:alcu-validation}
\end{figure}

Figure~\ref{fig:alcu-validation} shows the full population at both experimental times. The simulation gives $f_{\theta'}=1.55\%$, a mean diameter of 361.5 nm, a mean thickness of 8.79 nm and an aspect ratio of 41.2. The measured values (mean $\pm$ s.d.) are $1.0\pm0.5\%$, $342.5\pm46.8$ nm, $9.0\pm0.6$ nm and $39.0\pm6.6$, respectively \citep{RodriguezVeiga2018}. The standardized residuals, that is, the differences divided by the measured standard deviations, are $+1.10$, $+0.41$, $-0.36$ and $+0.33$. We repeated the simulation with 200 independently sampled initial populations. The resulting 95\% intervals for diameter, thickness and aspect ratio are 358.5--364.4 nm, 8.69--9.00 nm and 39.90--42.03. A separate sensitivity study varies three model assumptions. The final particle count is varied within one reported standard deviation. The fraction of the Cu released by $\theta''$ that is captured by $\theta'$ is varied from 50 to 100\%. The transferred shape response is varied by $\pm10\%$. The resulting ranges are 1.175--1.550\% in fraction, 286.2--434.8 nm in diameter, 6.85--10.79 nm in thickness and 36.54--46.09 in aspect ratio. The experimental values lie inside every range. The three-dimensional phase-field study of Ta et al. serves only as a comparison of mechanisms. In that study, anisotropic mobility increased the aspect ratio about threefold relative to isotropic mobility, but no absolute value was tabulated \citep{Ta2021}.

This quantitative agreement is limited in four respects. First, the loss of particles and the disappearance of $\theta''$ are imposed from the measured data (conditioned) rather than predicted. Second, the $\theta'$ inventory uses a stoichiometric transfer that assumes equal atomic volumes. Third, the particle positions do not produce resolved overlap of diffusion fields or elastic interaction between particles. Fourth, the published TEM study reports only the means and standard deviations of the particle diameters and thicknesses. It does not report the diameter and thickness measured together for each individual particle. The correlation between diameter and thickness is therefore reconstructed to first order from these summary statistics. For the same reason, complete measured size distributions are not available. We therefore do not construct artificial distributions. We also do not compute statistical distances between simulated and measured distributions, such as Kolmogorov--Smirnov or Wasserstein statistics. The calculation reports the comparison with experiment, but no residual threshold decides whether a run passes. A machine-readable protocol, the transcribed measurements, a sensitivity table and a SHA-256 provenance manifest accompany the outputs. The Supporting Information contains the short local calculation on $128^3$ cells and its mesh test from 96 to 128 cells. It also contains the conservation result and the topology check. Only the dimensionless shape response of this calculation is transferred to the population model.

Figure~\ref{fig:alcu-validation} separates three kinds of information: the initial data, the conditioned evolution and the quantities used for comparison. Panels (a) and (b) show the complete three-dimensional populations. The dense $\theta''$ population present at 120 h disappears. The number of $\theta'$ plates falls. The surviving plates become visibly wider, and all three crystallographic variants remain. Panel (c) shows how Cu released by $\theta''$ contributes to the $\theta'$ inventory. Panel (d) marks the measured final particle count used to condition the selection of survivors. Panels (e) and (f) compare phase fraction, diameter, thickness and aspect ratio with experiment. None of these quantities was imposed as a target. All four values lie within the sensitivity ranges. This agreement remains conditional, however, because the final count and the disappearance of $\theta''$ constrain the evolution. The figure demonstrates a quantitatively traceable application to a real alloy. It does not present the population model as a fully predictive model of nucleation, particle interaction or coarsening.
\FloatBarrier

\section{Discussion}
The results establish GPK as a general technique for modelling microstructure evolution, not only as a curvature estimator. GPK combines five ingredients. These are compact persistent labels, a temporary view of the geometry, explicit process thermodynamics, a sub-cell interface position and a shared topology update. This separation allows one program to represent processes with different driving forces. It also makes it possible to identify whether a prediction depends on geometry, constitutive physics, transport or discretization. The Gaussian projection, computed separately for each label, supplies smooth interface geometry. It never becomes part of the stored state or of the evolution law.

GPK combines these ingredients in a specific way. The Gaussian field is temporary and label conditioned. The sub-cell fraction carries displacement independently of the Gaussian width. The mobility $M$, the driving force $\Delta G$ and the interfacial energy $\gamma$ enter through one uniform resident--candidate interface. Every crossing is mediated by a digital-topology transaction that does not depend on the constitutive model. Gaussian smoothing, a sharp-interface law, label storage and a simple-point test do not provide this structure individually. Their combination permits additional process modules to reuse the same geometry, kinetics and topology engine while keeping their assumptions auditable.

GPK occupies a middle ground between diffuse-interface and label-based methods. Like phase field, it derives the local law from a free energy and accepts thermodynamic, anisotropic and transport couplings. Like a cellular automaton, it stores compact labels in the bulk. GPK differs from both by keeping the temporary geometry separate from the stored state and from the definition of the timestep. It does not replace phase field when the diffuse-interface structure, interfacial adsorption or coupled non-equilibrium order parameters are themselves the subject of study. It targets problems in which a sharp-interface law is known and in which large populations of moving interfaces must be coupled consistently to material fields.

The verification reports both results for the whole framework and limitations of individual modules. The correct curvature sign, the circle curvature, the topology rules and the integral coarsening laws are reproduced. The two- and three-dimensional galleries show the characteristic microstructures of isotropic growth, anisotropic growth, recrystallisation, multiphase competition and elastic coupling. Chemical, mechanical and capillary terms all reach the same update path. On the other hand, the geometry layer cannot recover features smaller than its kernel. When the kernel length is fixed relative to the feature size, grid refinement removes the sampling error but not the smoothing bias, as the pore-closure test shows. The curvature at corners and junctions remains dependent on the kernel scale. The interface-local stencils also become more expensive as the kernel radius or the fraction of interface cells in three dimensions grows.

Several implementation directions follow directly. The interface-local products could be accelerated with vector instructions (single-instruction multiple-data, SIMD) on central processing units or with graphics processing units (GPUs). Grouping active cells into tiles that fit the processor cache would help further. The computational cost of solving the diffusion equations can be reduced by implicit or semi-implicit time-integration schemes. These remain stable at larger timesteps than explicit methods. In addition, the coupling between interface motion and bulk diffusion can be written in a conservative flux form. This keeps the mass balance exact and permits larger timesteps for the transport calculation. Adaptive grid refinement near interfaces would reduce the curvature error without increasing the cost in the bulk. Each specific application should be calibrated against independent experimental data or high-fidelity reference simulations. The verification presented here demonstrates that the numerical implementation is internally consistent. For example, the curvature converges with grid refinement, and the kinetic laws produce the expected growth rates. It also demonstrates that the constitutive equations are correctly coupled. It does not, however, guarantee predictive accuracy for any particular material system without such calibration.

GPK is therefore best suited to problems with three characteristics. First, the connectivity and spatial arrangement of many grains or phases are important. Second, the physical width of the interface itself is not the quantity being studied. It would be, for example, in a phase-field model of interface structure. Third, several non-linearly interacting material processes, such as grain growth, diffusion and precipitation, must share a single, verifiable and modular computational framework. The resulting modelling architecture allows errors in the geometry reconstruction, in the physical constitutive laws and in the label-update algorithm to be examined independently. The framework states the boundaries between these modules and the resolution limit of each explicitly. It thereby provides a reproducible basis for extending sharp-interface microstructure simulation. Process-specific assumptions are not hidden in neighborhood rules.

\section{Methods}
\subsection{Persistent state and model interface}
The program follows the eight operations of Section~\ref{sec:timestep}. This section summarizes the numerical modules used within those operations. Their formulas are given in the following subsections. The computational domain is a regular Cartesian grid in two or three dimensions. Each cell stores a grain or phase label, a sub-cell fraction, the composition, the temperature and optional orientation or stored-energy variables. A cell is active when at least one neighbour that shares a face or an edge has a different label. The boundary conditions are periodic or problem specific, as selected in the input file.

\subsection{Transient Gaussian geometry module}
\label{sec:geometry-methods}
This section gives the discrete formulas behind the geometry measurement of Section~\ref{sec:geometry}. For the resident label of an active cell, the binary indicator $I$ is smoothed with a separable Gaussian. Separable means that the smoothing is applied along one coordinate direction at a time. Along one direction, the smoothed value $\widetilde I_g(i)$ at cell $i$ is a weighted average of the indicator values $I_g(i+k)$ of its neighbours:
\begin{equation}
\widetilde I_g(i)=\sum_{k=-r}^{r}w_k\,I_g(i+k),
\qquad
w_k\propto\exp\!\left(-\frac{k^2}{2\sigma_K^2}\right).
\label{eq:discrete-gaussian}
\end{equation}
Here $w_k$ is the Gaussian weight of the neighbour at a distance of $k$ cells. For brevity, we write $f=\widetilde I_g$ in the rest of this section. The weights are scaled so that they sum to one. The sum is truncated at the radius $r$, which is $3.5\sigma_K$ rounded up to a whole number of cells. The derivatives of $f$ are then approximated by the standard fourth-order central differences. Each derivative uses the values of up to two neighbouring cells on each side, and its error decreases with the fourth power of the grid spacing:
\begin{equation}
\frac{\partial f}{\partial x}\approx\frac{f(x-2\Delta x)-8f(x-\Delta x)+8f(x+\Delta x)-f(x+2\Delta x)}{12\,\Delta x},
\label{eq:discrete-curvature}
\end{equation}
\begin{equation}
\frac{\partial^2 f}{\partial x^2}\approx\frac{-f(x-2\Delta x)+16f(x-\Delta x)-30f(x)+16f(x+\Delta x)-f(x+2\Delta x)}{12\,\Delta x^2}.
\label{eq:discrete-second}
\end{equation} The other coordinate directions are treated in the same way. A mixed second derivative, such as $f_{xy}$, is obtained by applying the first-derivative formula twice, first along $x$ and then along $y$. Using the same order of accuracy for the pure and the mixed terms keeps the Hessian consistent in all directions. This was confirmed in a strict test on spheres in three dimensions. We write $f=\widetilde I_g$, and $f_x$, $f_{xx}$, $f_{xy}$ and so on denote its partial derivatives. In two dimensions, the curvature is then
\begin{equation}
\kappa=-\,\frac{f_{xx}f_y^2-2f_xf_yf_{xy}+f_{yy}f_x^2}{\left(f_x^2+f_y^2\right)^{3/2}}.
\label{eq:curvature-hessian}
\end{equation}
In three dimensions, the same quantity follows from $\kappa=-\nabla\cdot(\nabla f/|\nabla f|)$. This expression is expanded analytically, with the product rule, into first and second derivatives of $f$ with respect to $x$, $y$ and $z$. The code evaluates this expanded form directly from the derivatives of $f$. It therefore never differentiates a numerically normalised normal field, which would add noise.
The Gaussian weights satisfy the design criteria stated in Section~\ref{sec:geometry}: they are positive, symmetric and normalized. Positivity prevents oscillations (``ringing'') near the jump of the indicator. Symmetry removes any shift of the boundary. Normalization keeps $f$ between zero and one. Unlike a simple box average, the Gaussian damps short waves smoothly and monotonically, without side lobes in its spectrum. Unlike an oscillating high-order filter, it does not overshoot at a label jump. Truncation at $3.5\sigma_K$ keeps more than 99.95\% of the total weight of the one-dimensional Gaussian before the weights are renormalized.

The implementation never computes the full smoothed field $f$. Smoothing and differentiation can be applied in either order. The Gaussian weights $w_k$ are therefore combined once, before the simulation, with the coefficients of these finite-difference formulas. This gives a combined ``smooth-and-differentiate'' weight set for the first and for the second derivative. Any derivative at an active cell is then obtained by applying one set of weights along each coordinate direction. For example, $f_x$ uses the first-derivative weights along $x$ and the plain Gaussian weights along $y$ and $z$. Similarly, $f_{xy}$ uses the first-derivative weights along both $x$ and $y$ and the plain Gaussian weights along $z$. These products give exactly the same result as smoothing followed by finite differences, also with periodic boundaries. They are evaluated only for the resident label of each active cell. The work is therefore proportional to $N_\Gamma(2r+5)^d$, and no geometry arrays per label or per thread are needed. The support of the direct product grows with the cube of the kernel radius in three dimensions. Kernels with a combined radius above 18 cells therefore use a shared separable smoothing with four arrays instead. The choice is fixed when the solver is set up. Both options evaluate the same Gaussian and the same fourth-order derivatives.

The default Gaussian width is two cells. It is always reported in cell units, so that the physical measurement length is $\ell_K=\sigma_K\Delta x$. Refined runs scale $\sigma_K$ to keep $\ell_K$ fixed. Normals and curvature are evaluated only at relevant interface cells. Cells with $|\nabla\widetilde I_g|\Delta x\leq10^{-4}$ receive a zero normal and are not interpreted geometrically. The magnitude $|\kappa|$ is capped at $1/\ell_K$. The default candidate is chosen among all neighbours with a different label. The neighbour whose label is most frequent in the local $3^d$ block is selected. A tie is broken by the alignment of the neighbour direction with the outward normal. The first rule suppresses one-cell label branches near junctions. The second rule keeps the correct direction on a boundary between two labels. It also prevents a neighbour with the same label along a grid axis from stalling an oblique interface. Phase-selection models may replace this geometric choice by the thermodynamically selected candidate. The selected pair then supplies the interfacial energy, the mobility and the driving force. The sensitivity to kernel width and grid resolution is treated as part of the curvature verification, not merely as a solver setting.

\subsection{Shared kinetic and topology update}
Equation~\ref{eq:master} gives the normal velocity. The stable timestep is the smaller of two limits. The first limits how far an interface may move in one step. The second, when diffusion is enabled, is the stability limit of explicit diffusion. Both are derived in Section~\ref{sec:stability}. A timestep supplied by the user is reduced to the same limit. The sub-cell fraction is updated with Equation~\ref{eq:alpha}. When $\alpha^*\leq0$, the selected candidate becomes the resident label, and its fraction is reset to one. Competing events are applied in a second pass. In this way, all geometry within a step is evaluated from one consistent state at the start of the step. The implementation keeps two copies of the state (double buffering). Every trial fraction is computed before any stored label is changed.

In every process module, a proposed crossing is checked with the TCKT rules of Section~\ref{sec:timestep}. Removing the cell from the resident and adding it to the candidate are tested separately. The reason is that a relabelling can keep one label connected while splitting or merging the other. Connected pieces are defined through face neighbours. A standard digital-topology count, which also considers edge and corner neighbours, classifies the local configuration as simple or critical. Only for a critical configuration with several local branches is a search over the whole periodic domain started (a flood fill). The change is rejected if removing cell $i$ separates branches of the resident. It is also rejected if adding cell $i$ joins pieces of the candidate that were previously separate. Local branches that already belong to one connected piece may reconnect. Physical closure of cavities and tunnels and the exchange of neighbours at junctions (T1 events) are therefore not frozen.

Rejected events are deferred, not deleted. Their sub-cell fraction is set to $10^{-6}$, and the complete kinetic force is evaluated again in the next step. Before the time integration, the normal velocity is averaged separately for every label pair. The averaging width is $0.85$ cell in two dimensions. In three dimensions, it is $0.375\sigma_K$, limited to the range from 0.85 to 3.00 cells. The wider support in three dimensions makes the velocity consistent over both tangential directions. It remains bounded relative to the geometry scale. For the production kernel of eight cells, the averaging width is therefore $\sigma_v=3.00$ cells. Section~\ref{sec:timestep} gives the reasons for these widths. After one or more accepted crossings, BGR evaluates narrow label projections. Their width is $0.70$ cell in two dimensions. In three dimensions, it is $0.125\sigma_K$, limited to the range from 0.70 to 1.00 cell. For every label pair, BGR performs equal numbers of exchanges in both directions. Proposals are ranked by their gain, and ties are broken by the cell index, so the result is deterministic. Every exchange must strictly decrease the boundary-face count $N_{\rm face}$. Two BGR sweeps are used in three dimensions. The number of cells of every label is preserved exactly by each sweep and by both together. The implementation records the numbers of accepted, deferred and exchanged events. Independent checks of the saved states count connected pieces, cavities, tunnels and boundary faces. These checks are acceptance tests only. They never modify a result.

\subsection{Transport module}
\label{sec:transport}
We first recall the exact conservation condition at a moving phase boundary. Consider a conserved composition $c$ with the diffusion flux $\mathbf J=-D\nabla c$ (Fick's law). Conservation in the bulk reads $\partial c/\partial t+\nabla\cdot\mathbf J=0$. We integrate this over a thin box that encloses a piece of the moving boundary. This gives the interfacial condition
\begin{equation}
J_n^{+}-J_n^{-}=v_n\left(c^{+}-c^{-}\right).
\label{eq:stefan}
\end{equation}
Here the superscripts $+$ and $-$ denote the values just in front of and just behind the boundary. $J_n=\mathbf J\cdot\mathbf n$ is the flux component along the normal. In words, if the composition jumps at a moving boundary, the moving boundary must reject or absorb solute. The difference in the diffusion fluxes on the two sides must exactly supply this solute. The same result follows at the level of one cell. Consider an interface cell with the resident fraction $\alpha$, the resident concentration $c_p$ and the candidate concentration $c_q$. Its average concentration is
\begin{equation}
\bar c=\alpha c_p+(1-\alpha)c_q.
\end{equation}
Differentiating in time and using Equation~\ref{eq:alpha-ode} for $\dot\alpha$ gives
\begin{equation}
\frac{\mathrm d\bar c}{\mathrm dt}
=\alpha\dot c_p+(1-\alpha)\dot c_q
+\frac{v_n}{\Delta x}(c_p-c_q).
\label{eq:cell-stefan}
\end{equation}
The first two terms describe diffusion inside each phase. The last term is the solute rejected or absorbed because the boundary moves through the cell. It is the discrete form of Equation~\ref{eq:stefan}. An exact implementation would therefore store both $c_p$ and $c_q$ in every cut cell.

The generic transport model is simpler. It stores only one average concentration per cell and uses the explicit update
\begin{equation}
c_i^{n+1}=c_i^n+\Delta t\,D_{\mathrm{eff},i}\nabla_h^2c_i^n,
\qquad
D_{\mathrm{eff},i}=\begin{cases}
D\alpha_i,&i\ \text{is an interface cell},\\
D,&i\ \text{is a bulk cell}.
\end{cases}
\label{eq:effective-diffusivity}
\end{equation}
Here $\nabla_h^2$ is the discrete Laplacian. In interface cells, the diffusivity is reduced in proportion to the resident fraction $\alpha_i$. This model restricts diffusion to the resident phase. However, it multiplies a cell-centred diffusivity by the discrete Laplacian instead of using fluxes across cell faces. When $D_{\mathrm{eff}}$ varies in space, it therefore does not conserve the total solute exactly. It also cannot represent a jump in concentration between phases. It is thus only an approximation inspired by the Stefan condition, not an exact implementation of Equation~\ref{eq:stefan}. It is used only in the generic qualitative coupling tests. The Al--Cu module instead uses conservative face fluxes and the local Cu transfer described with Equation~\ref{eq:alcu-kinetics}. Regimes with strong partitioning or solute trapping would likewise need a conservative cut-cell or multi-field scheme.

The solidification tests in Figures~\ref{fig:qualitative-2d}(e) and \ref{fig:qualitative-3d}(e) use an optional partitioning variant of this transport model. It still stores one average concentration $c_i$ per cell. Let $s_i$ be the solid fraction of cell $i$. Solute is assumed to diffuse only in the liquid. With a partition coefficient $k$, the solid holds $k$ times the liquid concentration. The liquid concentration of the cell is then
\begin{equation}
c_{L,i}=\frac{c_i}{1-s_i\,(1-k)}.
\label{eq:liquid-concentration}
\end{equation}
When the solid fraction of a cell grows by $\Delta s_i$ in one step, the cell rejects the solute $(1-k)\,c_{L,i}\,\Delta s_i$. This amount is moved to its liquid face neighbours, in proportion to their liquid fractions. Diffusion then uses fluxes across the cell faces,
\begin{equation}
c_i^{n+1}=c_i^n+\frac{\Delta t\,D}{\Delta x^2}\sum_{j\in\mathcal N(i)}\min(\ell_i,\ell_j)\,\bigl(c_{L,j}-c_{L,i}\bigr),
\label{eq:partitioned-diffusion}
\end{equation}
where $\mathcal N(i)$ are the face neighbours of cell $i$ and $\ell_i=1-s_i$ is its liquid fraction. Every face flux leaves one cell and enters its neighbour, so the total solute is conserved exactly. The rejected solute piles up ahead of the moving front. This is the diffusion-limited feedback that makes a growing front unstable and produces dendrites.

\subsection{Timestep stability derivation}
\label{sec:stability}
An explicit scheme becomes unstable when the timestep is too large. Small errors then grow from step to step instead of decaying. We derive the two timestep limits used by GPK. The first concerns diffusion. For one-dimensional forward-Euler diffusion, we define the grid Fourier number $\mathrm{Fo}=D\Delta t/\Delta x^2$. The update of the concentration $c_j$ in cell $j$ reads
\begin{equation}
c_j^{n+1}=c_j^n+\mathrm{Fo}\,(c_{j+1}^n-2c_j^n+c_{j-1}^n).
\end{equation}
The most dangerous error is the shortest wave that the grid can hold. This is a checkerboard pattern, in which the error alternates in sign from cell to cell: $c_j^n=a_n(-1)^j$, with amplitude $a_n$ at step $n$. For this pattern, $c_{j+1}^n=c_{j-1}^n=-c_j^n$. Inserting it into the update gives
\begin{equation}
a_{n+1}=(1-4\,\mathrm{Fo})\,a_n.
\label{eq:amplification}
\end{equation}
The error does not grow only if $|1-4\,\mathrm{Fo}|\leq1$. Longer waves change less from one step to the next, so they remain bounded whenever the checkerboard does. Stability therefore requires $\mathrm{Fo}\leq1/2$ in one dimension and $\mathrm{Fo}\leq1/(2d)$ in $d$ dimensions. The diffusion module applies the safety factor $C_D=0.4$:
\begin{equation}
\Delta t_D=C_D\frac{\Delta x^2}{2dD_{\max}}.
\label{eq:diffusion-cfl}
\end{equation}
Here $D_{\max}$ is the largest diffusivity in the domain. In $d$ dimensions, the checkerboard amplification factor of Equation~\ref{eq:amplification} becomes $1-4d\,\mathrm{Fo}$. With $\mathrm{Fo}=C_D/(2d)$, it equals $1-2C_D=0.2$. The shortest grid wave therefore decays by a factor of five in every step and does not change sign. The value $C_D=1$ would be the stability limit itself, and any $C_D>0.5$ would make the shortest wave change sign from step to step. The value $C_D=0.4$ stays clear of both limits. The second limit concerns interface motion. Equation~\ref{eq:alpha} assumes that an interface crosses at most one cell per step. The change of the sub-cell fraction must therefore satisfy $|\Delta\alpha|=|v_n|\Delta t/\Delta x\leq1$. Requiring instead the stricter bound $|\Delta\alpha|\leq C_I$, with a safety factor $C_I\leq1$, gives
\begin{equation}
\Delta t_I=C_I\frac{\Delta x}{|v_n|_{\max}},
\qquad
\Delta t=\min(\Delta t_D,\Delta t_I,\Delta t_{\mathrm{remaining}}).
\label{eq:interface-cfl}
\end{equation}
Here $|v_n|_{\max}$ is the largest interface speed among all interface cells, and $\Delta t_{\mathrm{remaining}}$ is the time left until the requested end time. The implementation uses $C_I=1/8$, so an interface moves at most one eighth of a cell per step. This stricter limit reduces errors caused by the grid orientation for oblique and curved fronts. The value is a power of two. When this limit sets the timestep, a planar front at constant speed crosses a cell in exactly eight steps. The fraction $\alpha$ then changes by increments that are represented exactly in binary arithmetic. Over one eighth of a cell, the measured normal and curvature change only slightly. Treating $v_n$ as constant during the step (Equation~\ref{eq:alpha}) therefore remains accurate. At this limit, the five driven planar bicrystals reproduce $v=M\Delta G$ to within $1.1\times10^{-7}$. The last term in the minimum ensures that the final step ends exactly at the requested time.

\subsection{Process-specific constitutive modules}
Thermodynamic modules provide $M$, $\gamma$ and $\Delta G$ for each label pair. Their sources follow Table~\ref{tab:parameter-provenance}. A module first reads the current state $(c,T,\rho,\text{orientation},\ldots)$. It then evaluates its constitutive relation and returns the quantities in the physical units of Equation~\ref{eq:master}. No conversion factor is fitted inside the geometry or topology layers. The modules used here cover isotropic and anisotropic grain growth, recrystallisation driven by stored energy, multiphase competition of Gibbs energies and coherent elastic energies. All formulae are given in Section~\ref{sec:constitutive}. Numerical values, sources and input files are tabulated in the Supporting Information.

\subsection{Al--Cu precipitation module}
\label{sec:alcu-module}
The generic boundary-localized elastic regression is not used for the real precipitate morphology. For this module, a cell stores total Cu, $c$, in addition to phase label and sub-cell precipitate fraction $f$. Local partition is reconstructed as
\begin{equation}
c=f c_{\theta'}+(1-f)c_m,
\qquad c_{\theta'}=\frac13,
\qquad c_m^{\rm eq}=3.719\times10^{-4}.
\label{eq:alcu-partition}
\end{equation}
Only matrix Cu diffuses. The discrete form of the diffusion equation in Equation~\ref{eq:alcu-kinetics} uses, on the face shared by cells $i$ and $i+1$,
\begin{equation}
J_{i+1/2}=-D\left[1-\frac{f_i+f_{i+1}}2\right]
\frac{c_{m,i+1}-c_{m,i}}{\Delta x},\qquad
c_i^{n+1}=c_i^n-\frac{\Delta t}{\Delta x}
\left(J_{i+1/2}-J_{i-1/2}\right).
\label{eq:alcu-flux}
\end{equation}
Summing Equation~\ref{eq:alcu-flux} over a periodic domain cancels every face flux with equal magnitude and opposite sign, proving discrete Cu conservation apart from floating-point roundoff. We use $D=1.66\times10^{-20}\,\mathrm{m^2\,s^{-1}}$ at \SI{453}{K}, following the phase-field material study \citep{Ta2021}. For the homogeneous metastable FCC branch, pycalphad supplies $g_\alpha(c_m)$ and the COST-507 Al$_2$Cu phase supplies $g_{\theta'}$ at $c_{\theta'}=1/3$. Their grand-potential difference is
\begin{equation}
\Delta G^{\rm chem}(c_m)=\frac{1}{V_m}\left[
g_\alpha(c_m)+(c_{\theta'}-c_m)\frac{\partial g_\alpha}{\partial c_m}
-g_{\theta'}\right],
\qquad V_m=10^{-5}\ \mathrm{m^3\,mol^{-1}}.
\label{eq:alcu-chemical}
\end{equation}
The preprocessing script samples the homogeneous branch, differentiates its spline, and archives 300 values over $10^{-5}\le c_m\le0.05$; the runtime uses linear interpolation and therefore does not invoke pycalphad inside a timestep. The zero of Equation~\ref{eq:alcu-chemical} is $c_m^{\rm eq}=3.719\times10^{-4}$ at \SI{453}{K}.
Because $c_m$ is defined only in the matrix, assigning $c_m^{\rm eq}$ inside a precipitate would create two different chemical forces on opposite sides of one sharp interface. The Al--Cu module evaluates the one-sided matrix trace
\begin{equation}
c_m^\Gamma=\frac{K_{\sigma_c}*[(1-f_{\theta'})c_m]}
{K_{\sigma_c}*(1-f_{\theta'})},\qquad \sigma_c=0.75\ \text{cell},
\label{eq:alcu-composition-trace}
\end{equation}
and uses $c_m^\Gamma$ in Equation~\ref{eq:alcu-chemical} on both sides. This normalized projection changes neither $c_{\rm tot}$ nor the phase labels; it only supplies the thermodynamic trace required by the pair velocity.

Long-range coherency elasticity is evaluated by minimizing
\begin{equation}
F_{\rm el}=\frac12\int_\Omega
[\boldsymbol\epsilon(\mathbf u)-\boldsymbol\epsilon^*]:\mathsf C:
[\boldsymbol\epsilon(\mathbf u)-\boldsymbol\epsilon^*]~\mathrm dV,
\qquad
\boldsymbol\epsilon^*=\sum_{v=1}^{3} f_v\boldsymbol\epsilon_v^*.
\label{eq:alcu-elastic-energy}
\end{equation}
The three eigenstrains are cyclic permutations of $\operatorname{diag}(0.00746,0.00746,-0.051)$. For every nonzero Fourier mode, mechanical equilibrium $k_j\widehat\sigma_{ij}=0$ gives
\begin{equation}
A_{il}(\mathbf k)\widehat u_l=-\mathrm i k_j C_{ijkl}
\widehat\epsilon^*_{kl},
\qquad A_{il}=C_{ijkl}k_jk_k,
\qquad \frac{\delta F_{\rm el}}{\delta f_v}=-\boldsymbol\sigma:\boldsymbol\epsilon_v^*.
\label{eq:alcu-fourier}
\end{equation}
The zero mode follows the normal-expansion condition so that the free body carries no artificial macroscopic stress. Cubic Al uses $C_{11}=107.07$, $C_{12}=63.08$ and $C_{44}=28.52$ GPa \citep{Ta2021}. The solve is tensorial and long ranged but assumes homogeneous cubic-Al stiffness; spatially varying $\theta'$ stiffness is outside the model scope.

For variant $v$ with plate-normal axis $\mathbf a_v$, the interfacial energy is $\gamma_v(\mathbf n)=(\mathbf n\!\cdot\!\mathsf B_v\mathbf n)^{1/2}$, with eigenvalues corresponding to \SI{0.10}{J.m^{-2}} on the coherent face and \SI{0.245}{J.m^{-2}} at the rim. The plate velocity is Equation~\ref{eq:alcu-kinetics}, $v_n=M(\mathbf n)\left[\Delta G^{\rm chem}-\Delta E^{\rm el}-P_{\rm cap}\right]$. Its three terms are evaluated as follows. $\Delta G^{\rm chem}$ is Equation~\ref{eq:alcu-chemical}. The elastic term is $\Delta E^{\rm el}=\delta F_{\rm el}/\delta f_v$ from Equation~\ref{eq:alcu-fourier}, the increase in elastic energy when the plate fraction grows. The capillary term is the surface divergence of the Cahn--Hoffman vector,
\begin{equation}
\boldsymbol\xi_v=\frac{\partial\gamma_v}{\partial\mathbf n}=\frac{\mathsf B_v\mathbf n}{\gamma_v},
\qquad
P_{\rm cap}=\nabla_s\!\cdot\boldsymbol\xi_v,
\label{eq:alcu-cahn-hoffman}
\end{equation}
which replaces the non-variational product $\gamma(\mathbf n)\kappa$ \citep{CahnHoffman1974}. The mobility $M(\mathbf n)$ is $M_v(\mathbf n)$ of the plate variant $v$. With $q_v=|\mathbf n\!\cdot\!\mathbf a_v|$, an effective finite-resolution contrast $\beta_\mu=50$ and ledge-availability exponent $p_\mu=4$,
\begin{equation}
M_v(\mathbf n)=M_{\rm rim}\frac{1+\beta_\mu(1-q_v^2)^{p_\mu}}{1+\beta_\mu},
\qquad M_{\rm rim}=10^{-19}\ \mathrm{m^4\,J^{-1}\,s^{-1}},
\label{eq:alcu-mobility}
\end{equation}
so $M(q_v=0)=M_{\rm rim}$ at the rim and $M(q_v=1)=M_{\rm rim}/51$ on the coherent broad face. The experimental 2001:1 limiting ratio motivates the ledge-controlled form \citep{Ta2021}, but that limiting atomistic contrast is not resolved within a finite Gaussian angular support. The stated 51:1 ratio is therefore a resolution-regularised numerical closure, applied identically to every variant, rather than a fitted claim about atomistic broad-face mobility.

The label-conditioned projection retains a separate full-band normal $\mathbf n_v$ for every $\theta'$ variant. The capillary term of Equation~\ref{eq:alcu-cahn-hoffman} is evaluated independently for each full field and only then sampled on the resident--candidate interface of variant $v$. This separation prevents artificial jumps in $\boldsymbol\xi$ and opposite capillary signs on adjacent sides of a variant interface.

For this two-class material module, $v_n$ is stored in a single precipitate-outward frame on both sides of the interface: $v_n>0$ means precipitation and $v_n<0$ means dissolution. Pair-conditioned Gaussian projection therefore averages the same physical sign on the matrix and precipitate sides. This is equivalent to the generic resident-outward formulation after its label sign is transformed back to the canonical pair frame, but avoids applying that transformation twice. Total precipitate fraction is not forced to increase: coherency and Gibbs--Thomson terms shift the chemical lever-rule equilibrium, and competitive ripening may dissolve smaller plates. The acceptance tests instead enforce Cu conservation, mechanical equilibrium, the interface CFL, topology, and survival of every crystallographic variant represented initially.

The phase-change transaction uses a conservative Stefan transfer: forming $\Delta f$ of $\theta'$ adds $(c_{\theta'}-c_m)\Delta f$ locally and removes the identical Cu inventory from a normalized Gaussian neighbourhood in the matrix. Dissolution reverses the transfer, and the final floating-point residue is returned deterministically to the largest-weight donor. Particle-class interfaces then receive an equal-volume Gaussian upper-level remap after kinetic crossings. This collective transaction suppresses sub-kernel tunnels while preserving each particle's categorical volume and supplying no growth. Geometry is cached until a label event, conservative diffusion is subcycled at its own stability scale, and the complete variational force of Equation~\ref{eq:alcu-kinetics} determines the direction of every interface proposal. The same TCKT rule used by every GPK process accepts or defers crossings from the persistent label state, with cumulative activity stored as \texttt{topology\_deferred\_cells}. The $128^3$ production calculation completed in 469 s.

\subsection{Population closure for the Al--Cu ageing comparison}
\label{sec:alcu-population}
The closure of Equations~\ref{eq:alcu-population} and \ref{eq:alcu-population-scale} starts from a sampled 120 h population in the periodic volume of side $L=L_{\rm box}$.
For each phase at 120 h, diameter and thickness are sampled from a positive truncated bivariate normal. Although particle-resolved pairs are unavailable, the reported aspect-ratio standard deviation constrains their covariance. First-order propagation for the aspect ratio $r=\ell/H$ gives
\begin{equation}
s_r^2\simeq\frac{s_\ell^2}{\bar H^2}+\frac{\bar\ell^2s_H^2}{\bar H^4}
-\frac{2\bar\ell}{\bar H^3}C_{\ell H},\qquad
C_{\ell H}=\frac{\bar H^3}{2\bar\ell}\left(\frac{s_\ell^2}{\bar H^2}
+\frac{\bar\ell^2s_H^2}{\bar H^4}-s_r^2\right).
\label{eq:pop-covariance}
\end{equation}
Here $\bar\ell$ and $\bar H$ are the measured mean diameter and thickness, $s_\ell$, $s_H$ and $s_r$ the standard deviations of diameter, thickness and aspect ratio, and $C_{\ell H}$ the covariance of diameter and thickness. This yields correlations 0.909 for $\theta'$ and $-0.515$ for $\theta''$; both covariance matrices are positive-definite after rounding. If $\widetilde V_i$ denotes the sampled disk volume, one common factor
\begin{equation}
\Phi_0=\left(\frac{f^{\rm exp}L^3}{\sum_i\widetilde V_i}\right)^{1/3},
\qquad (\ell_i,H_i)=\Phi_0(\widetilde\ell_i,\widetilde H_i)
\label{eq:pop-initial-scale}
\end{equation}
enforces the independently measured phase inventory while preserving every sampled aspect ratio and rank. Centres are uniform random periodic points. Variants are stratified along the volume ranking so any survivor set differs by at most one particle among the three habit planes.

The coarsening closure follows the ledge-controlled mechanism discussed by Merle and Merlin \citep{MerleMerlin1981}; published phase-field calculations additionally show that anisotropic mobility substantially raises the plate aspect ratio \citep{Ta2021}. The capillary ordering is represented by retaining the $N_{\theta'}$ largest initial equivalent volumes. The particle number $N_{\theta'}$, the $\theta''$ fraction and the $\theta'$ fraction follow Equation~\ref{eq:alcu-population}, with the normalized time $\tau_s=(t-120\,\mathrm{h})/(48\,\mathrm{h})$. The decrease of $N_{\theta'}$ from 131 to 17 corresponds to a constant loss rate of $\ln(131/17)/(48\,\mathrm h)=\SI{0.04254}{h^{-1}}$. This rate is conditioned on the 168 h number density, not predicted. The $\theta''$ disappearance path is likewise conditioned on the endpoint. The $\theta'$ inventory is instead predicted from the disappearing Cu inventory. Under the equal-atomic-volume approximation, $x_{\rm Cu}^{\theta''}=1/4$ and $x_{\rm Cu}^{\theta'}=1/3$, which gives the factor $0.75$ in that equation. Thus $f_{\theta'}(168\,\mathrm{h})=1.55\%$ is not fit to the measured $1.0\pm0.5\%$. It remains a \emph{conditional} result because the endpoint count and $\theta''$ non-detection are inputs. Matrix-supplied or rejected Cu, phase-specific atomic-volume differences and residual unclassified precipitates are not included.

The independent local $128^3$ GPK case starts with aspect ratio 8.0 and gives 10.417 under the audited ledge-mobility law, hence $\Lambda_{\rm loc}=1.302$. This local response is fixed before the 168 h experimental comparison and transferred as $\Lambda=\Lambda_{\rm loc}^{\tau_s}$. The surviving particles are then scaled with Equation~\ref{eq:alcu-population-scale}. For the survivor set $\mathcal S(t)$, its common factor is
\begin{equation}
\Phi(t)=\left[\frac{f_{\theta'}(t)L^3}{\sum_{i\in\mathcal S(t)}V_i(120\,\mathrm{h})}\right]^{1/3}.
\label{eq:pop-survivor-scale}
\end{equation}
The factors $\Lambda^{1/3}$ and $\Lambda^{-2/3}$ of Equation~\ref{eq:alcu-population-scale} keep $\ell_i^2H_i$ unchanged, so Equation~\ref{eq:pop-survivor-scale} sets the total volume exactly, and the aspect ratio changes by $\Lambda$. The 168 h $\theta'$ fraction, diameter, thickness and aspect ratio are therefore conditioned on the measured data and used only as a check. None of them is inserted into these relations, but the measured endpoint count influences their values through the survivor inventory.

\subsection{Verification and statistical analysis}
The foundational tests compare measured slopes or signs with analytical results. Linear regressions use ordinary least squares. To enable direct comparison with published experimental analyses, a standalone post-processing script reads only the raw simulation output files. It does not use internal state variables of the solver. It fits $\ln[-\ln(1-X)]$ versus $\ln t$ to extract the Avrami exponent, exactly as is standard practice in the experimental recrystallisation literature. Statistics over repeated random seeds are used for the grain-growth rate check with pure-Al parameters. Nuclei whose initial radius spans fewer than two grid cells are excluded from the analysis, because they are not adequately resolved. They are not counted as successful nucleation events. No statistical hypothesis tests are applied to the deterministic verification runs. Instead, a tolerance is defined for each metric, and the numerical error ranges are reported.

\subsection{Software and reproducibility}
The reference implementation is written in C++ with OpenMP and CMake. Analysis and symbolic verification use Python. The release tests comprise compiled tests of real materials, foundational tests and comparisons between models. They also include symbolic, executable and literature-comparison regression tests. All of these tests pass under the acceptance limits and test totals stated in the Supporting Information.

\subsection{Use of AI tools}
Claude (Anthropic) was used for language editing of the manuscript. The authors reviewed all edited text and take full responsibility for the content of the manuscript.

\subsection{Code availability}
The GPK C++ source code, the Python analysis scripts, the input files and the automated tests are publicly available at \url{https://github.com/albert-hzbn/GPK} under the MIT licence.

\section*{Data availability}
The input files of every simulation, the processed data plotted in every figure and the scripts that draw the figures are available in the folder \texttt{paper} of the GPK repository (\url{https://github.com/albert-hzbn/GPK}). The large three-dimensional snapshot fields (about 29 GB) are available from the corresponding authors upon reasonable request. The repository lists their checksums and the commands that regenerate them.

\section*{Acknowledgements}
The authors acknowledge the computational and institutional environment of the Max Planck Institute for Sustainable Materials. A.L. acknowledges support from the Max Net Initiative SusMax -- The Max Planck Network on Sustainable Materials. D.R. is grateful for financial support from the European Union through the ERC Advanced Grant ROC (Grant Agreement No. 101054368). Views and opinions expressed are however those of the authors only and do not necessarily reflect those of the European Union or the ERC. Neither the European Union nor the granting authority can be held responsible for them. The funders played no role in study design, data collection, analysis and interpretation of data, or the writing of this manuscript.

\section*{Author contributions}
A.L. conceived the method, developed the software, performed the simulations and validation, analysed the results, and drafted the manuscript. D.R. supervised the research, contributed to interpretation and manuscript revision. Both authors approved the manuscript.

\section*{Competing interests}
All authors declare no financial or non-financial competing interests.

\section*{Additional information}
Supporting information accompanies this paper. Correspondence and requests for materials should be addressed to Albert Linda (a.linda@mpi-susmat.de) or Dierk Raabe (d.raabe@mpi-susmat.de).

\clearpage
\bibliographystyle{naturemag}
\bibliography{references}
\end{document}


\maketitle
\noindent\textbf{Guide to this document.} Supplementary Notes~1--4 provide derivation and numerical details that support the shared GPK framework. Note~5 lists the verification tests and their acceptance boundaries. Note~6 records process-specific assumptions and simulation definitions. Note~7 collects parameter provenance and representative outputs. Note~8 is a notation and terminology glossary, Note~9 maps every evidence class to its interpretation, and Note~10 states the claim boundaries. Readers seeking application definitions and parameters can begin with Notes~6 and 7; readers auditing the mathematics can begin with Notes~1--5; and readers checking the scope of a claim can begin with Notes~8--10.

\tableofcontents

\section{Matched-asymptotic normalization}
\begin{figure}[ht]
\centering\includegraphics[width=0.95\textwidth]{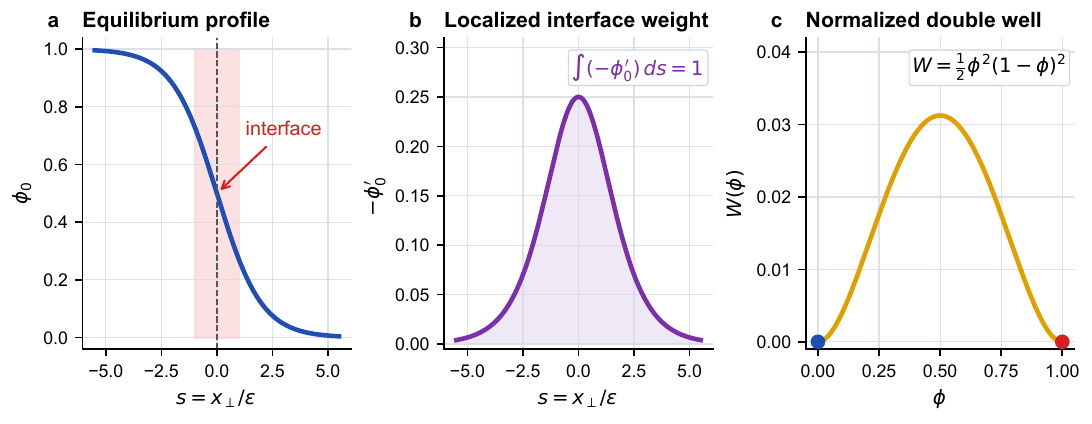}
\caption{Normalized Allen--Cahn profile and its connection to the sub-cell GPK representation.}
\label{fig:sprofile}
\end{figure}

The derivation of the velocity law from a diffuse-interface model is given in full in main-text Section~\ref{M-sec:thin-limit}, Equations~\ref{M-eq:free-energy}--\ref{M-eq:surface-normalization}. This note records only the symbolic audit of that derivation. The audit checks the profile equation (main-text Equation~\ref{M-eq:profile-ode}), its solution (Equation~\ref{M-eq:profile}), the integral $I_1=1/6$ (Equation~\ref{M-eq:profile-integral}), the curvature sign, the dimensions and the Gibbs--Thomson radius. The integral $\int_0^1\sqrt{2W(\phi)}\,\mathrm d\phi$ has the same value $1/6$. The factor $1/6$ therefore cancels, and the implemented law, main-text Equation~\ref{M-eq:master}, contains no hidden calibration coefficient. Figure~\ref{fig:sprofile} shows the profile, the double well and the normalization.

Figure~\ref{fig:sprofile} makes the cancellation used in the derivation visible. The normalized profile rises monotonically between its two bulk states, while the double-well factor vanishes in either bulk and is largest at the interface midpoint. Its derivative enters the energy balance across the layer, and the shaded area gives $I_1=1/6$. The GPK lineal coordinate $\alpha$ uses the same normal direction to record sub-cell displacement, but GPK neither stores nor evolves this diffuse equilibrium profile; the profile is used only to derive the sharp-interface coefficient and sign.

\section{Gaussian-kernel geometry and update details}
The projected indicator $\widetilde I_g$ is label conditioned at each interface cell. This matters at multi-grain junctions: resident geometry is reconstructed from its own support, not from an unsigned union of boundaries. The primary implementation composes the truncated one-dimensional Gaussian with the fourth-order first- and second-derivative coefficients and evaluates their tensor products directly over the resident-label stencil. By associativity this is algebraically the same projected gradient and Hessian as full Gaussian convolution followed by the stated finite differences, but it neither scans the full domain for every active label nor stores a projected field. When the composed radius exceeds 18 cells in three dimensions, the cubic tensor support is replaced by a shared four-array separable projection; the workspace is independent of thread count and the numerical operators are unchanged. Curvature is then evaluated directly from the Hessian identity in the main text. This avoids differentiating a normalized discrete normal field and keeps axial and mixed Hessian terms at the same order in the three-dimensional sphere audit. Main-text Figure~\ref{M-fig:geometry-pipeline} shows the reconstruction pipeline, and main-text Figure~\ref{M-fig:curvature-signs} fixes the sign convention used in every module.

For the physical kernel length $\ell_K=\sigma_K\Delta x$, convolution damps a wave of wavenumber $k$ by the factor of main-text Equation~\ref{M-eq:gaussian-spectrum}. Differentiation subsequently multiplies that mode by $|\mathbf k|$ for a normal component and by $|\mathbf k|^2$ for an unnormalized second derivative. Convolving first therefore limits the pixel-scale amplification that would otherwise dominate derivatives of a binary field.

The same operation introduces a controlled smoothing bias. Main-text Equation~\ref{M-eq:gaussian-moment} shows that its leading term is even and of order $\ell_K^2$. For a resolved feature of radius $R$, the curvature error is therefore of order $\ell_K^2/R^3$, or $(\ell_K/R)^2$ relative to the curvature itself. This estimate applies to smooth, resolved interfaces. It is not asserted at triple junctions, corners or during topology changes.

The normal is computed with main-text Equation~\ref{M-eq:geometry}. The implementation uses a hard gradient threshold rather than an additive regularization of the denominator: cells with $|\nabla\widetilde I_g|\Delta x\leq10^{-4}$ receive a zero normal. This dimensionless criterion preserves unit normalization wherever curvature is evaluated, remains unchanged under grid refinement and avoids assigning a direction to a uniform projected field. The Hessian identity in the main text is therefore exact in the accepted region. The numerical value is capped at $|\kappa|\leq1/\ell_K$ so that features below the Gaussian measurement scale neither dominate the velocity nor collapse the global timestep.

The executable geometry-path regression uses a $40^3$ sphere of radius 10 cells with $\sigma_K=2$ and obtains mean curvature 0.20923 against $2/R=0.200$ (4.62\% error). A separate $56^3$, radius-14 sphere with $\sigma_K=5$ forces the broad-kernel separable path and gives 0.14712 against 0.14286 (2.98\% error). The corresponding owned-array capacities are 81.01 bytes per cell for the compact two-dimensional path and 97.01 bytes per cell with the broad three-dimensional workspace. One- and four-thread runs give identical categorical states, sub-cell coordinates and curvature arrays in the deterministic thread-count audit.

The eight operations of each timestep, including the velocity law $v_{n,i}=M_{pq}(\Delta G_{q-p}-\gamma_{pq}\kappa_p)$ used at every interface cell, are listed in main-text Section~\ref{M-sec:timestep} and Figure~\ref{M-fig:timestep}.

All active cells advance on one accepted physical timestep. Asynchronous local clocks were not enabled because a topology certificate is state dependent: accepting a crossing changes the admissibility and candidate support of nearby crossings. Local subcycling without a conservative event-order proof would therefore exchange runtime for an uncontrolled ordering dependence. The implemented optimization removes bulk and inactive-label work while retaining the global interface-crossing and diffusion bounds used by every validation trajectory.

\section{Derivation and verification of the topology-aware Gaussian transaction}
The update couples a pair-conditioned velocity projection, a topology-certified kinetic transaction (TCKT) and a conservative geometric remap. The only topological declaration is label semantics: a persistent region identity must not split or merge spuriously, whereas a phase class may contain many disconnected particles and must permit physical coalescence and pinch-off. No operation receives a material name, phase name or crystallographic direction, and no post-kinetic closing, hole filling or line construction is permitted. Synthetic grain-network input is normalized separately before the $t=0$ state is written: seeds occupy unique cells, Voronoi distances use the periodic minimum image, the largest face-connected component of each identity is retained, and disconnected integer-tie cells attach in simultaneous layers to already connected face neighbours. Because each accepted target cell has a face path to the retained target component, this construction terminates with exactly one periodic component per initialized identity without altering an evolved interface.

The raw velocity has opposite resident conventions on the two sides of the same interface. The pair-averaged velocity of main-text Equation~\ref{M-eq:pair-velocity} therefore gives all cells on a $p$--$q$ boundary one common sign before averaging. Its width $\sigma_v$ is 0.85 cell in two dimensions and $0.375\sigma_K$, limited to 0.85--3.00 cells, in three dimensions (main-text Section~\ref{M-sec:timestep}).
Canonical signing prevents cancellation across the interface, and pair conditioning prevents a third grain from contaminating the velocity at a junction. A surface has two tangential directions; the wider three-dimensional projection prevents neighbouring voxels on one physical face from receiving mutually incoherent crossing times. The proportional rule prevents a narrow geometry kernel from inheriting an unrelated broad velocity support, while the upper cap keeps $\sigma_v$ smaller than half of the production geometry width $\sigma_K=8$ cells. This projection removes unresolved tangential velocity modes without changing the pair-conditioned constitutive force.

A proposed change $p\rightarrow q$ of a crossing cell $i$ must satisfy the resident and candidate rules of main-text Section~\ref{M-sec:timestep}. Both rules count the face-connected pieces $P_p$ and $P_q$ in the $3^d$ block around cell $i$, without the centre cell. A count $P_p=0$ is accepted only for a one-cell terminal extinction. If either count exceeds one, an early-exit periodic face-connected search tests whether the incident branches are already connected outside $N_i$. A genuine resident split or merger of disconnected candidate components is deferred; branches already connected through the wider domain are admitted as a resolved cavity, tunnel or T1 event. The search terminates as soon as all branches are reached, so an oblique stair usually visits only a few nearby cells. This local/global construction is informed by digital simple-point theory \citep{KongRosenfeld1989,Bertrand1994}, but does not impose a strict local binary simple-point condition because resolved physical contact and pinch-off must remain admissible.

Candidates are evaluated in lexicographic order, and each certificate sees all previously accepted events. The common case costs $O(N_\Gamma)$ with constant local memory; only locally multi-branch candidates invoke a connectivity search. The compact pair projection costs $O(N_\Gamma r_v^d)$ rather than one full-domain convolution per grain pair, which is essential for refined polycrystals.

After one or more kinetic crossings, the balanced Gaussian remap (BGR) ranks only face-neighbour relabelling proposals, using the narrow Gaussian score of width $\sigma_r$ described in main-text Section~\ref{M-sec:timestep}. For every label pair, the same number of highest-gain proposals is accepted in each direction. The cell count of every label is therefore unchanged (main-text Equation~\ref{M-eq:balanced-remap}). Thus a one-cell protrusion can be exchanged with an indentation without adding volume, a preferred matrix, a growth law or a physical diffuse thickness. Volume balance alone is insufficient, because a remote equal-count swap could lengthen the boundary. The boundary-face count $N_{\rm face}$ of main-text Section~\ref{M-sec:timestep}, the number of cell faces whose two cells carry different labels, measures this length. The two proposed relabellings are assessed as one atomic exchange, including their mutual face when they are adjacent, and are eligible only if they lower $N_{\rm face}$. They are then tested sequentially with the local topology certificate. A multi-branch local pattern is accepted only when the branches reconnect inside the surrounding $5^d$ box; the bounded proof admits an oblique boundary crossing the $3^d$ stencil but cannot authorize a distant topology event. If the area or either topology condition fails, both moves are rejected. Three-dimensional steps apply two such sweeps to remove ledges aligned with either tangential direction. Each sweep separately keeps every label count and lowers $N_{\rm face}$, so their composition is exactly volume neutral and does not lengthen the boundary. The compact implementation costs $O(N_\Gamma r_r^d+N_\Gamma\log N_\Gamma)$ per sweep and records both deferred crossings and exchanged cells.

\begin{figure}[ht]
\centering\includegraphics[width=0.98\textwidth]{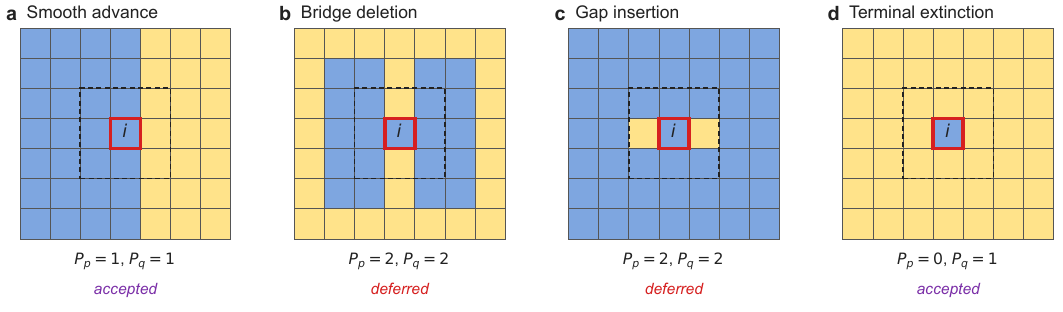}
\caption{Model-independent topology transactions. The centre cell $i$ (red outline) is the proposed crossing from the resident $p$ (blue) to the candidate $q$ (yellow). The dashed square is the $3\times3$ block, and $P_p$ and $P_q$ count the face-connected pieces of each label in this block without the centre cell. (a) Smooth boundary advance and (d) one-cell terminal extinction satisfy the certificate. (b) Bridge deletion would split the resident; (c) gap insertion would merge disconnected candidate components. The latter two events are deferred rather than repaired after the fact.}
\label{fig:stckt}
\end{figure}

Supplementary Figure~\ref{fig:stckt} illustrates the executable unit tests. Smooth advance and one-cell extinction are accepted; deletion of a one-cell bridge and insertion between disconnected candidate regions are deferred. The same tests are instantiated in two and three dimensions and do not construct a thermodynamic model beyond the neutral base interface. The full validation suite then exercises the rule under grain-growth, recrystallisation, multiphase and elastic modules. The transaction protects resolved components while allowing physically necessary topology events; it cannot make a sub-cell feature resolved.

Periodic initialization regressions audit the input contract independently of rendering. An 80-grain $96^2$ field contains 80 labels, 80 face-connected components and no component of one or two cells at $t=0$ and after a short evolution. The corresponding $48\times48\times24$ field requires two integer-tie cells to be reassigned and then gives the same 80-label/80-component result. The production starts of main-text Figure~\ref{M-fig:qualitative-2d} contain 360 components across 360 labels in both the isotropic and anisotropic cases. The starts of main-text Figure~\ref{M-fig:qualitative-3d} contain 240 labels and no cell without a same-label face neighbour; the complete production-volume audit is reported below.

The three-dimensional coherence width and remap count were selected on a $192\times192\times96$ calibration ensemble containing 80 grains, with $\sigma_K=8$ cells and a common \SI{0.25}{s} probe horizon. This ensemble is separate from the denser 240-grain production fields in main-text Figure~\ref{M-fig:qualitative-3d}. Table~\ref{tab:si-3d-interface-selection} reports categorical metrics before any rendering. $\Delta N_{\rm face}$ is the relative change of the boundary-face count; the low-support count is the number of interface cells having at most two same-label face neighbours, reported as its change from the common initialized field. The three-cell, two-sweep setting is the only candidate that lowers $N_{\rm face}$ in both the isotropic and the anisotropic run and also lowers the low-support count by at least 11\% in both runs. All candidates retain 79--80 grains and create no label adjacent to only one enclosing label. In the retained setting, the smallest final grains contain 3760 isotropic and 7875 anisotropic voxels, both above the 2145-voxel geometry-support volume. Four-cell velocity support reduces $N_{\rm face}$ but lowers the anisotropic low-support count by only 7.95\%; it is therefore not used. The production rule $\sigma_v=0.375\sigma_K$, limited to 0.85--3.00 cells, reproduces the selected three cells for $\sigma_K=8$ and is used with two remap sweeps in all three-dimensional runs (main-text Section~\ref{M-sec:timestep}). The two-dimensional width $\sigma_v=0.85$ cell was not selected in this ensemble. It is the width at which a face neighbour receives half the weight of the centre cell, $\sigma_v=1/\sqrt{2\ln2}$.

\begin{table}[ht]
\centering
\caption{Calibration of the three-dimensional pair-velocity width $\sigma_v$ and the number of balanced-remap sweeps on a $192\times192\times96$ ensemble with 80 grains, $\sigma_K=8$ cells and a common \SI{0.25}{s} horizon. $\Delta N_{\rm face}$ is the relative change of the boundary-face count, and ``low support'' is the relative change in the number of interface cells with at most two same-label face neighbours. Negative percentages denote improvement from the common initialized field. The selected setting, three cells with two sweeps, is the last row. Wall time is the mean of the isotropic and anisotropic runs on the same 24-thread workstation.}
\label{tab:si-3d-interface-selection}
\small
\begin{tabular}{@{}ccrrrc@{}}
\toprule
$\sigma_v$ [cells] & BGR sweeps & $\Delta N_{\rm face}$ iso./ani. & Low support iso./ani. & Final $N_g$ iso./ani. & Wall time \\
\midrule
2.00 & 1 & $-0.003/ +0.257\%$ & $-11.23/-6.41\%$ & 79/80 & \SI{280}{s} \\
3.00 & 1 & $-0.180/ +0.038\%$ & $-14.31/-8.82\%$ & 79/80 & \SI{332}{s} \\
4.00 & 1 & $-0.231/-0.067\%$ & $-12.67/-7.95\%$ & 79/80 & \SI{377}{s} \\
3.00 & 2 & $-0.253/-0.047\%$ & $-16.34/-11.58\%$ & 79/80 & \SI{373}{s} \\
\bottomrule
\end{tabular}
\end{table}

The accepted and deferred examples in Figure~\ref{fig:stckt} show what the certificate protects. The smooth advance in panel (a) preserves one connected resident and candidate neighbourhood, and the terminal deletion in panel (d) removes a one-cell component without leaving a fragment. By contrast, panel (b) would split the resident bridge and panel (c) would join previously disconnected candidate components. Deferring those two proposals before commitment prevents a locally plausible velocity from creating a global split or merge that the constitutive law did not prescribe.

\section{Kernel-width selection, numerical accuracy and stability}
The curvature error is controlled jointly by resolution and Gaussian width. Two competing requirements define the useful interval of main-text Equation~\ref{M-eq:kernel-window}.
The lower bound supplies enough support to regularize differentiation of a pixelated indicator; the upper bound limits geometric averaging over the smallest radius of interest. For smooth interfaces, the relative curvature error is of order $(\sigma_K\Delta x/R)^2$ away from grid-scale topology and junctions. The measured circular-interface comparison is shown in main-text Figure~\ref{M-fig:gaussian-accuracy}. A larger Gaussian width suppresses pixel noise but broadens the geometric support, so $\sigma_K=2$ cells is the baseline rather than a universal optimum. The $768^2$ galleries use $\sigma_K=4$ cells for grain networks and 8 cells for the multiphase case, and the $1024^2$ dendrite uses $\sigma_K=2$ cells at $\Delta x=\SI{0.5}{\micro m}$. All two-dimensional gallery runs therefore share the physical length $\ell_K=\sigma_K\Delta x=\SI{1}{\micro m}$. The three-dimensional inputs use the same fixed-physical-kernel principle. Quantitative studies should report $\Delta x$, $\sigma_K$, the truncation radius and $R_{\min}/\Delta x$ together.

\begin{figure}[ht]
\centering\includegraphics[width=0.96\textwidth]{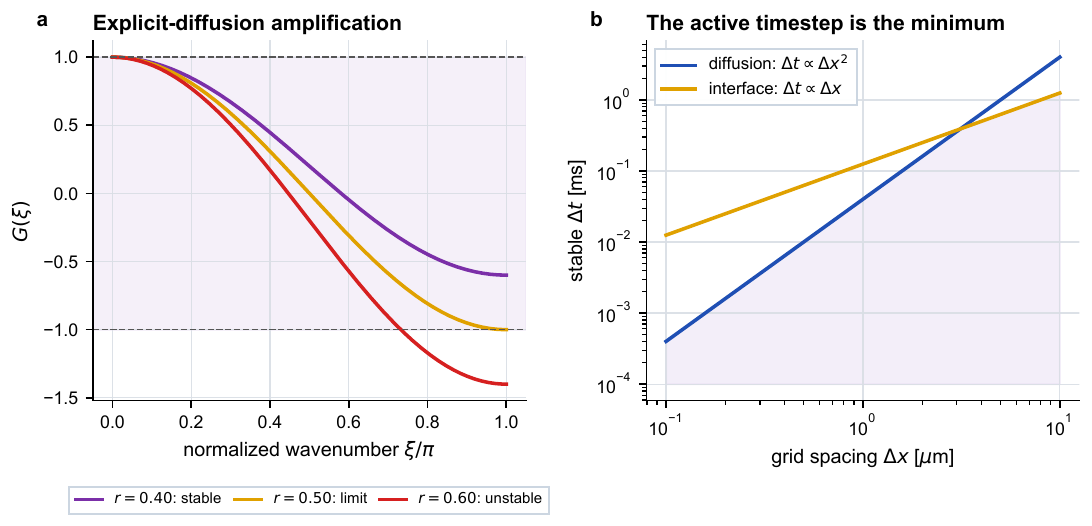}
\caption{Explicit-diffusion stability analysis and numerical verification.}
\label{fig:sstability}
\end{figure}

The two timestep limits, for explicit diffusion and for interface motion, are main-text Equations~\ref{M-eq:diffusion-cfl} and \ref{M-eq:interface-cfl}. They are derived in main-text Section~\ref{M-sec:stability}. The implementation uses $C_D=0.4$ and $C_I=1/8$ and clips manual timestep requests to the same bound. Under this crossing limit, the signed fractional velocity deviations $v_{\rm GPK}/v_{\rm exact}-1$ of the five driven flat-front regressions lie between $-1.10\times10^{-7}$ and $+3.49\times10^{-8}$. Figure~\ref{fig:sstability} reports the von Neumann analysis and numerical check.

Figure~\ref{fig:sstability} distinguishes the two timestep restrictions used by the coupled update. The analytical amplification factor remains inside the unit circle below the diffusion bound, and the numerical Fourier-mode amplitudes decay at the predicted rate in that interval. Once the explicit bound is crossed, alternating growth appears rather than physical decay. The interface-CFL condition separately limits the fractional boundary crossing per update. Satisfying both bounds is therefore required: diffusion stability alone does not prevent an interface from skipping cells, and the interface bound alone does not stabilize a conserved bulk field.

\section{Verification inventory and quantitative boundaries}
\subsection{Equation-consistency audit}
\begin{table}[ht]
\centering
\caption{Analytical and implementation consistency checks for the reported formulation.}
\small
\begin{tabular}{@{}p{0.23\textwidth}p{0.49\textwidth}p{0.17\textwidth}@{}}
\toprule
Check & Consistency requirement & Outcome \\
\midrule
Free-energy normalization & $6\int_0^1\sqrt{2W}\,\mathrm d\phi=1$ & $E_\Gamma/A=\gamma$ \\
Sharp-interface dimensions & $[M][\Delta G]=[M][\gamma\kappa]=\mathrm{m\,s^{-1}}$ & Consistent \\
Energy sign & $\Delta G_{q-p}=\psi_q-\psi_p<0$ makes resident $p$ recede at a flat interface & Consistent \\
Curvature sign & Convex resident: $\kappa_p>0$; resident matrix around an inclusion: $\kappa_p<0$ & Consistent \\
Gaussian normal & Unit normalization above the code threshold; zero below it & Matched to code \\
Sub-cell update & $\dot\alpha=v_n/\Delta x$ and negative $v_n$ consumes resident fraction & Matched to code \\
Critical radius & $R_{\rm crit}=(d-1)\gamma/\mathcal D$, $\mathcal D=-\Delta G_{q-p}>0$ & Sign and units verified \\
Diffusion closure & Cell-centered $D_{\rm eff}\nabla_h^2c$ is not a conservative face-flux discretization & Limitation retained \\
Stability bounds & $\Delta t_D\leq\Delta x^2/(2dD)$ and $\Delta t_I\leq\Delta x/|v_n|$ before safety factors & Consistent \\
Grain-growth units & $10^{-9}\,\mathrm{m^2\,s^{-1}}=1\,\mathrm{\mu m^2\,ms^{-1}}$ & Conversion verified \\
\bottomrule
\end{tabular}
\end{table}

\subsection{Test inventory}
\begin{table}[ht]
\centering
\caption{Foundational tests used to define the numerical envelope.}
\label{tab:foundations}
\small
\begin{tabular}{p{0.24\textwidth}p{0.31\textwidth}p{0.35\textwidth}}
\toprule
Test & Reference result & Measured result and interpretation \\
\midrule
Gaussian--Hessian circle curvature & $\kappa=1/R$ & 3.51\% error at $R/\Delta x=15$; 0.89\% at 60 for fixed $\ell_K/R$.\\
Gauss--Bonnet area loss & $\dot A=-2\pi M\gamma$ for any simple closed curve & Circle, ellipse and square errors are 2.30\%, 2.58\% and 0.93\%.\\
Von Neumann--Mullins & $\dot A_i=(\pi/3)M\gamma(n_i-6)$ & Five $640^2$, 300-grain networks give 1165 topology-stable grains, 19.95\% slope error and $r=0.914$; $n=9,10$ contain 29 and 5 grains.\\
Perturbation decay & Exponential decay of a sinusoidal mode & Nine-point refined rate error is 3.78\%; log-amplitude fit has $1-R^2=2.0\times10^{-4}$.\\
Interfacial energy proxy & Non-increasing between topological events & Discrete events cause excursions; net relaxed-window drift is $-4.65\%$ and remains below the declared 15\% no-runaway bound.\\
Fixed-physical-kernel refinement & Consistent $\mathrm dR^2/\mathrm dt$ with fixed $\ell_K$ & Errors remain 14.17--14.30\%; the retained plateau separates kernel bias from local curvature convergence.\\
\bottomrule
\end{tabular}
\end{table}

For the von Neumann--Mullins audit in main-text Figure~\ref{M-fig:foundations}(b), Voronoi distances are evaluated periodically on the same torus used by the evolution operator. Five seeds (123, 42017, 73013, 91009 and 120011) each initialize 300 grains on a $640^2$ grid with $\Delta x=\SI{1}{\micro\metre}$ and $\sigma_K=4$ cells. After 400 relaxation steps, area change is measured over 32 steps. Grains below 200 cells, grains that vanish and grains whose face-neighbour set changes during this window are excluded because the von Neumann--Mullins law assumes one fixed topological class. The resulting 1165 accepted grains include 29 with $n=9$ and 5 with $n=10$. Their measured/theoretical mean rates are $11.631/10.179$ and $12.115/13.572$, respectively, in units of $10^{-9}\,\mathrm{m^2\,s^{-1}}$. The all-grain least-squares coefficient with an intercept is $4.070\times10^{-9}\,\mathrm{m^2\,s^{-1}}$ per unit $(n-6)$, 19.95\% above theory, and the individual-grain correlation is $r=0.914$. The all-grain regression, rather than a regression through class means, supplies the quantitative gate; classes with fewer than eight grains remain unconnected diagnostics.

\subsection{Dedicated three-dimensional and multiphase audit}
The strict dimensional audit tests orientation dependence that a mean-curvature value alone could conceal. Axial and mixed Hessian derivatives are all fourth order. Table~\ref{tab:dimensional-audit} reports mean, directional and pointwise diagnostics at the four-cell Gaussian scale used by the simulations. The sphere test samples every resident-side interface cell; its coefficient of variation and 95th-percentile local error expose rough directional sectors even when the mean is accurate. The cylinder separates one non-zero principal curvature from one zero principal curvature, and the plane tests for spurious out-of-plane curvature.

\begin{table}[ht]
\centering
\caption{Dimension-specific geometry and multiphase tests. All fourteen executable gates passed.}
\label{tab:dimensional-audit}
\small
\begin{tabular}{@{}p{0.34\textwidth}p{0.25\textwidth}p{0.30\textwidth}@{}}
\toprule
Test & Exact/invariant result & Measured result \\
\midrule
3-D sphere, $R/\Delta x=18$ & $\kappa\Delta x=2/18$ & Mean error 2.41\%; directional CV 2.24\%; 95th-percentile local error 5.78\% \\
3-D periodic cylinder & $\kappa\Delta x=1/18$ & Mean error 2.71\% \\
2-D and 3-D planar fronts & $\kappa=0$ & Maximum $|\kappa|=0$ in both dimensions \\
Known-Gibbs flat front & Identical displacement in 2-D and 3-D & Both advance 10 cells; displacement difference is exactly 0 \\
Equal-energy products & Mirror symmetry & $|f_1-f_2|=0$ in both dimensions \\
Categorical integrity & $f_0+f_1+f_2=1$ and no invalid labels & Machine-precision closure; zero invalid labels \\
\bottomrule
\end{tabular}
\end{table}

The saved high-resolution qualitative multiphase fields receive a separate raw-state audit at their final \SI{8}{ms} (2-D) and \SI{6}{ms} (3-D) states. Product fractions increase monotonically at every saved interval. The final 2-D fractions are $(f_0,f_1,f_2)=(0.1410,0.3663,0.4927)$ and the 3-D fractions are $(0.1106,0.3235,0.5659)$. Fraction sums close to machine precision and no invalid phase identifier occurs. The 3-D product contains no isolated periodic face-neighbour voxel; the 2-D field contains one single-cell phase-1 remnant, equal to $1.04\times10^{-5}$ of product cells and below the declared $2\times10^{-5}$ sub-kernel diagnostic tolerance. It is reported rather than removed because it is smaller than the Gaussian support and has no effect on the displayed isocontour. The smallest non-periodic product components contain one cell in 2-D and 18 cells in 3-D. The raw-grid serration indices are 3.356 and 2.792, respectively, and are categorical-grid diagnostics rather than interface curvature. The Gaussian isocontour or isosurface used for kinetics and display remains the continuous geometry representation.

The dimensional initializer uses $z=0$ and $\Delta z=0$ when $N_z=1$, so a requested 2-D disk has its declared in-plane radius. Every model uses the same TCKT foreground-connectivity test, with the one-cell extinction exception of Supplementary Note~3; phase-specific island deletion and grain-only completion operations are not part of the core. Topology semantics are therefore independent of the \texttt{phase\_selection} and \texttt{grain\_network} flags.

\begin{table}[ht]
\centering
\caption{Focused constitutive isolation tests retained in the manuscript.}
\label{tab:crossmodel}
\small
\begin{tabular}{llll}
\toprule
ID & Constitutive module & Isolated quantity & Result \\
\midrule
F1 & Isotropic growth & Flat, zero-force boundary & Exactly 0 changed cells \\
F2 & Anisotropic growth & Orientation-dependent boundary response & Correct directional trend \\
F3 & Recrystallisation & 2-D/3-D site-saturated transformation & $n=1.528/1.975$, $R^2=0.978/0.980$ \\
F4 & Multiphase growth & Lower-free-energy candidate selection & Favored-phase gain 0.761 \\
F5 & Elastic coupling & Eigenstrain-energy scaling & $E(2\delta)/E(\delta)=4$ \\
\bottomrule
\end{tabular}
\end{table}

\section{Application-specific information}
\subsection{Grain growth}
\begin{figure}[ht]
\centering\includegraphics[width=0.98\textwidth]{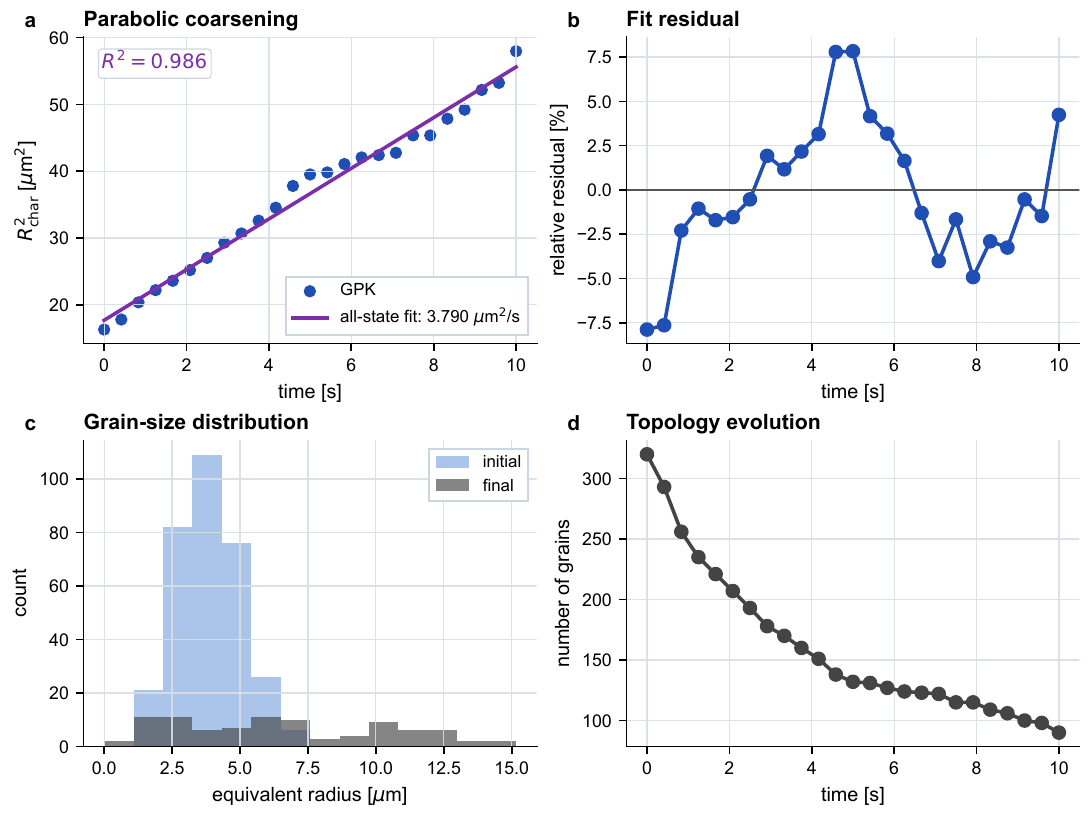}
\caption{High-resolution grain-growth validation using 25 states from a 320-grain run: full-series parabolic fit, residuals, grain-size distributions and topology evolution. All 25 states define the fit; no fitting-window selection, universal network prefactor or curve smoothing is applied.}
\label{fig:sgrain}
\end{figure}

For isotropic growth, $\Delta G=0$ and $v_n=-M\gamma\kappa$. Normal grain growth is expected to give a parabolic mean-size trend, but the network prefactor is not universal: it depends on dimension, topology and the chosen radius statistic. We therefore define the characteristic mean-area radius in two dimensions as $R_{\rm char}=\sqrt{A_{\rm domain}/(\pi N_g)}$, test linearity of $R_{\rm char}^2(t)$, and report $K/(M\gamma)$ rather than comparing with an unsupported universal coefficient. The dedicated $512^2$ validation run in Figure~\ref{fig:sgrain} uses 25 saved states and evolves from 320 to 90 grains. Its characteristic radius increases from 4.04 to \SI{7.61}{\micro\metre}; the initial value is 16.15 cells, or 4.04 times the four-cell Gaussian width. The mean equivalent-grain radius changes from $(3.88\pm1.13)\,\si{\micro\metre}$ to $(6.58\pm3.83)\,\si{\micro\metre}$, showing the broadened size distribution also visible in panel (c). The declared coarsening regression uses all 25 saved states from 0 to \SI{10}{s}; this predeclared full-series choice avoids selecting a fitting window after inspecting the discrete grain-extinction events. The result is $R^2=0.9855$ with slope \SI{3.790}{\micro\metre\squared\per\second}. The fitted residuals span $-7.88$ to $+7.83\%$, and their largest adjacent change is 5.71 percentage points. These fluctuations are retained because $R_{\rm char}^2=A_{\rm domain}/(\pi N_g)$ changes discretely when grains vanish. No smoothing filter is applied. Local kinetic prefactors are constrained more sharply by the Gauss--Bonnet and von Neumann--Mullins tests.

The panels in Figure~\ref{fig:sgrain} expose complementary parts of the coarsening claim. The full-series $R_{\rm char}^2$ fit tests the expected parabolic trend, while the adjacent residual panel shows every deviation instead of hiding discrete extinction events in a smoothed curve. The grain-size distributions demonstrate that the network broadens as small grains disappear, and the topology panel records the accompanying decrease from 320 to 90 identities. Using all 25 states makes the regression auditable and prevents the fitted coefficient from being mistaken for a universal material constant.
\FloatBarrier

\subsection{Matched GPK--phase-field protocol and grain-wise metrics}
The matched-comparison executable uses a periodic $192\times192$ grid with $\Delta x=64/192\,\si{\micro\metre}\simeq\SI{0.333}{\micro\metre}$, 80 Voronoi grains and seed 42017 on a \SI{64}{\micro\metre} square. Labels are identical at $t=0$, and the initial characteristic radius is 12.1 cells compared with $\sigma_K=2$ cells. GPK uses the isotropic module with $M=5.0\times10^{-11}\,\mathrm{m^4\,J^{-1}\,s^{-1}}$ and $\gamma=\SI{0.324}{\joule\per\square\metre}$. The PF reference uses one continuous order parameter per grain, including grains that later vanish; no order-parameter reuse or colour-class merging is permitted. Its explicit step is $\Delta\tau=0.18/(\kappa_{\rm PF}L_{\rm PF})=0.045$ for $\kappa_{\rm PF}=4$ and $L_{\rm PF}=1$.

Physical PF time is fixed before the polycrystal run. A two-order-parameter circle of radius 24 cells gives $-\mathrm d(R/\Delta x)^2/\mathrm d\tau=7.9948$. Equating this with $-2M\gamma/\Delta x^2$ gives $\mathrm dt/\mathrm d\tau=\SI{0.027417}{s}$ and a physical PF step of \SI{1.2338}{ms}. Thus no polycrystal point is used to tune time. Categorical PF labels are the pointwise $\arg\max_i\eta_i$, matching GPK's exact label partition for area and overlap comparisons.

\subsubsection*{Accuracy-matched resolution selection}
The primary performance audit does not impose the same cell count on two methods with different interface representations. A single set of 80 seed coordinates in the continuous periodic \SI{64}{\micro m} square is rasterised independently at $N=96$, 128, 160, 192, 224, 256, 320 and 384. Each method is evolved to \SI{0.25}{s}. The physical regularisation length is held fixed at \SI{0.667}{\micro m}: GPK uses $\sigma_K=2N/192$ cells and PF uses $\sqrt{\kappa_{\rm PF}}=2N/192$ cells. The isolated-circle PF clock is recalibrated for every $\kappa_{\rm PF}$, so neither the physical interface width nor the mobility changes as the grid is refined.

Each result is nearest-cell projected to the $384^2$ lattice and compared with that method's own $384^2$ result. The gates were fixed before the sweep: total grain-area MAE $\leq0.030$, grain-area correlation $\geq0.995$, pointwise label agreement $\geq0.950$, and boundary F1 $\geq0.900$ with a two-reference-cell tolerance. Comparing each method with its own fine-grid result measures discretisation error rather than mixing it with the constitutive difference between GPK and PF. The $384^2$ solution is a finite numerical reference, not an assertion of exact continuum truth.

For PF, a categorical $\arg\max_i\eta_i$ boundary may look converged even when the diffuse field is under-resolved. A two-phase, one-dimensional periodic planar interface is therefore relaxed at every $\kappa_{\rm PF}$. The discrete excess energy of either of the two interfaces is evaluated with the same centered gradient used in the audit,
\begin{equation}
 \widehat\gamma_h=\frac{1}{2\sqrt{\kappa_{\rm PF}}}
 \sum_j\left[f(\boldsymbol\eta_j)-f_{\rm bulk}
 +\frac{\kappa_{\rm PF}}{2}\sum_q
 \left(\frac{\eta_{q,j+1}-\eta_{q,j-1}}{2}\right)^2\right],
 \label{eq:si-pf-planar-energy}
\end{equation}
and compared with the continuum value $\widehat\gamma=\sqrt{2}/3$ for $\omega_{\rm PF}=1.5$. Full PF accuracy additionally requires relative surface-energy error $\leq1\%$. This is a field-resolution requirement, not a penalty assigned from a nominal number of points across the interface.

\begin{table}[ht]
\centering
\caption{Fixed-physics resolution audit against the $384^2$ method-specific references. ``Morph.'' applies all four categorical morphology gates; ``full'' additionally applies the PF 1\% planar-energy gate. GPK has no diffuse free-energy profile, so its full gate is its morphology gate.}
\label{tab:si-accuracy-resolution}
\small
\begin{tabular}{@{}rrrrrrrrr@{}}
\toprule
$N$ & Width [cells] & GPK MAE & GPK agree & GPK full & PF MAE & PF agree & PF energy err. & PF full \\
\midrule
96  & 1.00 & 0.0366 & 0.929 & No  & 0.0219 & 0.957 & 7.118\% & No \\
128 & 1.33 & 0.0251 & 0.938 & No  & 0.0129 & 0.968 & 3.785\% & No \\
160 & 1.67 & 0.0221 & 0.946 & No  & 0.0079 & 0.975 & 2.409\% & No \\
192 & 2.00 & 0.0152 & 0.951 & Yes & 0.0073 & 0.980 & 1.670\% & No \\
224 & 2.33 & 0.0142 & 0.954 & Yes & 0.0069 & 0.982 & 1.226\% & No \\
256 & 2.67 & 0.0133 & 0.955 & Yes & 0.0064 & 0.982 & 0.939\% & Yes \\
320 & 3.33 & 0.0106 & 0.959 & Yes & 0.0046 & 0.988 & 0.600\% & Yes \\
384 & 4.00 & 0 & 1 & Yes & 0 & 1 & 0.417\% & Yes \\
\bottomrule
\end{tabular}
\end{table}

GPK first passes at $192^2$; PF passes the categorical morphology gates already at $96^2$ but first passes the additional diffuse-field gate at $256^2$. The primary timing comparison therefore uses $192^2$ GPK and $256^2$ PF. It should be interpreted as a full-field-accuracy comparison. For a study that uses PF only as a categorical interface locator and does not require the diffuse profile or interfacial free energy, the morphology-only $96^2$ PF selection is the relevant counterfactual and no GPK coarser-mesh advantage follows from this audit.

\subsubsection*{Accuracy-matched system-size cost protocol}
The selected spacings, $\Delta x_{\rm GPK}=\SI{0.333}{\micro m}$ and $\Delta x_{\rm PF}=\SI{0.250}{\micro m}$, are transferred to square domains of side 32, 48, 64, 80, 96, 128, 160, 192 and \SI{256}{\micro m}. Grain density is fixed at $80/(\SI{64}{\micro m})^2$, producing 20--1280 grains and PF order parameters. The common physical timing horizon is \SI{0.05}{s}; constitutive parameters, physical Gaussian/PF width, compiler, hardware and 24-thread setting are fixed. This shorter horizon still contains 19--31 accepted GPK steps and 72 PF steps while making three repetitions of the dense $1024^2\times1280$-field endpoint tractable. Each size uses one deterministic continuum Voronoi construction and three fresh solver states. Evolution order alternates by repetition. Only evolution kernels are timed; seed generation, rasterisation, state construction, PF clock calibration, output and metrics are excluded.

\begin{table}[ht]
\centering
\caption{Extended accuracy-matched system-size scaling to \SI{256}{\micro m}. All runs use a \SI{0.05}{s} horizon. Times are three-run medians with measured minima--maxima in parentheses. Memory is allocated numerical-array capacity in MiB.}
\label{tab:si-accuracy-cost}
\footnotesize
\setlength{\tabcolsep}{3.6pt}
\begin{tabular}{@{}rrrrrrrr@{}}
\toprule
$L$ [$\mu$m] & Grains & Grids G/P & Steps G/P & GPK [s] & PF [s] & PF/GPK & Memory G/P \\
\midrule
32  & 20   & 96/128   & 19/72 & 0.0256 (0.0249--0.0275) & 0.1252 (0.1184--0.1340) & 4.895 & 0.71/2.56 \\
48  & 45   & 144/192  & 24/72 & 0.0384 (0.0373--0.0421) & 0.2830 (0.2796--0.2929) & 7.378 & 1.60/12.80 \\
64  & 80   & 192/256  & 27/72 & 0.0693 (0.0654--0.0693) & 0.5888 (0.5877--0.5949) & 8.501 & 2.85/40.25 \\
80  & 125  & 240/320  & 30/72 & 0.1000 (0.0942--0.1047) & 1.0322 (1.0285--1.1573) & 10.325 & 4.45/98.05 \\
96  & 180  & 288/384  & 24/72 & 0.1161 (0.1079--0.1422) & 1.9444 (1.9093--1.9666) & 16.749 & 6.41/203.06 \\
128 & 320  & 384/512  & 30/72 & 0.2459 (0.2358--0.2619) & 4.3108 (4.1431--4.4504) & 17.530 & 11.39/641.00 \\
160 & 500  & 480/640  & 29/72 & 0.3092 (0.2595--0.3475) & 8.2318 (7.5268--9.1346) & 26.624 & 17.80/1564.06 \\
192 & 720  & 576/768  & 26/72 & 0.3240 (0.3142--0.4275) & 18.9194 (17.9861--18.9916) & 58.395 & 25.63/3242.25 \\
256 & 1280 & 768/1024 & 31/72 & 0.7581 (0.6922--0.8409) & 72.6765 (72.5347--72.9907) & 95.867 & 45.56/10244.00 \\
\bottomrule
\end{tabular}
\end{table}

Fits against physical domain area give $t_{\rm GPK}\propto A^{0.8065}$ ($R^2=0.9846$) and $t_{\rm PF}\propto A^{1.5022}$ ($R^2=0.9810$). The GPK array-capacity exponent differs from unity by $-1.0\times10^{-4}$, while the PF capacity scales as $A^{1.9950}$. GPK is faster at all nine sizes; the PF/GPK median ratio increases from 4.90 at \SI{32}{\micro m} to 95.87 at \SI{256}{\micro m}. At the largest case GPK allocates 45.56 MiB and PF 10,244 MiB, a factor of 224.83. These are allocated-array counts rather than peak resident-set measurements. The finite-range fits describe these two codes and do not rank other PF time integrators or sparse/GPU implementations.

The machine-readable files \path{accuracy_matched_resolution.csv}, \path{accuracy_matched_selection.csv}, \path{accuracy_matched_cost_scaling.csv} and \path{accuracy_matched_cost_scaling_repeats.csv} retain every resolution metric, selection threshold, summary value and individual timing repetition used in main-text Figure~\ref{M-fig:benchmark} and main-text Table~\ref{M-tab:gpk-pf-cost}.

\subsubsection*{Equal-grid implementation diagnostic}
The cost audit is embedded in the same executable and uses the same solver objects, numerical parameters and physical-time mapping as the morphology comparison. It was compiled in Release mode with GCC 15.2.0 using \texttt{-O3 -DNDEBUG -fopenmp} and run on an Intel Core Ultra 9 285 with 24 OpenMP threads (\texttt{OMP\_DYNAMIC=FALSE}). An empty parallel region creates the thread team before timing. Each of five repetitions constructs fresh solver states outside the timed interval, then follows the production output schedule at 0.25, 0.50, 0.75 and \SI{1}{s}; execution order alternates between GPK-first and PF-first. Retaining these checkpoints matters because GPK clips its final accepted step to each requested output time. The timed interval contains only calls to the evolution kernels. Voronoi generation, state construction, isolated-circle PF calibration, CSV output, categorical-label extraction and all overlap metrics are excluded. A separate single trajectory records cumulative evolution time at every 0.25 s morphology snapshot through \SI{4}{s}.

Array memory is measured as the sum of the capacities of solver-owned numerical arrays. For PF this includes both buffers for all order parameters and the scalar sum-of-squares field. For GPK it includes persistent fields, update buffers, compact active-cell indices, block-prefix arrays and the one-dimensional Gaussian and derivative kernels; broad three-dimensional kernels additionally allocate one shared four-array separable workspace. The count excludes C++ allocator metadata, solver-object scalars, the thermodynamic-model object, OpenMP/runtime-library allocations and operating-system overhead; it is therefore an auditable array-capacity measure, not peak resident-set size. The GPK count is independent of thread number because neither geometry path uses per-thread full-domain workspaces. PF memory scales with its order-parameter count, while the GPK persistent label state does not; neither the measured memory ratio nor the wall-time ratio is asserted to be universal.

\begin{table}[ht]
\centering
\caption{Raw evolution-only timings for the matched \SI{1}{s} cost audit. The solver executed first was alternated; the table preserves the measured runs rather than reporting only the median.}
\label{tab:si-gpk-pf-cost}
\small
\begin{tabular}{@{}rrrrrr@{}}
\toprule
Repeat & GPK steps & PF steps & GPK time [s] & PF time [s] & PF/GPK \\
\midrule
1 & 419 & 811 & 3.4466 & 5.8376 & 1.694 \\
2 & 419 & 811 & 3.5869 & 5.8517 & 1.631 \\
3 & 419 & 811 & 3.4313 & 5.8092 & 1.693 \\
4 & 419 & 811 & 3.4709 & 5.8124 & 1.675 \\
5 & 419 & 811 & 3.3435 & 5.7839 & 1.730 \\
\midrule
Median & 419 & 811 & 3.4466 & 5.8124 & 1.686 \\
\bottomrule
\end{tabular}
\end{table}

At \SI{1}{s}, GPK and PF allocate 2,986,404 and 23,740,416 array bytes (2.85 and 22.64 MiB), respectively. The five-run median gives 8.23 ms per accepted GPK step and 7.17 ms per explicit PF step. GPK nevertheless reaches the physical horizon in about half as many accepted steps. The independent \SI{4}{s} trajectory takes 1577 GPK and 3242 PF steps and records 12.852 and 23.468 s of cumulative kernel time, giving a PF/GPK ratio of 1.826.

\subsubsection*{Equal-grid domain-size scaling protocol}
The size sweep is a controlled system-size scaling experiment rather than a mesh-refinement or parallel weak-scaling study. It fixes $\Delta x=\SI{0.333}{\micro\metre}$, $M=5.0\times10^{-11}\,\mathrm{m^4\,J^{-1}\,s^{-1}}$, $\gamma=\SI{0.324}{\joule\per\square\metre}$, Gaussian width, PF parameters, physical horizon \SI{0.25}{s}, hardware and thread count. The grain density is fixed at $80/192^2$ grains per cell and rounded to the nearest integer at each $N$. Consequently the characteristic radius varies by only 0.7\%, from 12.04 to 12.12 cells, while the domain contains progressively more grains. Each size uses one deterministic periodic Voronoi state and three fresh timing repetitions; solver order alternates between repetitions. The timed and excluded operations are identical to the production-case protocol above.

\begin{table}[ht]
\centering
\caption{Matched controlled system-size scaling results at fixed thread count. Times are three-run medians; the complete minima, maxima and individual repeats are stored in the machine-readable outputs. Memory entries are allocated-array capacity in MiB.}
\label{tab:si-gpk-pf-scaling}
\small
\begin{tabular}{@{}rrrrrrrr@{}}
\toprule
$N$ & Grains & $R_{\rm char}/\Delta x$ & Steps G/P & GPK [s] & PF [s] & PF/GPK & Memory G/P \\
\midrule
64  & 9   & 12.04 & 69/203  & 0.106 & 0.162 & 1.530 & 0.32/0.30 \\
96  & 20  & 12.11 & 89/203  & 0.180 & 0.406 & 2.252 & 0.71/1.44 \\
128 & 36  & 12.04 & 101/203 & 0.315 & 0.592 & 1.881 & 1.27/4.56 \\
160 & 56  & 12.06 & 100/203 & 0.429 & 0.979 & 2.284 & 1.98/11.04 \\
192 & 80  & 12.11 & 105/203 & 0.525 & 1.479 & 2.816 & 2.85/22.64 \\
224 & 109 & 12.10 & 115/203 & 0.743 & 2.159 & 2.905 & 3.88/41.92 \\
256 & 142 & 12.12 & 112/203 & 1.060 & 3.079 & 2.905 & 5.06/71.25 \\
288 & 180 & 12.11 & 116/203 & 1.271 & 4.222 & 3.323 & 6.41/114.22 \\
320 & 222 & 12.12 & 126/203 & 1.589 & 5.739 & 3.611 & 7.91/173.83 \\
\bottomrule
\end{tabular}
\end{table}

Log--log least-squares fits against total cell count give runtime exponents $p=0.8446$ for GPK ($R^2=0.9914$) and $p=1.0890$ for PF ($R^2=0.9919$); the GPK array-capacity exponent differs from unity by $-3.0\times10^{-4}$, and the PF exponent is 1.9819. GPK is faster at all nine equal-grid sizes, and the median PF/GPK ratio reaches 3.61 at $320^2$. The distinct slopes are consistent with interface-local GPK reconstruction and dense order-parameter PF updates. This result is retained as an implementation diagnostic and is not used for the primary accuracy-matched performance claim.

{\sloppy The raw files \path{computational_cost_scaling.csv} and \path{computational_cost_scaling_repeats.csv} contain the equal-grid summary and all 27 size-repeat records. Together with \path{computational_cost_repeats.csv}, \path{computational_cost_summary.csv} and \path{computational_cost_trajectory.csv}, they preserve the secondary diagnostics in Tables~\ref{tab:si-gpk-pf-cost}--\ref{tab:si-gpk-pf-scaling}.\par}

\begin{table}[ht]
\centering
\caption{Grain-resolved GPK--PF comparison. Area MAE is the sum of absolute per-grain area-fraction differences; boundary F1 accepts a two-cell positional tolerance. The \SI{1}{s} row is the declared quantitative gate; later rows are ungated diagnostics after additional method-specific topology events.}
\label{tab:gpk-pf}
\small
\begin{tabular}{@{}rrrrrrr@{}}
\toprule
$t$ [s] & Mean IoU & Area MAE & Area $r$ & Label agreement & Boundary F1 & Grains GPK/PF\\
\midrule
0.00 & 1 & 0 & 1 & 1 & 1 & 80/80\\
0.25 & 0.883 & 0.0263 & $1-2.0\times10^{-3}$ & 0.959 & $1-1.0\times10^{-3}$ & 79/78\\
0.50 & 0.893 & 0.0464 & $1-5.0\times10^{-3}$ & 0.949 & $1-1.0\times10^{-3}$ & 71/71\\
0.75 & 0.865 & 0.0718 & $1-1.0\times10^{-2}$ & 0.934 & 0.987 & 66/66\\
1.00 & 0.853 & 0.0979 & 0.982 & 0.923 & 0.973 & 60/62\\
1.25 & 0.822 & 0.1272 & 0.976 & 0.907 & 0.958 & 58/60\\
1.50 & 0.786 & 0.1617 & 0.973 & 0.889 & 0.935 & 56/59\\
1.75 & 0.757 & 0.1826 & 0.970 & 0.879 & 0.915 & 53/58\\
2.00 & 0.760 & 0.2126 & 0.963 & 0.862 & 0.883 & 50/54\\
2.25 & 0.727 & 0.2406 & 0.957 & 0.848 & 0.848 & 46/53\\
2.50 & 0.758 & 0.2647 & 0.947 & 0.833 & 0.817 & 42/48\\
2.75 & 0.768 & 0.2842 & 0.942 & 0.820 & 0.779 & 42/45\\
3.00 & 0.764 & 0.3059 & 0.936 & 0.807 & 0.748 & 39/43\\
3.25 & 0.737 & 0.3350 & 0.929 & 0.792 & 0.725 & 37/43\\
3.50 & 0.725 & 0.3653 & 0.920 & 0.780 & 0.697 & 35/43\\
3.75 & 0.730 & 0.3885 & 0.912 & 0.772 & 0.672 & 30/41\\
4.00 & 0.724 & 0.4026 & 0.909 & 0.768 & 0.655 & 29/40\\
\bottomrule
\end{tabular}
\end{table}

The executable gates require area correlation $\geq0.98$, area MAE $\leq0.15$, mean IoU $\geq0.62$, label agreement $\geq0.84$ and boundary F1 $\geq0.65$ at \SI{1}{s}; all pass. The later rows are ungated diagnostics. Once one method resolves a marginal junction or terminal grain first, neighboring curvatures and subsequent extinction times no longer represent the same discrete state, even though ensemble and boundary correlations remain high. No morphology repair or benchmark-only topology rule is used at any time.
\FloatBarrier

\subsection{Recrystallisation}
\begin{figure}[ht]
\centering\includegraphics[width=0.98\textwidth]{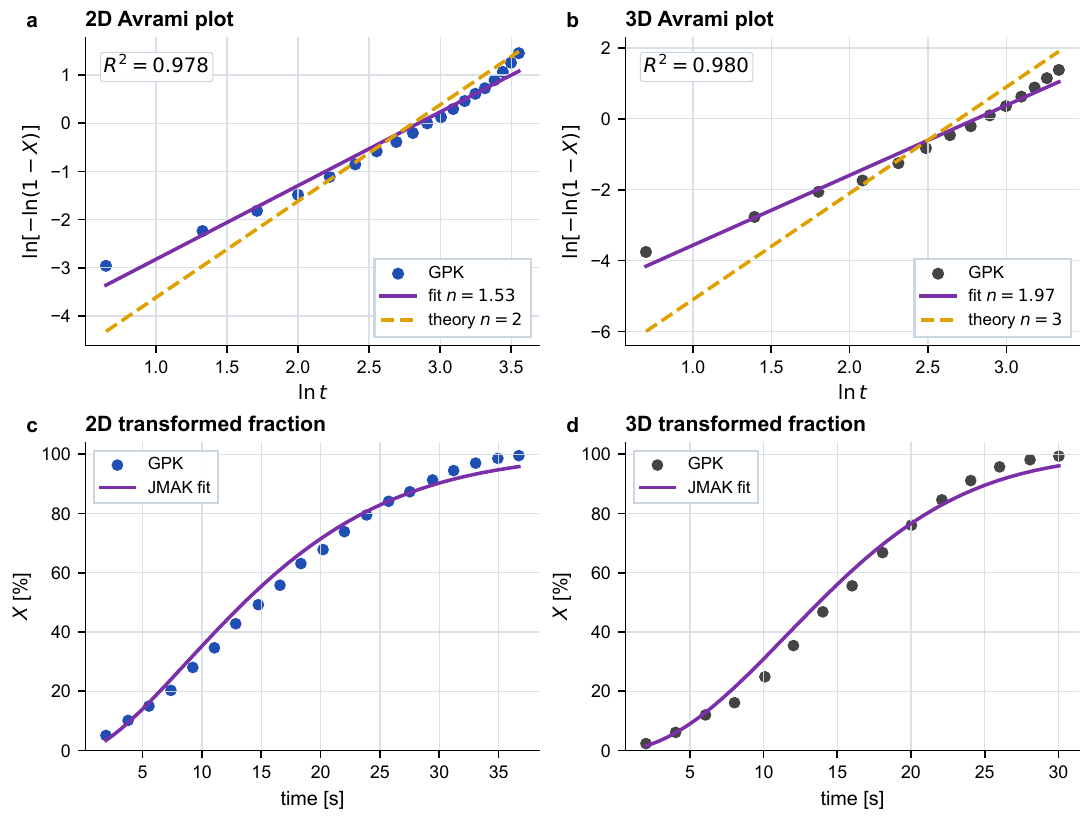}
\caption{Two- and three-dimensional recrystallisation validation. Avrami plots and transformed-fraction curves use 19 and 14 accepted fit points, respectively; dashed lines show ideal site-saturated slopes for comparison.}
\label{fig:srex}
\end{figure}

The product driving force contains the stored deformation energy. Nuclei are prescribed at $t=0$ for this site-saturated audit; subsequent interfaces follow the common kinetic law. Figure~\ref{fig:srex} uses 21 saved 2-D states ($0$--\SI{36.71}{s}) and 16 saved 3-D states ($0$--\SI{30.03}{s}). The accepted transformed ranges contain 19 and 14 points and give $n=1.5283$, $R^2=0.9783$ in 2-D and $n=1.9745$, $R^2=0.9798$ in 3-D. The final transformed fractions are 0.9952 and 0.9935. The exponents lie below the ideal site-saturated values 2 and 3 because finite nuclei, grid-resolved fronts, junction impingement and the finite fitting window violate the point-nucleus assumptions; they are reported as numerical outcomes rather than calibrated material constants.

In Figure~\ref{fig:srex}, the transformed-fraction panels show monotonic impingement and saturation in two and three dimensions, whereas the Avrami panels test whether the accepted intermediate ranges are approximately linear after the logarithmic transformation. The dashed ideal slopes provide a geometric reference for site-saturated point nuclei. The smaller measured slopes are consistent with finite resolved nuclei, junction impingement and finite fitting windows; they are not corrected to the ideal values. Reading the fraction and Avrami panels together prevents a visually linear transformed plot from obscuring the actual saturation interval used in the fit.
\FloatBarrier

\subsection{Bicrystal regression suite}
\begin{figure}[ht]
\centering\includegraphics[width=0.98\textwidth]{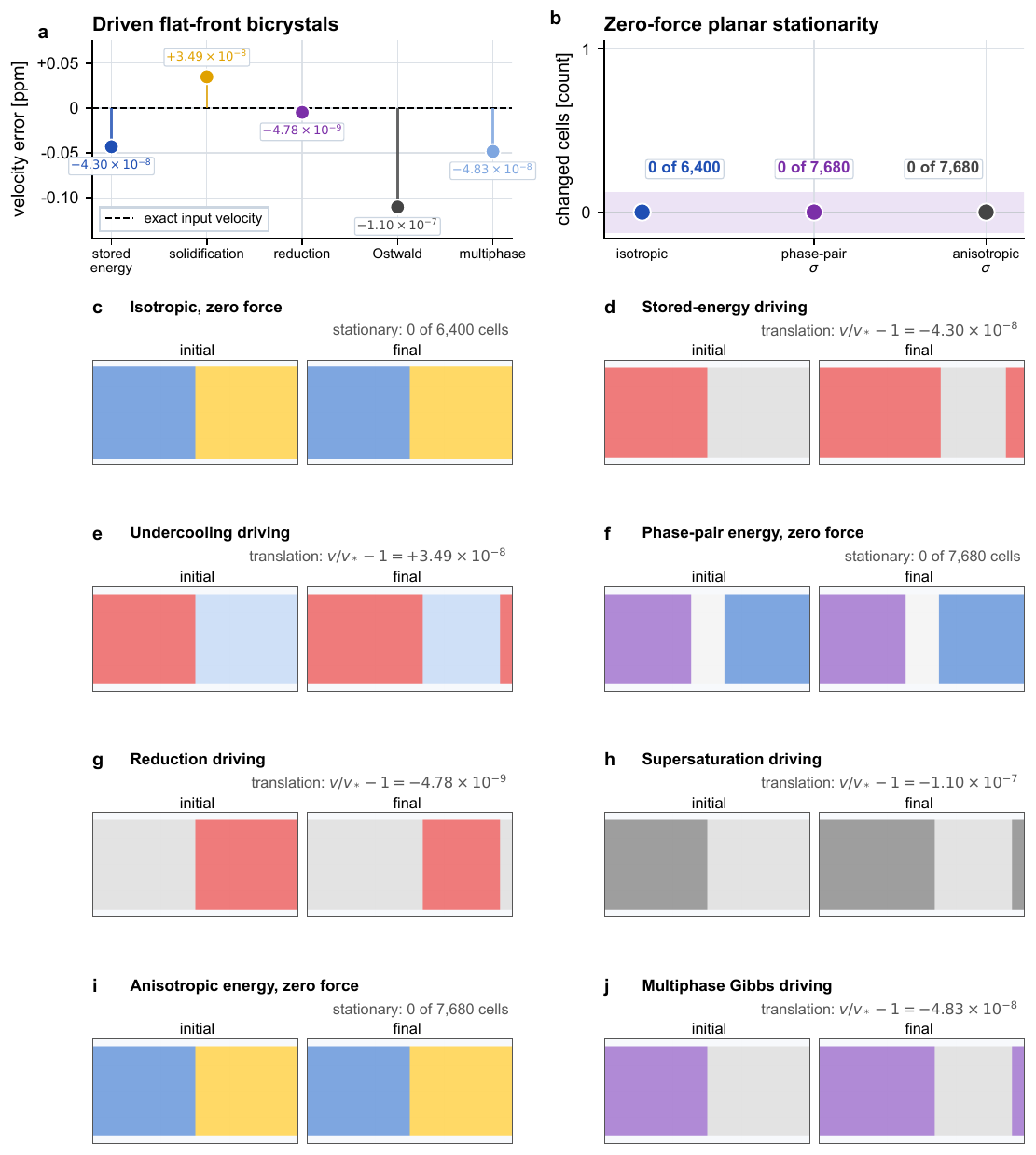}
\caption{Planar bicrystal validation. (a) Signed velocity errors in parts per million for five driven planar fronts; annotations give the signed fractional deviations $v/v_*-1$ in scientific notation. (b) Exact changed-cell counts for three zero-force boundaries. (c--j) True initial and final label fields for isotropic stationarity, stored-energy translation, undercooled solidification, phase-pair-energy stationarity, reduction, supersaturation, anisotropic-energy stationarity and multiphase Gibbs selection. The fields are PyVista renderings of the raw VTK cell images without line overlays, display smoothing or cell resampling; only the uniform canvas margin above and below each flat-front field is omitted. Every metric uses the same raw labels.}
\label{fig:sbicrystal}
\end{figure}

The bicrystal suite uses the same executable and update path as the polycrystal and gallery results. Twenty-two of twenty-two gates pass. For stored-energy growth (G2), solidification-type Gibbs driving (G5), reduction (G7), Ostwald-type driving (G8) and multiphase Gibbs selection (G10), the fitted signed fractional velocity deviations $v/v_*-1$ are $-4.30\times10^{-8}$, $+3.49\times10^{-8}$, $-4.78\times10^{-9}$, $-1.10\times10^{-7}$ and $-4.83\times10^{-8}$, respectively. Zero-force isotropic (G1), phase-pair-energy (G6) and anisotropic-energy (G9) planar boundaries change exactly 0 of 6,400, 0 of 7,680 and 0 of 7,680 cells, respectively. The remaining tests cover critical-radius crossover, grooving-angle relaxation and curved-boundary kinetics; the G12 supercritical growth-rate audit has 0.04\% error. Figure~\ref{fig:sbicrystal} shows the actual initial and final fields for all eight planar cases. All categorical fields in this suite are rendered from the saved VTK data with PyVista \citep{SullivanKaszynski2019}.

\begin{figure}[ht]
\centering\includegraphics[width=0.98\textwidth]{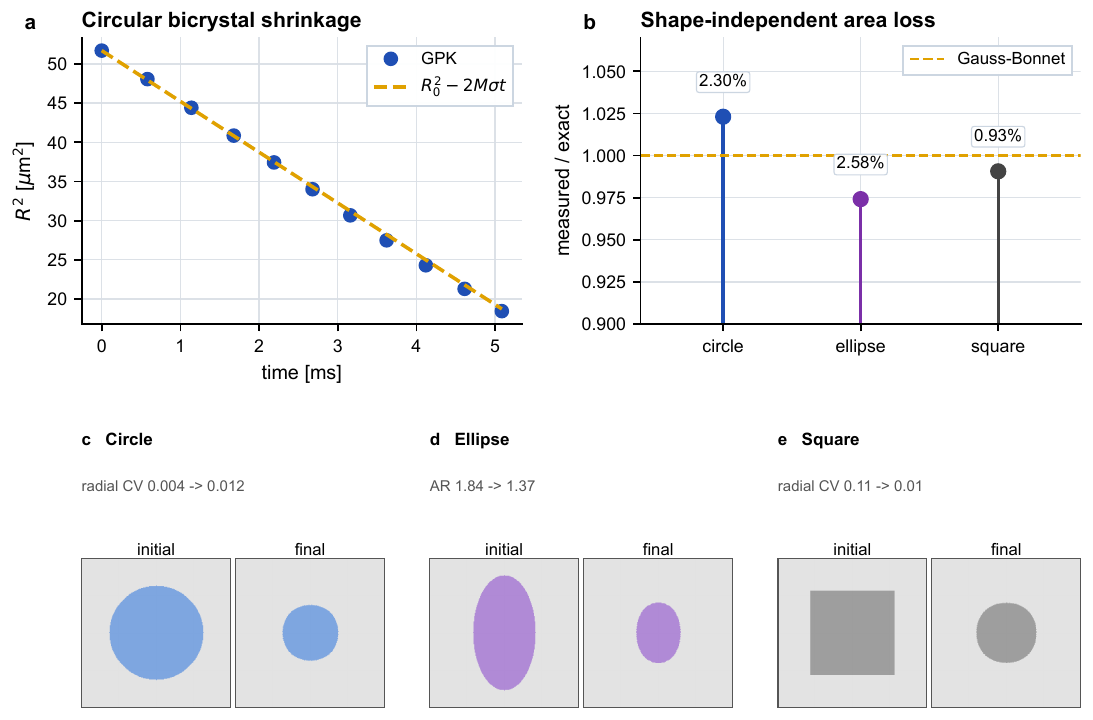}
\caption{Curved bicrystal validation on $360^2$ grids. (a) Circular-grain $R^2(t)$ evolution against the exact capillary law. (b) Measured area-loss rates for circle, ellipse and square normalized by $-2\pi M\gamma$; annotations give absolute relative errors. (c--e) PyVista renderings of the raw initial and final VTK cell images. The ellipse relaxes toward unit aspect ratio and the square regularises its corners. No line overlay or display smoothing is applied; regressions and descriptors use the same raw labels and subcell phase volumes.}
\label{fig:scurved-bicrystal}
\end{figure}

The G13--G15 tests are curved bicrystals: label 1 is a closed grain embedded in the label-0 grain. A circle, ellipse and square are discretised on $360^2$ grids at $\Delta x=\SI{0.1}{\micro\metre}$ and evolved by the same isotropic capillary operator without a bulk driving force. The runs retain 35.7\%, 37.4\% and 40.8\% of their initial areas, respectively, so the final images represent substantial evolution. The circular $R^2$-slope error is 2.30\%; ellipse and square area-rate errors are 2.58\% and 0.93\%. The circle remains close to rotational symmetry, with radial coefficient of variation changing from 0.004 to 0.012 as it shrinks. The ellipse second-moment aspect ratio decreases from 1.84 to 1.37, and the square boundary radial coefficient of variation decreases from 0.11 to 0.01. Figure~\ref{fig:scurved-bicrystal} therefore tests both exact kinetics and visible relaxation across constant, smoothly varying and initially corner-singular curvature.

\begin{table}[ht]
\centering
\caption{Bicrystal regression matrix. Planar cases isolate zero-curvature bulk terms; curved cases test capillary kinetics and shape relaxation.}
\label{tab:si-bicrystal-matrix}
\small
\begin{tabular}{@{}llll@{}}
\toprule
Case & Isolated constitutive term & Exact response & Measured response\\
\midrule
G1 & Isotropic capillarity, $\Delta G=0$ & Stationary & 0 of 6,400 cells changed\\
G2 & Stored deformation energy & Constant translation & $v/v_*-1=-4.30\times10^{-8}$\\
G5 & Solidification undercooling & Constant translation & $v/v_*-1=+3.49\times10^{-8}$\\
G6 & Phase-pair interfacial energy, $\Delta G=0$ & Stationary & 0 of 7,680 cells changed\\
G7 & Oxide-reduction free energy & Constant translation & $v/v_*-1=-4.78\times10^{-9}$\\
G8 & Supersaturation free energy & Constant translation & $v/v_*-1=-1.10\times10^{-7}$\\
G9 & Orientation-dependent $\gamma$, $\Delta G=0$ & Stationary & 0 of 7,680 cells changed\\
G10 & Multiphase Gibbs-energy difference & Constant translation & $v/v_*-1=-4.83\times10^{-8}$\\
\addlinespace
G13 & Circular isotropic boundary, $\Delta G=0$ & $\mathrm dR^2/\mathrm dt=-2M\gamma$ & 2.31\% error\\
G14 & Elliptical isotropic boundary, $\Delta G=0$ & $\dot A=-2\pi M\gamma$ & 2.58\% error\\
G15 & Square isotropic boundary, $\Delta G=0$ & $\dot A=-2\pi M\gamma$ & 0.93\% error\\
\bottomrule
\end{tabular}
\end{table}

Figure~\ref{fig:sbicrystal} separates driven motion from numerical drift. Panel (a) resolves the small signed regression residuals that would be hidden by decimal ratios close to one: the fractional velocity deviations span $-1.10\times10^{-7}$ to $+3.49\times10^{-8}$. The common mobility and sign convention are therefore recovered independently of the physical origin of $\Delta G$. Panel (b) reports integer changed-cell counts because the categorical label field has no meaningful subcell percentage: all three counts are exactly zero, not rounded estimates. The raw initial/final pairs in panels (c)--j then reveal what those scalar metrics mean: driven boundaries translate without a spurious change of thickness or orientation, whereas stationary boundaries remain at the same location. This combination rules out obtaining the correct average speed from compensating forward and backward cell events.

Figure~\ref{fig:scurved-bicrystal} adds curvature variation to the planar suite. Panel (a) tests the complete time law for a circle, and panel (b) compares the integrated area-loss rate for a circle, ellipse and initially sharp-cornered square. Panels (c)--(e) show that the same runs also evolve plausibly in shape: the circle remains nearly rotationally symmetric, the ellipse approaches unit aspect ratio and the square rapidly loses its corners. Agreement of the integral rates together with those raw morphologies is important because an area-only statistic could remain accurate while a boundary develops an unphysical local topology.
\FloatBarrier

\subsection{Multiphase boundary growth}
The multiphase module evaluates candidate Gibbs energies and phase-pair interfacial penalties through the same resident--candidate contract used for grain boundaries. In the self-contained regression, the lower-energy phase gains a volume fraction of 0.7611 over the run. The polynomial Gibbs-energy functions are replaceable: a material study can supply tabulated or CALPHAD values while retaining the GPK geometry and topology update.

\subsection{Elastic coupling}
The coherent elastic term is a boundary-localized accommodation penalty added to the chemical free-energy difference. Doubling the prescribed misfit produces the expected fourfold energy scaling. This test verifies the coupling path but should not be interpreted as a full Eshelby or spectral-elasticity solution.

\FloatBarrier
\clearpage
\subsection{Al--Cu material case, literature transcription and timestep audit}
\begin{figure}[ht]
\centering\includegraphics[width=0.98\textwidth]{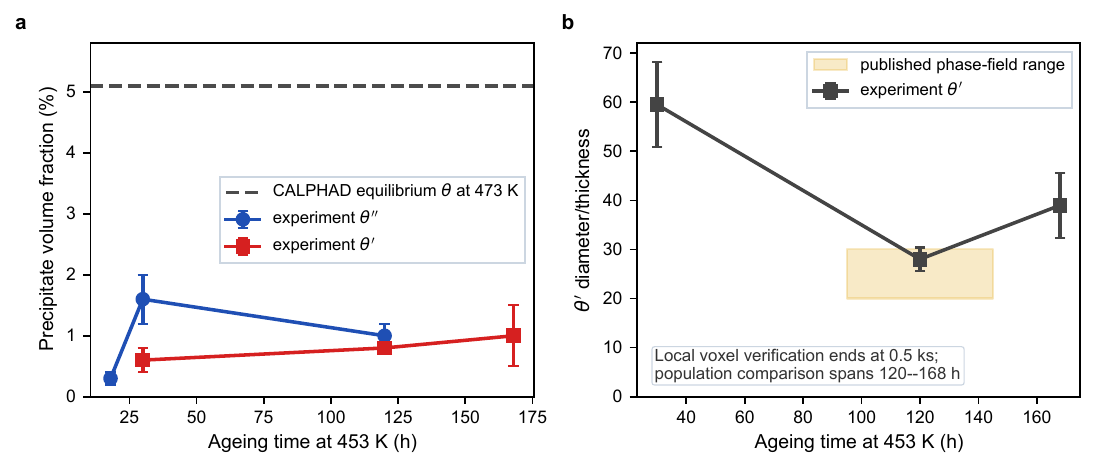}
\caption{Experimental and thermodynamic precipitation targets for Al--4 wt.\% Cu. (a) Experimental metastable $\theta''$ and $\theta'$ fractions at \SI{453}{K}, compared with the stable CALPHAD $\theta$ equilibrium fraction at \SI{473}{K}. (b) Experimental $\theta'$ aspect ratios and the published phase-field range. The local voxel calculation ends at \SI{0.5}{ks}; the population-scale comparison spans 120--168 h and is treated separately.}
\label{fig:salcu}
\end{figure}

The Al--Cu equilibrium calculation uses $x_{\rm Cu}=0.017384$, corresponding to 4 wt.\% Cu, $P=\SI{101325}{Pa}$ and temperatures from 423.15 to \SI{823.15}{K}. Pycalphad 0.11.2 minimises the corrected public COST-507 assessment over 
\texttt{FCC\_A1}, \texttt{ALCU\_THETA} and, in a separate metastable calculation, \texttt{ALCU\_PRIME}. The TDB is archived with the inputs so the numerical result does not depend on an undocumented commercial database. At \SI{473.15}{K}, the stable $\theta$ fraction is 0.050972; this is the horizontal equilibrium reference in Figure~\ref{fig:salcu}. It must not be compared as an ageing-time prediction because $\theta''$ and $\theta'$ are metastable and diffusion controlled.

The material recrystallisation input is \texttt{al4cu\_recrystallisation\_2d.ini}. It uses a $192^2$ grid, \SI{1}{\micro m} spacing, 40 site-saturated nuclei of radius three cells, $\gamma=\SI{0.30}{J.m^{-2}}$, $E_\rho=\SI{1.0e6}{J.m^{-3}}$ and an effective $M=1.224\times10^{-13}\,\mathrm{m^4\,J^{-1}\,s^{-1}}$. The experimental datum used for calibration is only ``fully recrystallised after five minutes'' for conventionally solidified, 91\%-reduced Al--4.5 wt.\% Cu at \SI{623.15}{K} \citep{Matos2020AlCu}. The simulation gives $X=0.9954$ at \SI{300.77}{s}. No intermediate experimental $X(t)$ values were digitised from micrographs, and no goodness-of-fit statistic is reported.

The precipitation points in Figure~\ref{fig:salcu} are direct transcriptions of Table~1 of Rodríguez-Veiga et al. \citep{RodriguezVeiga2018}. At 30, 120 and 168 h, the reported $\theta'$ diameter/thickness ratios are $59.5\pm8.6$, $28.0\pm2.4$ and $39.0\pm6.6$. The shaded 20--30 range represents the equilibrium aspect-ratio scale reported by multiscale phase field when interfacial and elastic anisotropies are both included \citep{Vaithyanathan2004}; it does not represent a separate phase-field run. These long-age targets delimit the present early-growth calculation rather than serving as contemporaneous validation points.

The two panels in Figure~\ref{fig:salcu} define different comparison scales. Panel (a) separates the experimentally observed metastable $\theta''/\theta'$ ageing fractions from the stable $\theta$ CALPHAD equilibrium reference; equality with the horizontal equilibrium line is therefore neither expected nor used as a kinetic target. Panel (b) shows that the measured $\theta'$ aspect ratio is non-monotonic with ageing time and places the published phase-field range beside the data. These long-time observations motivate the population calculation and anisotropic shape coupling, but they do not calibrate the local \SI{0.5}{ks} voxel result.

\subsection{Material-specific Al--Cu precipitation formulation}
The formulation of this module, Equations~\ref{M-eq:alcu-partition}--\ref{M-eq:alcu-mobility}, is given in the Methods of the main text (Section~\ref{M-sec:alcu-module}). This section reports the production run and its acceptance audit.

The three-dimensional production run uses $128^3$ cells at \SI{0.50}{nm} and three spherical nuclei, one for each tetragonal variant. States are saved at 0, 100, 250 and \SI{500}{s}. Its median aspect-ratio sequence is exactly 1 at initialization, followed by 1.273, 1.445 and 1.769. Final particle aspect ratios span 1.72--1.82, with major dimensions 12.88--13.51 nm and minor dimensions 7.28--7.84 nm. Thus the reported change measures model-generated habit selection rather than persistence of an imposed plate. The 64-nm three-dimensional box is comparable to the 70-nm reference fine-grid domain \citep{Ta2021}. For display only, the 3-D label support is projected with a one-cell Gaussian kernel and its continuous 0.45 isosurface is triangulated; periodic particles are minimum-image unwrapped before rendering. No connected component is deleted for display, and all reported morphology and topology statistics use the unsmoothed categorical volume with periodic connectivity.

\begin{figure}[ht]
\centering\includegraphics[width=0.98\textwidth]{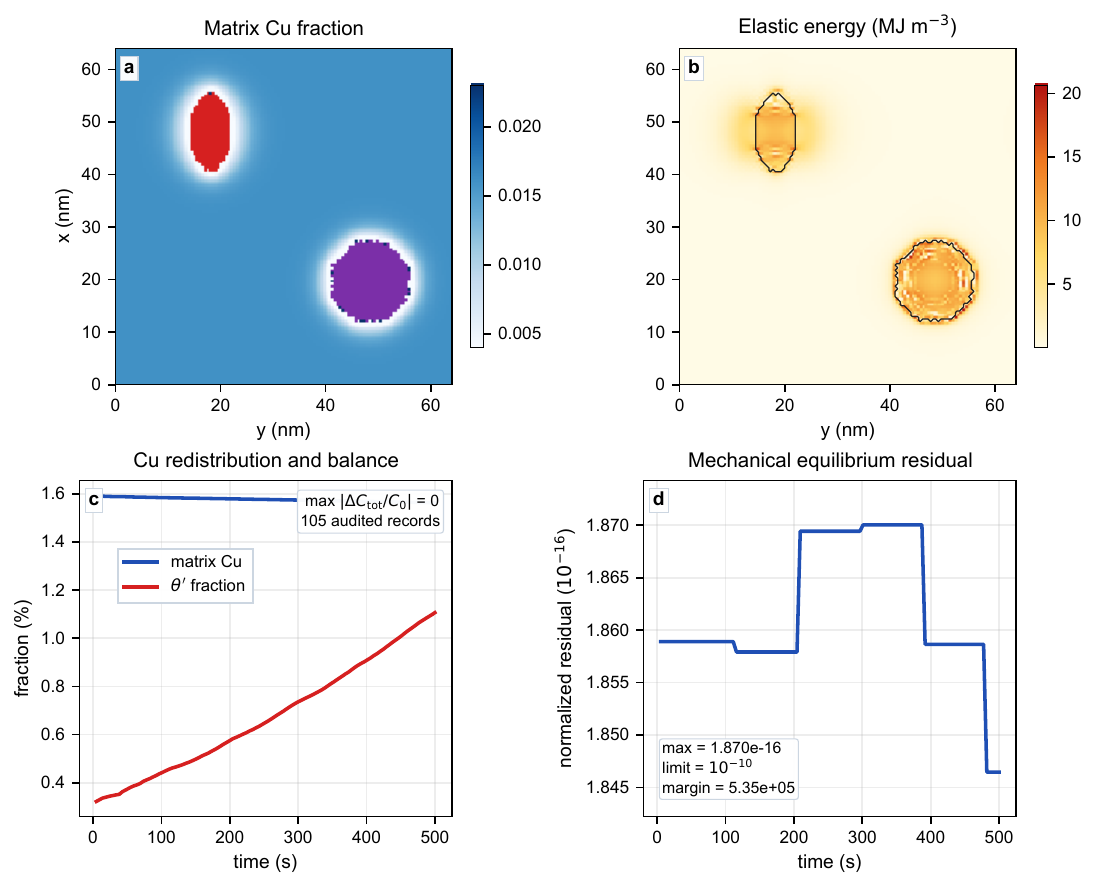}
\caption{Three-dimensional fields and numerical audits for the Al--Cu module. (a) and (b) Central sections of the final matrix-Cu and elastic-energy fields, drawn cell by cell from the saved fields without interpolation. In (a), the matrix Cu fraction is defined only in the matrix, so the $\theta'$ plates are filled with their variant colours, as in main-text Figure~\ref{M-fig:alcu-validation}. The white halo around each plate is the Cu-depleted zone. In (b), a thin line marks the plate boundaries. (c) Matrix-Cu and $\theta'$ fractions during solute redistribution; the inset reports the exact saved total-Cu balance result over all 105 audited records. (d) The non-zero variation of the normalized $\nabla\cdot\boldsymbol\sigma=0$ residual on a resolved linear scale; the inset gives its maximum, the $10^{-10}$ acceptance limit and their ratio.}
\label{fig:salcu-audits}
\end{figure}

Acceptance is based on an independent saved-label audit rather than on TCKT or the smooth render. Because this prescribed-nucleus module contains no nucleation law, each initialized variant must remain a single connected component and no state may acquire an enclosed matrix cavity. The $128^3$ endpoint contains one connected particle per variant, zero cavities and zero noncontiguous habit-normal lines. Across 105 saved-state and mesh-audit records, the maximum total-Cu error is zero to the reported precision; the largest normalized Fourier force-balance residual is $1.87\times10^{-16}$, and the maximum interface displacement is 0.35 cell. These conservation, mechanics, displacement and topology checks are independent acceptance gates (Figure~\ref{fig:salcu-audits}).

Figure~\ref{fig:salcu-audits} connects the final morphology to its coupled fields and numerical acceptance gates. The matrix-Cu section in panel (a) shows the one-sided solute gradients supplying growth, while the elastic-energy section in panel (b) localizes the coherency penalty around the anisotropic precipitates. Panel (c) shows that the precipitate fraction increases as the matrix Cu fraction decreases, while the independently accumulated total-Cu diagnostic remains exactly zero in all 105 saved records. This exact integer-like floating-point result is stated directly rather than plotted at an arbitrary logarithmic floor. Panel (d) expands the narrow $1.846\times10^{-16}$ to $1.870\times10^{-16}$ force-balance range so its small time variation is visible; the maximum remains $5.35\times10^{5}$ times below the $10^{-10}$ acceptance limit. The field maps alone would not establish correctness, and the audits alone would not establish a physically resolved morphology; their joint presentation tests both interpretations.

\begin{table}[ht]
\centering
\caption{Audited three-dimensional Al--Cu mesh comparison using the CALPHAD-coupled model and spherical nuclei. Percentage differences compare the physically matched $96^3$ and $128^3$ endpoints at \SI{500}{s}. The production row uses smaller prescribed nuclei and is not included in that convergence estimate.}
\label{tab:alcu-mesh}
\small
\begin{tabular}{@{}lrrrrr@{}}
\toprule
Case & $\Delta x$ [nm] & Time [ks] & $f_{\theta'}$ [\%] & Aspect ratio & Relative difference [\%]\\
\midrule
3-D $96^3$ matched & 0.667 & 0.5 & 2.7466 & 1.5195 & Reference\\
3-D $128^3$ matched & 0.500 & 0.5 & 2.5643 & 1.6778 & 6.64 / 10.42\\
3-D $128^3$ & 0.500 & 0.5 & 1.1061 & 1.7687 & Production\\
\bottomrule
\end{tabular}
\end{table}

For the matched 3-D pair, the median major dimension changes from 16.92 to 17.24 nm (1.89\%), whereas the median minor dimension changes from 11.20 to 10.68 nm (4.64\%). The phase-fraction and derived aspect-ratio differences are 6.64\% and 10.42\%. The pair therefore passes all conservation, force-balance and topology gates but does not satisfy a 2\% direct-measure convergence gate. The $128^3$ production morphology is identified as a distinct smaller-nucleus calculation, not as a third point in the matched-time convergence estimate.

\begin{figure}[ht]
\centering\includegraphics[width=0.98\textwidth]{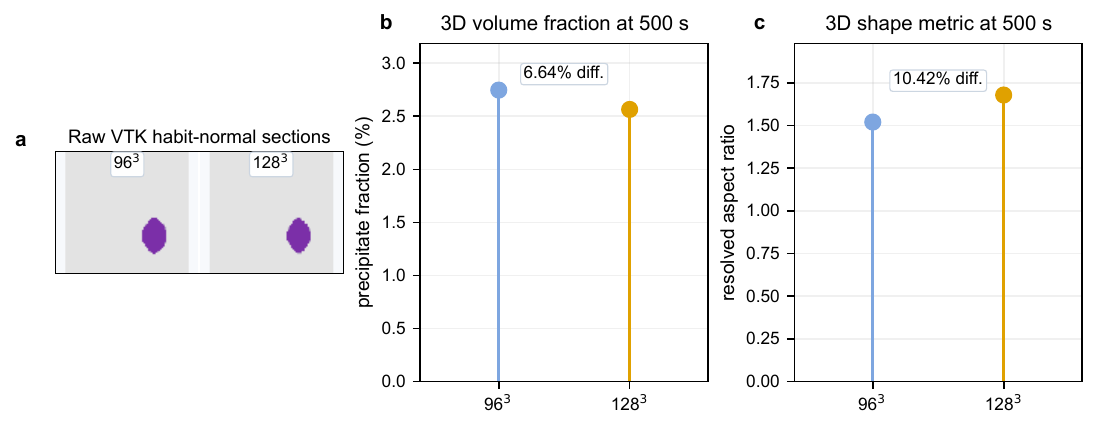}
\caption{Three-dimensional matched-domain mesh audit at \SI{500}{s}. (a) Side-by-side PyVista renderings of the raw VTK habit-normal section through the periodic centroid of variant 3 on the $96^3$ (left) and $128^3$ (right) meshes. (b,c) Matched-time precipitate fraction and aspect ratio. The measured differences are retained as numerical uncertainty.}
\label{fig:salcu-mesh}
\end{figure}

Figure~\ref{fig:salcu-mesh} makes the unresolved part of that audit explicit. Panel (a) shows broadly similar habit-normal sections at the two matched physical resolutions, so both meshes recover the same plate-forming mechanism. Panels (b) and (c) nevertheless quantify 6.64\% and 10.42\% differences in precipitate fraction and aspect ratio. Those differences exceed the 2\% direct-measure gate and are retained as numerical uncertainty rather than being hidden by the visually similar sections. The production morphology is therefore reported as a topology-valid calculation, but not as a mesh-converged prediction of fraction or aspect ratio.

\subsection{Population-scale 3-D ageing comparison}
The local voxel calculation above verifies interface geometry, conservative coupling, elasticity and topology; it is not extrapolated from \SI{500}{s} to the experimental ageing times. A direct \SI{1}{\micro m^3} calculation at the production spacing would require $2000^3$ dense cells. The experimentally matched calculation therefore uses a dilute population closure in which every precipitate remains an explicit three-dimensional sharp object with a persistent identity, centre $\mathbf x_i$, variant $v_i$, diameter $\ell_i$, thickness $H_i$ and thin-disc volume $V_i=\pi\ell_i^2H_i/4$, as in the main text.
The periodic population volume has side length $L=\SI{1}{\micro m}$. Resolved diffusion-field overlap, collision and long-range particle--particle elasticity are not evaluated in this closure; these are model-form limitations rather than hidden fit parameters.

Table~\ref{tab:si-pop-targets} transcribes the complete quantitative precipitate table of Rodr\'iguez-Veiga et al. \citep{RodriguezVeiga2018}; the machine-readable transcription is written beside the simulation outputs. Number densities have units $\mu\mathrm{m}^{-3}$, so they equal particle counts in the chosen volume. The 120 h state contains 131 $\theta'$ and 2883 $\theta''$ plates, or 3014 precipitates in total. At 168 h, $\theta''$ was not detected and 17 $\theta'$ plates remain. The 30 h $\theta'$ population is reported separately because those plates nucleated heterogeneously at dislocations and grain boundaries; it is not treated as the ancestor population of the 120 h homogeneous plates.

\begin{table}[ht]
\centering
\caption{Experimental precipitate statistics for Al--4 wt.\% Cu aged at \SI{453}{K}. Uncertainties are reported standard deviations; ``--'' denotes an unreported or undetected quantity.}
\label{tab:si-pop-targets}
\small
\begin{tabular}{@{}llrrrrr@{}}
\toprule
Time & Phase & $N$ [$\mu\mathrm{m}^{-3}$] & $f$ [\%] & $\ell$ [nm] & $H$ [nm] & $\ell/H$\\
\midrule
18 h & GP zones & $92997\pm9543$ & $0.4\pm0.1$ & $14.6\pm0.8$ & $0.256$ & $57.1\pm3.3$\\
18 h & $\theta''$ & $13298\pm1793$ & $0.3\pm0.1$ & $13.1\pm0.6$ & $1.4\pm0.1$ & $9.3\pm0.8$\\
30 h & $\theta''$ & $15229\pm1183$ & $1.6\pm0.4$ & $31.3\pm1.4$ & $1.6\pm0.2$ & $19.4\pm2.8$\\
30 h & $\theta'$ & $86\pm11$ & $0.6\pm0.2$ & $171.9\pm13.5$ & $2.9\pm0.2$ & $59.5\pm8.6$\\
120 h & $\theta''$ & $2883\pm170$ & $1.0\pm0.2$ & $45.1\pm2.1$ & $2.1\pm0.2$ & $21.6\pm2.7$\\
120 h & $\theta'$  & $131\pm63$ & $0.8\pm0.1$ & $142.5\pm25.1$ & $4.9\pm0.6$ & $28.0\pm2.4$\\
168 h & $\theta'$  & $17\pm6$ & $1.0\pm0.5$ & $342.5\pm46.8$ & $9.0\pm0.6$ & $39.0\pm6.6$\\
\bottomrule
\end{tabular}
\end{table}

The sampling of the 120 h population and the closure that carries it to 168 h, Equations~\ref{M-eq:pop-covariance}--\ref{M-eq:pop-survivor-scale}, are given in the Methods of the main text (Section~\ref{M-sec:alcu-population}). Table~\ref{tab:si-pop-history} lists the resulting primary realization.

\begin{table}[ht]
\centering
\caption{Primary population realization. Intermediate states are closure predictions because no corresponding TEM population statistics were used.}
\label{tab:si-pop-history}
\small
\begin{tabular}{@{}rrrrrrr@{}}
\toprule
Time [h] & $N_{\theta'}$ & $f_{\theta'}$ [\%] & $d_{\theta'}$ [nm] & $h_{\theta'}$ [nm] & $d/h$ & $N_{\theta''}$\\
\midrule
120 & 131 & 0.800 & 127.67 & 4.42 & 28.72 & 2883\\
132 & 79  & 1.128 & 177.73 & 5.57 & 31.90 & 1622\\
144 & 47  & 1.363 & 232.70 & 6.68 & 34.84 & 721\\
156 & 28  & 1.503 & 293.88 & 7.80 & 37.68 & 180\\
168 & 17  & 1.550 & 361.46 & 8.79 & 41.16 & 0\\
\bottomrule
\end{tabular}
\end{table}

Relative to the conditional 168 h measurements, the $\theta'$ fraction, diameter, thickness and aspect-ratio residuals are $+0.55$ percentage points, $+18.96$ nm, $-0.21$ nm and $+2.16$, or $+1.10$, $+0.41$, $-0.36$ and $+0.33$ reported experimental standard deviations. Two hundred independent initial samples give 95\% finite-population intervals of 358.45--364.42 nm, 8.69--9.00 nm and 39.90--42.03 for the three morphology means. These intervals quantify random initialization only.

Model-form sensitivity is reported separately. The final count is swept from 11 to 23 $\mu\mathrm{m}^{-3}$ (the measured mean $\pm1$ standard deviation), capture of the Cu released by $\theta''$ is varied from 50 to 100\%, and the transferred local aspect response is perturbed by $\pm10\%$. The full $5\times3\times3=45$ factorial table is written to \texttt{parameter\_sensitivity.csv}; its envelope is summarized in Table~\ref{tab:si-pop-sensitivity}. This is a robustness calculation, not a posterior credible interval, because no prior distribution is assigned to these structural choices.

\begin{table}[ht]
\centering
\caption{Structural sensitivity envelope for the 168 h comparison with experiment. Experimental intervals are mean $\pm1$ reported standard deviation.}
\label{tab:si-pop-sensitivity}
\small
\begin{tabular}{@{}lrr@{}}
\toprule
Quantity & GPK sensitivity envelope & Experiment\\
\midrule
$f_{\theta'}$ [\%] & 1.175--1.550 & 0.5--1.5\\
$d_{\theta'}$ [nm] & 286.21--434.75 & 295.7--389.3\\
$h_{\theta'}$ [nm] & 6.85--10.79 & 8.4--9.6\\
$d/h$ & 36.54--46.09 & 32.4--45.6\\
\bottomrule
\end{tabular}
\end{table}

The experimental means lie inside every structural envelope, while the phase-fraction envelope reaches the upper edge of the experimental one-standard-deviation interval. This supports a conditional reconstruction of mean morphology, not a parameter-free prediction of nucleation or coarsening kinetics. Particle-resolved TEM pairs were not published, so distribution-level tests such as Kolmogorov--Smirnov distance, Wasserstein distance or interval coverage cannot be computed without inventing pseudo-data; none is reported.

The primary realization uses NumPy's PCG64 generator with seed 453120. The sampling interval uses 200 complete, independent realizations with consecutive seeds 453120--453319 and reports the empirical 2.5 and 97.5 percentiles of the predicted population means. Standardized residuals use the experimental standard deviations in Table~\ref{tab:si-pop-targets}, not the sampling intervals. The executable asserts only internal invariants: endpoint conditioning, inventory identities and finite outputs. Experimental residuals are report-only; there is no ``within two standard deviations'' pass gate. The output directory contains the CSV history, compressed particle coordinates and variants, full literature transcription, an evidence ledger, factorial sensitivity table, JSON protocol and validation report, and a SHA-256 manifest. The protocol is explicitly labelled a post-hoc audit rather than a preregistration. These details fix the statistical estimand, random-number stream, uncertainty convention and output provenance required for an exact comparison.

The timestep sweep uses the same 96-cell periodic circle, radius 24 cells, $\kappa_{\rm PF}=4$, $L_{\rm PF}=1$ and $\omega_{\rm PF}=1.5$ as the independent clock calibration. A diffuse phase fraction, $A_1=\sum_x\eta_1^2/(\eta_0^2+\eta_1^2)$, avoids quantisation by the categorical $\arg\max$ contour. A step is accepted only when every order parameter remains finite with $\max|\eta|\leq10$ and the final $A_1/\pi$ differs by at most 5\% from the $\Delta\tau=0.005625$ reference. Table~\ref{tab:timestep-sweep} contains every tested PF point. GPK was then requested to take 0.001, 0.003, 0.01, 0.03, 0.10 and \SI{0.30}{s} steps to \SI{0.10}{s}; the maximum accepted values were 0.001, 0.003, 0.010, 0.030, 0.03008 and \SI{0.03008}{s}, respectively. All GPK runs completed because \texttt{step()} clips the request to $C_I\Delta x/\max|v_n|$. The accepted GPK maximum is 17.6 times the first inaccurate explicit-PF step at \SI{1.7136}{ms}, or 19.9 times the last accepted explicit-PF step at \SI{1.5079}{ms}; these ratios apply only to this explicit matched-grid circle audit.

\begin{table}[ht]
\centering
\caption{Explicit phase-field timestep sweep. Error is relative to the reference diffuse $R^2=A_1/\pi$. ``Unbounded'' means the order parameter exceeded the declared bound before the target time.}
\label{tab:timestep-sweep}
\small
\begin{tabular}{@{}rrrrr@{}}
\toprule
$\Delta\tau$ & $\Delta t$ [ms] & $R^2$ [cell$^2$] & Relative error [\%] & Result\\
\midrule
0.005625 & 0.1542 & 540.706 & 0 (reference) & Accepted\\
0.01125 & 0.3084 & 540.639 & 0.012 & Accepted\\
0.0225 & 0.6169 & 540.594 & 0.021 & Accepted\\
0.045 & 1.2338 & 540.499 & 0.038 & Accepted\\
0.055 & 1.5079 & 540.381 & 0.060 & Accepted\\
0.0625 & 1.7136 & 602.056 & 11.346 & Inaccurate\\
0.07 & 1.9192 & -- & -- & Unbounded\\
0.08 & 2.1934 & -- & -- & Unbounded\\
0.09 & 2.4675 & -- & -- & Unbounded\\
0.12 & 3.2900 & -- & -- & Unbounded\\
\bottomrule
\end{tabular}
\end{table}
\FloatBarrier

\subsection{Qualitative-gallery run definitions}
The two qualitative figures in the main text (Figures~\ref{M-fig:qualitative-2d} and \ref{M-fig:qualitative-3d}) were generated from cell-wise CSV fields written by the reported simulations. Table~\ref{tab:gallery-runs} records the grids, final times and timestep counts; every case shows its actual initialized field and final field. The 2-D figure shows raw categorical VTK cells: $768\times768$ cells for the first four cases, with the same persistent label-to-colour map as the GPK--phase-field comparison, and the central $800\times800$ of $1024\times1024$ cells for the dendrite. The 3-D grain-growth calculation starts from 240-grain Voronoi fields and evolves on $192\times192\times96$ cells to \SI{4.0}{s}. Labels and cell-wise orientations are then injected by nearest-cell prolongation, without modal filtering or interpolation, onto $256\times256\times128$ cells; a \SI{0.10}{s} GPK relaxation removes the categorical prolongation transient before measurement. The domain remains $64\times64\times32~\si{\micro m^3}$ and $\sigma_K$ changes from 8 to 10.667 cells, so $\ell_K=\SI{2.667}{\micro m}$ is fixed. The initial mean equivalent grain radii are 4.96 and \SI{4.97}{\micro m} for the isotropic and anisotropic seeds, respectively. Grain-growth panels (a) and (b) extract the complete specimen exterior directly from the raw VTK cell image. Every saved label owning a camera-facing exterior cell contributes to the contiguous three-face polycrystal view; there is no display subset. Recrystallisation panel (c) renders all 64 nonzero initial labels as individual nuclei; because the endpoint has $X_{\mathrm{RX}}=0.9949$, its complete categorical exterior is rendered to expose the impinged grain network. Counts, volumes, boundary areas and topology audits use every label in both states. No modal filter, interface-line overlay, Gaussian display isosurface, contour smoothing or label interpolation is applied. Persistent grain or phase colours are retained in initial and final states; parallel-projection cameras and specimen extents are identical within each row. A common matte material, two soft scene lights and a weak headlight provide depth without cast-shadow wedges and do not modify CSV data or numerical analysis.

The complete 2-D raw fields contain 152 isotropic and 156 anisotropic grains at \SI{10}{s}, with changed-cell fractions of 0.6178 and 0.6031. Both initialized fields have zero component excess across 360 labels, and both endpoints retain zero component excess and no one- or two-cell component. The recrystallisation field reaches $X_{\rm RX}=0.9915$ at \SI{20.02}{s}. At \SI{8}{ms}, the multiphase fractions are $(f_0,f_1,f_2)=(0.06483,0.28974,0.64543)$ and 0.9274 of the cells differ from their initialized labels. These values are calculated from the same CSV snapshots used to render main-text Figure~\ref{M-fig:qualitative-2d}, without masking, resampling or display reconstruction. The three-dimensional recrystallisation run (64 nuclei) and multiphase run (320 seeds) use the same $256\times256\times128$ grid and $\Delta x=\SI{0.25}{\micro m}$ as the grain-growth runs. They reach $X_{\rm RX}=0.9949$ at \SI{20.02}{s} and $(f_0,f_1,f_2)=(0.1328,0.2645,0.6027)$ at \SI{6}{ms}.

The elastic-coupled dendrites use the partitioning transport of main-text Equation~\ref{M-eq:partitioned-diffusion} with $k=0.3$, $D=3.0\times10^{-9}~\si{m^2.s^{-1}}$, $M=2.0\times10^{-9}\,\mathrm{m^4\,J^{-1}\,s^{-1}}$, $\gamma_0=\SI{0.20}{J.m^{-2}}$, $\chi_4=0.04$, $c_0=0.02$ and an undercooling of \SI{10}{K}. Both use $\Delta x=\SI{0.5}{\micro m}$ and $\sigma_K=2$ cells. Their final states are selected before the crystal can interact with a periodic face. At \SI{80}{ms}, after 9603 timesteps, the $1024^2$ run has a solid fraction of 0.14544 and raw categorical margins of 222, 221, 221 and 220 cells on the negative and positive $x$ and $y$ edges. At \SI{60}{ms}, after 10805 timesteps, the $512^3$ run has a solid fraction of 0.14626 and six-face margins of 31, 27, 28, 26, 38 and 26 cells on the negative and positive $x$, $y$ and $z$ faces. These margins are measured from the unfiltered labels. Boundary-interacting states are excluded from both galleries and from the reported fractions.

The topology rule distinguishes resolved grains from objects smaller than the geometry support. Resolved grains receive the connectivity certificate; a boundary-inward move may complete extinction when $V_g<\omega_d\sigma_K^d$, because treating such an object as a permanently protected component would make a sub-kernel remnant immortal. The coarse \SI{4}{s} states contain 237 isotropic and 239 anisotropic labels; nearest-cell prolongation and fine-grid relaxation to \SI{4.10}{s} preserve those counts. Their minimum final raw volumes are 6526 and 7123 voxels, both above the 5085-voxel spherical support associated with $\sigma_K=10.667$ cells. Each label forms exactly one periodic face-connected component, no component contains one or two cells, and neither endpoint contains a label adjacent to one or fewer other labels.

Table~\ref{tab:si-3d-production-audit} reports boundary diagnostics on the raw final fields. Multiplying unlike-face counts by $\Delta x^2$ makes the coarse initial and fine final measures comparable: they decrease by 0.44\% for isotropic and 1.54\% for anisotropic growth. A low-support interface cell has at most two same-label periodic face neighbours. Such cells constitute 0.69\% of both final interface populations, compared with 1.97\% and 1.98\% in the corresponding coarse initialized states. The exterior-fragment diagnostic counts connected single-label components smaller than 25 pixels over all six specimen faces and is reported before rendering; because its pixel threshold is resolution dependent, it is not used for coarse--fine acceptance. The complete initialized, prolonged and fine-evolved audit is stored in \texttt{figure7\_3d\_raw\_quality.json}.

\begin{table}[ht]
\centering
\caption{Raw-label audit of the \SI{4.10}{s} fine-grid three-dimensional grain-growth endpoints used in main-text Figure~\ref{M-fig:qualitative-3d}. No display reconstruction is applied.}
\label{tab:si-3d-production-audit}
\small
\begin{tabular}{@{}lrrrrrr@{}}
\toprule
Case & $N_g$ & $V_{\min}$ [voxels] & $N_{\rm face}$ & Low support & Exterior $<25$ px & One-neighbour labels \\
\midrule
Isotropic & 237 & 6526 & 1103426 & 9370/1357104 (0.69\%) & 55 & 0 \\
Anisotropic & 239 & 7123 & 1083622 & 9301/1342426 (0.69\%) & 54 & 0 \\
\bottomrule
\end{tabular}
\end{table}

\begin{table}[ht]
\centering
\caption{Simulation outputs used in the qualitative two- and three-dimensional galleries.}
\label{tab:gallery-runs}
\small
\begin{tabular}{@{}p{0.30\textwidth}p{0.19\textwidth}p{0.18\textwidth}p{0.20\textwidth}@{}}
\toprule
Case & Grid & Final time (timesteps) & Displayed measure \\
\midrule
2-D isotropic grain growth (360 grains) & $768\times768$ & \SI{10.0}{s} & $N_g$ \\
2-D anisotropic grain growth (360 grains) & $768\times768$ & \SI{10.0}{s} & $N_g$ \\
2-D recrystallisation (180 nuclei) & $768\times768$ & \SI{20.02}{s} & $X_{\mathrm{RX}}$ \\
2-D multiphase boundary growth (108 seeds) & $768\times768$ & \SI{8.00}{ms} & $f_1,f_2$ \\
2-D elastic-coupled dendrite & $1024\times1024$ & \SI{80}{ms} (9603) & Solid fraction \\
3-D isotropic grain growth (240 grains) & \shortstack{$192\times192\times96$\\$\rightarrow256\times256\times128$} & \SI{4.10}{s} & $N_g$ \\
3-D anisotropic grain growth (240 grains) & \shortstack{$192\times192\times96$\\$\rightarrow256\times256\times128$} & \SI{4.10}{s} & $N_g$ \\
3-D recrystallisation (64 nuclei) & $256\times256\times128$ & \SI{20.02}{s} & $X_{\mathrm{RX}}$ \\
3-D multiphase boundary growth (320 seeds) & $256\times256\times128$ & \SI{6.00}{ms} & $f_1,f_2$ \\
3-D elastic-coupled dendrite & $512\times512\times512$ & \SI{60}{ms} (10805) & Solid fraction \\
\bottomrule
\end{tabular}
\end{table}
\FloatBarrier

\section{Parameter provenance and representative outputs}
The entries below separate material data, computed constitutive quantities, numerical controls, calibrations and conditioning data. A value labelled ``benchmark input'' is prescribed to verify an operator and is not presented as a measured material constant. A value labelled ``calibrated'' is estimated from the stated datum, so reproducing that datum is not counted as validation. Quantities labelled ``computed'' are evaluated from the current state or a preprocessing calculation and are not independently fitted.

\begin{longtable}{p{0.18\textwidth}p{0.25\textwidth}p{0.46\textwidth}}
\caption{Value-by-value provenance of the principal parameters and constitutive quantities used in the reported calculations.}\label{tab:parameter-provenance-si}\\
\toprule
Module or quantity & Value or relation & Origin and status \\
\midrule
\endfirsthead
\toprule
Module or quantity & Value or relation & Origin and status \\
\midrule
\endhead
Gaussian geometry & $\ell_K=\sigma_K\Delta x$; baseline $\sigma_K=2$ cells; refined galleries use the values reported in Table~\ref{tab:gallery-runs} and their input decks & Numerical measurement scale selected by the inequality $\Delta x\lesssim\ell_K\ll R_{\min}$ and the circle/refinement audits. It is neither a physical interface thickness nor a fitted material parameter. \\
Derivative acceptance & $|\nabla\widetilde I_g|\Delta x>10^{-4}$; $|\kappa|\leq1/\ell_K$ & Dimensionless round-off guard and resolution cap fixed by the geometry definition and tested in the foundational suite. \\
Time integration & $C_I=1/8$, $C_D=0.4$ & Numerical safety factors used in the interface-crossing and explicit-diffusion bounds derived above; selected by stability tests, not by experimental fitting. \\
Isotropic grain growth and GPK--PF comparison & $M=5.0\times10^{-11}\ \mathrm{m^4\,J^{-1}\,s^{-1}}$, $\gamma=0.324\ \mathrm{J\,m^{-2}}$, $\Delta G=0$ & Declared pure-Al representative inputs shared by both solvers. The PF physical clock is fixed independently by the two-grain circle law before the polycrystal comparison; no polycrystal trajectory is used for tuning. \\
Anisotropic grain growth & $M=M_0f_{\rm RS}(x)\exp[-Q/(RT)]$; $x=\Delta\theta/\theta_m$; declared $\gamma(\mathbf n)$ & $M_0$, $Q$, $\theta_m$, $\gamma_0$ and anisotropy strength are declared constitutive inputs. Orientation and misorientation are stored state variables; local $M$ and $\gamma$ are computed at the interface. The gallery is a qualitative coupling test unless a material-specific parameter source is stated. \\
Generic recrystallisation & $E_\rho=\tfrac12\beta_d\mu b^2\rho$ and $\Delta G=E_\rho(q)-E_\rho(p)$ & Computed from the initialized/evolved dislocation-density field and declared $\beta_d$, $\mu$ and $b$. Mobility is an input or an explicitly identified calibration; the Gaussian supplies only geometry. \\
Al--Cu recrystallisation & $\gamma=0.30\ \mathrm{J\,m^{-2}}$, $E_\rho=1.0\times10^6\ \mathrm{J\,m^{-3}}$, $M=1.224\times10^{-13}\ \mathrm{m^4\,J^{-1}\,s^{-1}}$ & $\gamma$ and $E_\rho$ are held declared inputs. The effective $M$ is calibrated only to complete recrystallisation after five minutes at 623.15 K in the experiment of Matos et al.~\citep{Matos2020AlCu}; intermediate kinetics are therefore calibrated predictions, not independent validation. \\
Generic multiphase module & $\Delta G_{q-p}=G_q(c,T)-G_p(c,T)$ & Computed from the declared polynomial branches in the coupling regression. These branches are synthetic verification inputs and are not represented as an assessed alloy database. \\
Al--Cu chemical driving force & COST-507, $x_{\rm Cu}=0.017384$, $T=453$ or 473.15 K as stated; $c_{\theta'}=1/3$, $V_m=10^{-5}\ \mathrm{m^3\,mol^{-1}}$ & Pycalphad preprocessing evaluates and tabulates the FCC/Al$_2$Cu grand-potential difference; runtime interpolation supplies $\Delta G^{\rm chem}$. The archived TDB and composition/temperature are inputs, while the energy difference and $c_m^{\rm eq}$ are computed outputs. \\
Al--Cu diffusion and elasticity & $D=1.66\times10^{-20}\ \mathrm{m^2\,s^{-1}}$; $\epsilon_a=0.00746$, $\epsilon_c=-0.051$ (three cyclic variants); $C_{11}=107.07$, $C_{12}=63.08$, $C_{44}=28.52$ GPa & Literature-transcribed material inputs from the phase-field study of Ta et al.~\citep{Ta2021}. Fluxes, displacement, stress and $\delta F_{\rm el}/\delta f_v$ are computed by the conservative transport and Fourier mechanical solvers. \\
Al--Cu interfacial and kinetic anisotropy & $\gamma_{\rm face}=0.10$, $\gamma_{\rm rim}=0.245\ \mathrm{J\,m^{-2}}$; $M_{\rm rim}=10^{-19}\ \mathrm{m^4\,J^{-1}\,s^{-1}}$; effective mobility contrast 51:1 & Surface energies and anisotropy form are literature-informed model inputs. The 51:1 contrast is a stated finite-resolution closure motivated by, but not equated with, the larger atomistic limiting contrast; it is not fitted to the final displayed plate dimensions. \\
Elastic-coupled dendrites & $M=2.0\times10^{-9}\ \mathrm{m^4\,J^{-1}\,s^{-1}}$, $\gamma_0=0.20\ \mathrm{J\,m^{-2}}$, $\chi_4=0.04$, $D=3.0\times10^{-9}\ \mathrm{m^2\,s^{-1}}$, $k=0.3$, $c_0=0.02$, undercooling \SI{10}{K}; elastic modulus \SI{5.0e9}{J.m^{-3}}, eigenstrain coefficient 0.05 & Benchmark inputs chosen to produce diffusion-limited growth on a resolvable grid. They are not presented as a calibrated alloy. Solute rejection and liquid-only diffusion follow main-text Equations~\ref{M-eq:liquid-concentration} and \ref{M-eq:partitioned-diffusion}. \\
Population ageing closure & 120 h population; 168 h $N_{\theta'}$ and $\theta''$ non-detection & Experimental conditioning data from Rodr\'iguez-Veiga et al.~\citep{RodriguezVeiga2018}. The 168 h fraction and mean dimensions are used only as a check, but the result is not a parameter-free nucleation or coarsening prediction. \\
\bottomrule
\end{longtable}

\begin{longtable}{p{0.20\textwidth}p{0.28\textwidth}p{0.39\textwidth}}
\caption{Representative processes and outputs used by the validation suite. Values are case definitions, not a universal database.}\label{tab:parameters}\\
\toprule
Process & Representative material/input & Principal checked observable \\
\midrule
\endfirsthead
\toprule
Process & Representative material/input & Principal checked observable \\
\midrule
\endhead
Isotropic grain growth & Pure Al; $\gamma_{\mathrm{HAGB}}=0.324\ \mathrm{J\,m^{-2}}$ & $120^\circ$ junction, parabolic coarsening, repeated seeds \\
Anisotropic grain growth & Orientation-dependent $M$ and $\gamma$ & Boundary selection and morphology \\
Recrystallisation & AA1050-like; activation energy $176\ \mathrm{kJ\,mol^{-1}}$ & Stored-energy direction and Avrami fit \\
Multiphase growth & Three polynomial Gibbs-energy branches & Candidate selection and phase fractions \\
Elastic coupling & Coherent eigenstrain penalty & $E\propto\delta^2$ \\
Al--Cu case & 4 wt.\% Cu; COST-507; 623 K recrystallisation & Equilibrium fractions, calibrated completion time, explicit-PF timestep boundary \\
Al--Cu precipitation & 1.69 at.\% Cu; 453 K; $\theta''/\theta'$ populations & Local 3-D audits plus 3014-particle 120--168 h comparison \\
\bottomrule
\end{longtable}

\section{Notation and terminology glossary}
Symbols are listed by meaning and lifecycle because identical letters can otherwise acquire different interpretations across sharp-interface, phase-field and population descriptions.
\begin{longtable}{@{}>{\raggedright\arraybackslash}p{0.14\textwidth}>{\raggedright\arraybackslash}p{0.47\textwidth}>{\raggedright\arraybackslash}p{0.29\textwidth}@{}}
\caption{Symbols and terms used in the manuscript. ``Persistent'' means stored in $\mathcal S^n$; ``temporary'' means reconstructed from the unchanged step-entry state and discarded after a step.}\label{tab:glossary}\\
\toprule
Symbol or term & Definition & Units or status \\
\midrule
\endfirsthead
\toprule
Symbol or term & Definition & Units or status \\
\midrule
\endhead
$p,q$ & Resident and candidate labels for one local interface evaluation & Pair convention, not phase names \\
$g_i$ & Persistent grain, region or phase-class identity in cell $i$ & Integer label; persistent \\
$\alpha_i$ & Lineal fraction of the current resident along the reconstructed normal; not a general volume-of-fluid geometry & Dimensionless; persistent \\
$\mathbf u_i$ & Enabled material fields such as composition, temperature, dislocation density and orientation & Persistent \\
$I_g$ & Binary indicator of label $g$ & Temporary, categorical \\
$\widetilde I_g$ & Label-conditioned Gaussian projection $K_{\sigma_K}*I_g$ & Temporary, dimensionless \\
$\sigma_K$ & Gaussian standard deviation in grid-cell units & Numerical measurement scale \\
$\ell_K$ & Physical Gaussian length $\sigma_K\Delta x$ & Length \\
$\mathbf n_p$ & Outward unit normal from resident $p$ toward candidate $q$ & Temporary \\
$\kappa_p$ & $\nabla\!\cdot\!\mathbf n_p$; summed principal curvature in three dimensions & $\mathrm{m^{-1}}$ \\
$v_n$ & Boundary velocity along $\mathbf n_p$; positive values expand resident $p$ & $\mathrm{m\,s^{-1}}$ \\
$M_{pq}$ & Sharp-interface mobility returned by the pair constitutive module & $\mathrm{m^4\,J^{-1}\,s^{-1}}$ \\
$\gamma_{pq}$ & Interfacial energy for pair $p,q$; orientation dependent where declared & $\mathrm{J\,m^{-2}}$ \\
$\Delta G_{q-p}$ & Candidate-minus-resident volumetric free-energy difference, $\psi_q-\psi_p$ & $\mathrm{J\,m^{-3}}$ \\
TCKT & Topology-certified kinetic transaction applied to a proposed zero crossing & Shared update rule \\
BGR & Equal-count, pairwise balanced Gaussian exchange after kinetic crossings & Geometry anti-alias step \\
Region identity & A label whose resolved connected component must not split or merge spuriously & For example, a grain \\
Phase class & A label semantics that permits distinct particles and physical coalescence or pinch-off & Topology may change kinetically \\
Verification & Comparison with an analytic identity or controlled numerical reference & Method evidence \\
Validation & Comparison with an observation not used to determine the tested output & External evidence \\
Calibration & Parameter estimation from a declared datum & Agreement is not validation \\
Conditioning & Use of boundary or endpoint observations to fix part of a trajectory & Remaining outputs are conditional \\
Qualitative demonstration & Visual and directional capability test without a numerical accuracy claim & Diagnostic evidence \\
Gate & Predeclared numerical condition that can fail an executable test & Pass/fail \\
Diagnostic & Reported result that does not decide executable acceptance & Report-only \\
Display smoothing & Interpolation or isosurface extraction used only to render a label field & Never used in metrics \\
\bottomrule
\end{longtable}

\section{Evidence hierarchy and reviewer traceability}
Table~\ref{tab:evidence-map} states what each result establishes and, equally importantly, what it cannot establish. This prevents a qualitative morphology, an implementation regression and an external material comparison from being conflated.
\begin{longtable}{@{}>{\raggedright\arraybackslash}p{0.21\textwidth}>{\raggedright\arraybackslash}p{0.24\textwidth}>{\raggedright\arraybackslash}p{0.27\textwidth}>{\raggedright\arraybackslash}p{0.20\textwidth}@{}}
\caption{Reviewer-facing interpretation map for the manuscript evidence.}\label{tab:evidence-map}\\
\toprule
Evidence block & Input/reference & Establishes & Does not establish \\
\midrule
\endfirsthead
\toprule
Evidence block & Input/reference & Establishes & Does not establish \\
\midrule
\endhead
Main Figures~\ref{M-fig:overview} and \ref{M-fig:timestep} & Declared state contract and eight-step update & Architecture, data flow and permitted state changes & Numerical accuracy \\
Main Figures~\ref{M-fig:velocity-subcell} and \ref{M-fig:tckt-bgr} & Drawn label arrays at a junction, a receding cell and four topology cases & How velocity averaging, the sub-cell update, the TCKT rules and the balanced remap act on single cells & Numerical accuracy \\
Main Figures~\ref{M-fig:geometry-pipeline}--\ref{M-fig:gaussian-accuracy} & Binary labels, Gaussian projection and exact circle geometry & Normal sign, curvature construction and kernel-scale bias & Junction or topology accuracy \\
Main Figure~\ref{M-fig:foundations}(a) and Supplementary Figure~\ref{fig:scurved-bicrystal} & Circle, ellipse and square bicrystals & Integrated capillary rate and shape relaxation & Polycrystal statistics \\
Main Figure~\ref{M-fig:foundations}(b) & Five periodic 300-grain networks & Von Neumann--Mullins trend for topology-stable grains & Universal network prefactor \\
Main Figure~\ref{M-fig:foundations}(c) and (d) & Sinusoidal interface and circle refinement & Linear response and grid/kernel error separation & High-order convergence \\
Main Figures~\ref{M-fig:qualitative-2d} and \ref{M-fig:qualitative-3d} & Saved 2-D and 3-D production outputs & Capability and topology-consistent qualitative morphology & Material-specific predictive accuracy \\
Main Figure~\ref{M-fig:benchmark} & Fixed-physics $96^2$--$384^2$ resolution audit followed by nine accuracy-matched domains up to $768^2/1024^2$ cells and 1280 grains on one 24-thread build & The grid selected by declared morphology and PF field-energy gates, measured large-domain runtime and array scaling & A cost advantage when only categorical PF morphology is required, or a universal ranking over other PF formulations \\
Main Figure~\ref{M-fig:gpk-pf} & Identical 80-grain Voronoi state in GPK and explicit Fan--Chen PF & Grain-resolved agreement and divergence through time & A general ranking over all PF formulations \\
Supplementary Figures~S4--S7 & Executable gates, saved raw label states and exact geometry laws & Process kinetics, topology provenance and regression coverage & Independent alloy validation \\
Supplementary Figures~S8--S10 & Literature transcription, local Al--Cu fields and mesh audit & Material inputs, local conservation/topology and discretisation sensitivity & Mesh convergence of every Al--Cu observable \\
Main Figure~\ref{M-fig:alcu-validation} & TEM-conditioned 120 h population and 168 h quantities used only as a check & Conditional consistency of mean fraction and morphology & Parameter-free nucleation/coarsening prediction \\
CALPHAD equilibrium points & Corrected COST-507 TDB queried by pycalphad & Reproducible equilibrium phase inventory & Metastable ageing kinetics \\
Explicit-PF timestep sweep & One matched-grid forward-Euler Fan--Chen implementation & Stability boundary for the stated solver and observable & Failure of implicit, spectral or adaptive PF \\
\bottomrule
\end{longtable}

\paragraph{Minimum information needed to interpret a run.}
A quantitative GPK result is incomplete unless it reports the physical domain and boundary conditions, grid spacing $\Delta x$, Gaussian width $\sigma_K$ and physical length $\ell_K$, smallest quantitative feature relative to $\ell_K$, interface and transport timestep limits, label semantics, constitutive expressions and units for $M$, $\Delta G$ and $\gamma$, initialization and random seed, stopping time, raw-state measurement definition, and whether any datum was used for calibration or conditioning. Tables~\ref{tab:parameters} and \ref{tab:gallery-runs}, the input decks and the machine-readable validation reports provide these items for the calculations in this work.

\paragraph{How a failed or deferred event should be interpreted.}
A deferred TCKT crossing is not evidence that the thermodynamic force vanished; it means the proposed one-cell transaction would violate the declared resolved-region connectivity at that step. The fraction remains at $10^{-6}$ and the complete force is reevaluated. A curvature cap is likewise not a physical saturation law; it records that the requested radius lies below the Gaussian measurement scale. An excluded sub-two-cell nucleus is therefore reported as unresolved, not as a successful prediction of stability. Finally, a smooth rendered contour is not evidence of numerical convergence: convergence is assessed only from raw categorical states, sub-cell volumes and the explicitly reported mesh audits.

\section{Scope exclusions and claim boundaries}
The following claims are excluded:
\begin{itemize}
\item a universal speedup relative to all phase-field implementations;
\item high-order convergence of the present discretization;
\item online CALPHAD equilibrium solves or mobility assessment at every Al--Cu interface step (the solver interpolates a 300-point COST-507 grand-potential table generated offline);
\item heterogeneous-modulus microelasticity between Al and $\theta'$ (the tensor solution uses homogeneous cubic-Al stiffness);
\item a parameter-free nucleation or long-time coarsening prediction (the 120--168 h result conditions population loss and $\theta''$ disappearance on the endpoint observations, computes $\theta'$ inventory by an approximate Cu balance, transfers a local GPK shape response, and treats the 168 h $\theta'$ fraction and morphology as quantities conditioned on the measured data and used only as a check);
\item quantitative junction angles without a grid/window sensitivity study;
\item scale-independent curvature at corners, junctions or interfaces smaller than the Gaussian support;
\item predictive alloy behavior without independently calibrated thermodynamics and mobilities;
\end{itemize}

\bibliographystyle{unsrtnat}
\bibliography{references}